\documentclass[final,5p,times,twocolumn,authoryear]{elsarticle}

\usepackage{amssymb}
\usepackage{amsmath}
\usepackage{siunitx}
\usepackage{nicematrix}
\usepackage{caption}
\usepackage{subcaption}
\usepackage{float}
\usepackage{makecell}
\usepackage[colorlinks=true,linkcolor=black,filecolor=blue,citecolor=red,bookmarksnumbered=true]{hyperref}
\def\bm#1{\mbox{\boldmath{$#1$}}}

\journal{International Journal of Multiphase Flow}

\makeatletter
\def\ps@pprintTitle{%
  \let\@oddhead\@empty
  \let\@evenhead\@empty
  \def\@oddfoot{\reset@font\hfil\thepage\hfil}
  \let\@evenfoot\@oddfoot
}
\makeatother

\begin{document}

\begin{frontmatter}


\title{PLIC-Net: A Machine Learning Approach for 3D Interface Reconstruction in Volume of Fluid Methods}

\author[Cornell]{Andrew Cahaly\corref{cor1}}
\ead{ajc428@cornell.edu}
\cortext[cor1]{Corresponding Author.}
\author[Illinois]{Fabien Evrard}
\author[Cornell]{Olivier Desjardins}

\affiliation[Cornell]{organization={Sibley School of Mechanical and Aerospace Engineering, Cornell University},
            city={Ithaca},
            postcode={14853}, 
            state={NY},
            country={USA}}
\affiliation[Illinois]{organization={Department of Aerospace Engineering, University of Illinois Urbana-Champaign},
            city={Urbana},
            postcode={61801}, 
            state={Il},
            country={USA}}

\begin{abstract}
The accurate reconstruction of immiscible fluid-fluid interfaces from the volume fraction field is a critical component of geometric Volume of Fluid methods. A common strategy is the Piecewise Linear Interface Calculation (PLIC), which fits a plane in each mixed-phase computational cell. However, recent work goes beyond PLIC by using two planes or even a paraboloid. To select such planes or paraboloids, complex optimization algorithms as well as carefully crafted heuristics are necessary. Yet, the potential exists for a well-trained machine learning model to efficiently provide broadly applicable solutions to the interface reconstruction problem at lower costs. In this work, the viability of a machine learning approach is demonstrated in the context of a single plane reconstruction. A feed-forward deep neural network is used to predict the normal vector of a PLIC plane given volume fraction and phasic barycenter data in a $3\times3\times3$ stencil. The PLIC plane is then translated in its cell to ensure exact volume conservation. Our proposed neural network PLIC reconstruction (PLIC-Net) is equivariant to reflections about the Cartesian planes. Training data is analytically generated with $\mathcal{O}(10^6)$ randomized paraboloid surfaces, which allows for the sampling a broad range of interface shapes. PLIC-Net is tested in multiphase flow simulations where it is compared to standard LVIRA and ELVIRA reconstruction algorithms, and the impact of training data statistics on PLIC-Net's performance is also explored. It is found that PLIC-Net greatly limits the formation of spurious planes and generates cleaner numerical break-up of the interface. Additionally, the computational cost of PLIC-Net is lower than that of LVIRA and ELVIRA. These results establish that machine learning is a viable approach to Volume of Fluid interface reconstruction and is superior to current reconstruction algorithms for some cases.\medskip

\noindent © 2024. This manuscript version is made available under the CC-BY-NC-ND 4.0 license.\\
\noindent \url{http://creativecommons.org/licenses/by-nc-nd/4.0}
\end{abstract}




\end{frontmatter}

\section{Introduction}
\label{Introduction}
The accurate representation of the interface that separates immiscible fluids is a key challenge in the simulation of multiphase flows. The difficulty arises from needing to capture detailed interface features and handle topology changes robustly while also conserving volume. Out of the major approaches to this problem, interface capturing methods have shown a great ability to handle complex topology changes, such as in liquid atomization, by using an implicit interface representation. The Volume of Fluid (VOF) method \citep{Hirt} is a popular choice among interface capturing methods since it can be designed to be exactly volume conservative. 

VOF starts by representing the phasic distribution using a binary indicator function, which assigns one of the fluid regions (e.g., the gas) a value of 0 and the other region (e.g., the liquid) a value of 1. When discretized in a finite volume method, the normalized zeroth order spatial moment of the indicator function becomes the liquid volume fraction $\alpha$, which represents in each cell the ratio of liquid volume to the volume of the cell. While each Eulerian grid cell possesses a volume fraction such that $0\leq\alpha\leq1$, cells with $0 < \alpha < 1$ have both phases present and must therefore contain the interface. To transport the volume fraction field, the material transport equation
\begin{equation} \label{eq:alpha}
\frac{\partial \alpha}{\partial t}+\bm{u}\cdot\nabla \alpha=0
\end{equation}
can be used in most circumstances, where $t$ is time and $\bm{u}$ is the fluid velocity assumed to be continuous at the interface. The numerical integration of Eq.~\ref{eq:alpha} is susceptible to various errors such as numerical diffusion, which is equivalent to numerical mixing between the liquid and gas and ultimately leads to the loss of the concept of an interface. To prevent these numerical diffusion errors from smearing the $\alpha$ field while avoiding oscillations, geometric VOF methods first reconstruct a sharp interface from the local $\alpha$ field. The most common technique to accomplish this is the Piecewise Linear Interface Calculation (PLIC) method \citep{Youngs}. PLIC fits linear functions (planes in 3D) in each interfacial cell such that the volume fraction in the cell is exactly conserved. The accuracy of the reconstructed interface, and of the overall simulation, is thus dependent on the accurate calculation of these planes from the volume fractions.

There are many existing optimization algorithms to recover a linear reconstruction from the volume fractions. Two of the most common are the LVIRA and ELVIRA algorithms proposed by \cite{Pilliod}. LVIRA reconstructs the PLIC plane in the center cell of a stencil (typically a $3\times3\times3$ cell stencil). The cost function for the optimization is the $L^2$ loss between the actual volume fraction field in the stencil and the volume fractions created by a plane cutting through the stencil cells. The algorithm imposes the constraint that the plane must reproduce the center cell's volume fraction exactly, and rotates the plane until the loss function is minimized. ELVIRA also uses a stencil of cells to reconstruct the center cell plane. Instead of rotating a plane in the stencil to minimize the $L^2$ volume fraction loss, ELVIRA chooses a normal vector from a list of candidates. The candidate normal vectors are found with the backward, central, and forward differences of the stencil volume fractions in each direction, and the one that produces the lowest $L^2$ volume fraction loss is selected. The LVIRA and ELIVRA methods are frequently used in the literature for high fidelity multiphase simulations (e.g., \cite{Vu} and \cite{Kuhn}).

However, there are limitations to PLIC methods. Aside from the inherent inaccuracy of representing curved surfaces with planes, it is impossible to resolve sub-grid features with a single plane per cell, leading to artificial break-up of ligaments and sheets once their thickness approaches the mesh size. Furthermore, due to the linear nature of the reconstruction, the calculated curvature displays limited convergence with grid size \citep{Han2}. To address these limitations, alternative methods have been developed. The Reconstruction with 2 Panes (R2P) method \citep{Chiodi,Han3} can resolve sub-grid features such as thin films by optimally placing two planes inside a single cell. This technique has been successfully used in cases such as drop breakup in turbulent cross-flows \citep{Han}. Meanwhile, the new Piecewise Parabolic Interface Calculation (PPIC) method of \cite{Evrard} uses a paraboloid representation of the interface in each cell to approximate films and ligaments at the sub-grid scale and provide converging curvature within a single cell. The optimal placement of two planes or of a paraboloid in a cell is a challenging task that requires advanced optimization techniques and carefully crafted heuristics. Instead, a well-trained machine learning model has the potential to provide broadly applicable solutions at lower costs.

Machine Learning (ML) has recently seen increased interest for use in multiphase flow simulations, including in VOF and PLIC. Research by \cite{Qi} used ML to predict interface curvatures from the volume fractions for 2D cases. Using a feed-forward neural network with two hidden layers, a $3\times3$ stencil of volume fractions was used to predict the center cell curvature. The network was trained using data generated from circular interfaces. This work highlighted that neural networks can provide highly accurate predictions of interface quantities in VOF methods. \cite{Patel} expanded on this work by implementing the approach in 3D, using again a $3\times3\times3$ stencil of cells as input and spherical data for network training. The model was shown to achieve similar accuracy to the conventional Height Function approach to interface curvature \citep{Hirt}. \cite{Ondor} further developed the 2D curvature ML method by invoking symmetries to create a neural network that was invariant to symmetry transformations. 

Meanwhile, there have also been attempts to predict properties of a PLIC plane with ML. \cite{Ataei} proposed using a neural network to predict the position of a PLIC plane in its cell given the plane's normal vector and the cell's volume fraction. Like with the other ML approaches discussed, they made use of a feed-forward neural network but only considered data from a single cell as the input. To find the PLIC normal vector $\bm{n}$ for the input, the researchers used the gradient of the volume fraction field $\bm{n}=-\frac{\nabla\alpha}{||\nabla\alpha||}$, which provides an approximate solution. The approach of \cite{Ataei} was successfully tested in a flow solver. Predictions of the plane location given its normal vector may provide computational cost benefits in some applications, but on Cartesian meshes, analytical methods that directly conserve the cell phasic volume with the plane location are fast and guarantied to be volume conservative \citep{Scardovelli}. In contrast, an accurate prediction of the PLIC normal vector is a much more challenging problem. \cite{Nakano} used a graphical neural network to predict interface normal vectors on unstructured meshes with promising accuracy, although they did not demonstrate its use in a flow solver. They used paraboloid training data to more accurately represent the curved nature of the interface. Meanwhile, \cite{Svyetlichnyy} trained a neural network to predict PLIC normal vectors on structured grids. They first considered a 2D case trained with circular data. Svyetlichnyy then extended the approach to 3D by projecting the 3D data onto two 2D coordinate planes and using the 2D neural network to predict the two projected vectors. Then, the 3D vector was calculated from the projections. Svyetlichnyy reported good accuracy with this approach in 2D but did not test the model with 3D multiphase simulations. Thus, if neural networks are to be a viable alternative to techniques such as LVIRA, and be considered for more complicated interface reconstruction methods, their ability to accurately predict usable interface normal vectors for complex 3D simulations remains to be clearly demonstrated.

In the present work, a 3D machine learning interface reconstruction method is developed and used for VOF simulations. This ML approach is referred to as PLIC-Net. PLIC-Net is trained to predict the normal vector of PLIC planes. The resulting planes are then translated in their cells to conserve the phasic volume using analytical relations. A deep feed-forward artificial neural network with 3 hidden layers and 100 neurons per hidden layer is used. The inputs are the phasic volume fractions of a $3\times3\times3$ stencil of cells and the phasic barycenters (i.e., the normalized first spatial moments of the indicator function). This is a notable difference from typical VOF reconstruction algorithms, which usually only consider zeroth order moments, and is a similarity to Moment of Fluid (MOF) methods \citep{Dyadechko}. While MOF uses higher order moments from only a single cell to reconstruct interfaces, PLIC-Net considers neighboring cells in the input data. The training data is generated with analytical paraboloid data, with the average surface normal vector in the cell being the target. The neural network is equivariant to reflections about the Cartesian planes. After evaluating the accuracy of PLIC-Net on test data, it is applied to advection test cases in the Interface Reconstruction Library\footnote{https://github.com/robert-chiodi/interface-reconstruction-library} (IRL) \citep{Chiodi2} and then deployed in the NGA2 flow solver\footnote{https://github.com/desjardi/NGA2} \citep{Desjardins} for several multiphase simulation cases. The results of these simulations are comparable with LVIRA and ELVIRA in accuracy, but show significant reduction in the tendency to generate spurious planes, leading to cleaner numerical break-up of the interface. PLIC-Net is also found to have a lower computational cost than LVIRA and ELVIRA. 

Section 2 describes the methods used to construct PLIC-Net, generate the training data, enforce symmetry, train the neural network, and deploy it in a flow solver. Section 3 presents a range of tests and a discussion of the performance of PLIC-Net. Section 4 contains the conclusions of this work.

\section{Methods}
\label{Methods}

\subsection{Machine Learning Model}
\begin{figure*}
   \centering
       \includegraphics[width=\linewidth]{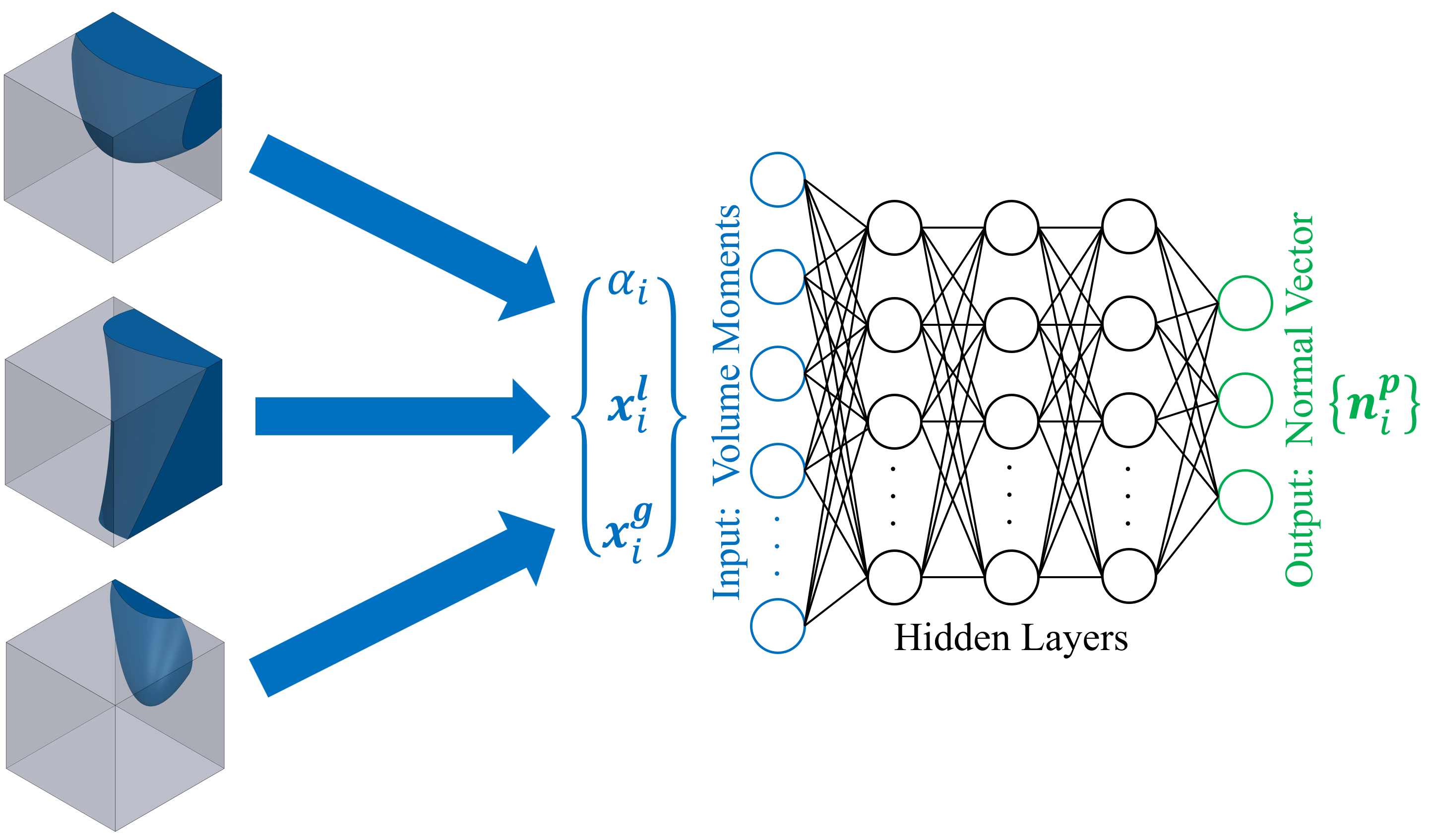}
  \caption{Schematic of the artificial neural network. Three sample paraboloids used to generate the input data ($\alpha$, $\bm{x}^l$, $\bm{x}^g$) are shown on the left. The grey box represents the $3^3$ stencil of cells.}
  \label{network}
\end{figure*}

An artificial Neural Network (NN) is constructed to predict the PLIC normal vector by using the phasic volume fractions and barycenters as input. A $3\times3\times3$ cell stencil is used, providing 27 volume fractions ($\alpha$) as input. Moreover, each stencil cell also provides two phasic barycenters with three coordinates each, leading to 189 total inputs to the neural network. These phasic barycenters correspond to the normalized first order volume moments of each phase in the cell, $\bm{x^l}$ and $\bm{x^g}$. They are transported alongside the volume fraction field, which is commonly done when using geometric VOF advection schemes~\citep{Dyadechko,Le Chenadec,Owkes2}. The NN input layer is followed by three fully connected hidden layers with 100 neurons each. The output layer contains three neurons for the three components of the interface normal vector in the central cell of the stencil. The output layer is linearly connected with the last hidden layer, while the hidden layers and input layer all have the ReLu activation function applied between them to introduce non-linearity. A Mean Squared Error (MSE) loss function is used on the normal vector components, and the Adam optimizer is used for back-propagation. The MSE loss is defined as
\begin{equation}
\begin{aligned}
MSE=&\frac{1}{3N}\sum_{i=1}^{N}\sum_{j=1}^{3}(n^t_{ij}-n^p_{ij})^2,
\end{aligned}
\end{equation}
where $N$ is the number of data, $n^t_{ij}$ is the $j^{th}$ component of the $i^{th}$ target normal vector, and $n^p_{ij}$ is the $j^{th}$ component of the $i^{th}$ predicted normal vector. 
The NN itself can be represented as
\begin{equation}
\begin{aligned}
n^p_{ij}=&\sum_{n=1}^{100}w_{jn}^oReLu\Bigg[\sum_{m=1}^{100}w_{nm}^{h3}ReLu\Bigg[\sum_{k=1}^{100}w_{mk}^{h2}ReLu\Bigg[\sum_{l=1}^{189}(w_{kl}^{h1}X_{il})\\
&+b_k^{h1}\Bigg]+b_m^{h2}\Bigg]+b_n^{h3}\Bigg]+b_j^o,
\end{aligned}
\end{equation}
where $w^o$ are the weights for the output layer, $w^{h1}$, $w^{h2}$, and $w^{h3}$ are the weights for the first, second, and third hidden layers, respectively, $b^o$ are the biases for the output layer, $b^{h1}$, $b^{h2}$, and $b^{h3}$ are the biases for the first, second, and third hidden layers, respectively, and $\bm{X}$ is the input data. $\bm{X}$ was constructed as a vector of the volume fractions and barycenters:  $X_i=[\alpha_{i,1}, \bm{x^l}_{i,1}, \bm{x^g}_{i,1}, \dots, \alpha_{i,27}, \bm{x^l}_{i,27}, \bm{x^g}_{i,27}]$. A schematic of the NN is shown in Figure~\ref{network}. 

As will be discussed later in more detail, the architecture of PLIC-Net has not been optimized for computational efficiency. The purpose of this work is to establish the viability of the machine learning reconstruction approach, and the optimization of the architecture will be the subject of future work. The proposed neural network is likely larger than necessary for this particular problem, including the output layer. Strictly speaking, when predicting a normalized normal vector, only two outputs are required: the vector's two angles. Outputting the three components of the vector allows the neural network to predict the magnitude of the vector if provided with training data that is not normalized. One of the future intended uses of PLIC-Net is to predict such a magnitude that can be used to decide between a one-plane or two-plane reconstruction in a given cell, which explains the choice to predict all three components in this work. However, this will also be the subject of future work and only normalized vectors are considered in this work. 

\subsection{Training Data}
In order to capture a broad range of interface shapes, paraboloids are randomly generated in the $3\times3\times3$ cell stencils to provide training data. While there is no expectation that all phase interfaces are contained in the subset of paraboloid surfaces, paraboloids provide for a sound local approximation of any surface and should greatly outperform spherical training data since the two principal curvatures can be independently varied. Morevover, paraboloids allow for the representation of highly curved shapes that resemble films and ligaments, two commonly found shapes in multiphase flows. The paraboloid surfaces are defined with eight coefficients: two curvature coefficients $a$ and $b$, three orientation angles $\theta$, $\phi$, and $\gamma$, and the three coordinates of an origin position $\bm{d}$. The angle $\theta$ is defined as rotation about the $x$-axis, $\phi$ as rotation about the $y$-axis, and $\gamma$ as rotation about the $z$-axis. The origin is defined as the point of maximum curvature for elliptic paraboloids and as the saddle point for hyperbolic paraboloids. A paraboloid's reference frame $\left(x',y',z'\right)$ can be defined by transforming the Cartesian coordinates $\left(x,y,z\right)$ using
\begin{multline}
x'=(x-d_x)\cos{\theta}\cos{\phi}+(y-d_y)(\sin{\gamma}\sin{\theta}\cos{\phi}\\-\cos{\gamma}\sin{\phi})+(z-d_z)(\cos{\gamma}\sin{\theta}\cos{\phi}+\sin{\gamma}\sin{\phi}),
\end{multline}
\begin{multline}
y'=(x-d_x)\cos{\theta}\sin{\phi}+(y-d_y)(\sin{\gamma}\sin{\theta}\sin{\phi}\\+\cos{\gamma}\cos{\phi})+(z-d_z)(\cos{\gamma}\sin{\theta}\sin{\phi}-\sin{\gamma}\cos{\phi}),
\end{multline}
\begin{multline}
z'=(x-d_x)\sin{\theta}+(y-d_y)\sin{\gamma}\cos{\theta}\\+(z-d_z)\cos{\gamma}\cos{\theta}.
\end{multline}
In this transformed reference frame, the paraboloid is then given by 
\begin{equation}
\begin{aligned}
z'=ax'^2+by'^2.
\end{aligned}
\end{equation}
For each randomly generated paraboloid, $\bm{d}$ is placed within the center cell of the stencil with a uniform random distribution. The cells are given $1\times 1\times1$ dimension. The values of $\theta$, $\phi$, and $\gamma$ are all randomly selected between $0$ and $2\pi$ with a uniform distribution. The choice for the ranges and distribution of $a$ and $b$ can have a large impact on the behavior of PLIC-Net. Thus, multiple datasets are created with different choices: one with the values of $a$ and $b$ randomly selected with a normal distribution with a mean of $0$ and a standard deviation of $0.3$ (leading to Neural Network 1, or NN1), one with $a$ and $b$ values randomly chosen between $-2$ and $2$ with uniform distribution (NN2), and one with $a$ and $b$ set to $0$ to create purely planar data (NN3). The performance of the three networks for PLIC-Net is compared hereinafter. It must be noted that the curvature ranges for NN2 include high curvature cases with corresponding volume moments that could not be produced by a static PLIC calculation. However, during the transport of volume moments in complex simulations with high curvature features, such volume fraction data can be encountered and must be handled. Thus, although PLIC reconstructions become inaccurate for such high curvatures, preparing PLIC-Net for these extreme cases is desirable.

The phasic volume fractions and their first moments resulting from the paraboloids clipping the stencil cells are analytically calculated with the method of \cite{Evrard}. The first moments are normalized by the volume fractions such that they represent the actual barycenter local of their respective phase in a given cell. Each barycenter has its reference origin at the center of its respective cell, meaning that the data and resulting neural network is invariant to translations. Each data-point contains the 27 volume fractions from the stencil along with the liquid and gas barycenters from each stencil cell. The analytical average surface normal vector of the paraboloid in the center cell is calculated to use as the target in the training data. Note that the normal vector magnitude is normalized to 1.

It was found that introducing perturbations to the barycenters in the training data improves the performance of the neural network when used in a flow solver by reducing the number of spurious planes. As a result, for each training paraboloid, there is an 87.5\% chance that the position of the barycenters is perturbed. That perturbation is selected from a random uniform distribution, at most $\pm20\%$, and capped so that the barycenters remain in their respective cell. The volume fractions and corresponding normal vectors are not perturbed in this process. 

\subsection{Reflection Symmetry}
In order to improve the robustness of PLIC-Net, enhance its training on smaller data-sets, and make it equivariant to reflections about the Cartesian planes, the symmetries of the data are considered. Symmetry with respect to the fluid-fluid interface is applied by switching the volume fractions to the opposite phase and swapping the order of the liquid and gas barycenters in the data when the center cell volume fraction is larger than 0.5. The direction of the normal vector is also flipped. To invoke reflection symmetries about the Cartesian planes, the global phase barycenter in the stencil is calculated. First order phasic geometric moments of the whole stencil are calculated given the phasic volume fractions and barycenters from the individual stencil cells. The global phase barycenter is then found by normalizing these stencil geometric moments by the total phasic volume in the stencil. Since the reference frame origin is located at the center of the stencil, the signs of the components of this global barycenter are checked to determine in which of the eight Cartesian octants it is located. If it is not in the first octant (all positive components), then the volume fractions and the phasic barycenters are reflected about the required Cartesian planes such that it brought back into the first octant. For example, if the global barycenter has a negative $x$ component, then the data is reflected about the $y-z$ plane. The same transformations can then be applied to the corresponding normal vector, which results in input data for only the first octant but with simple mappings back to the original octant.

\subsection{Neural Network Training}
The C++ distribution of PyTorch (LibTorch) is used to train PLIC-Net. A million paraboloids are randomly generated, with 70\% used for training, 15\% for validation, and 15\% for testing. The learning rate is set initially to \num{e-4}. A warm restart is performed at 2000 epochs, where the learning rate is decreased to \num{e-5}. Training is executed until the validation loss converged in order to prevent over-fitting to the training data. Training is done synchronously on a high performance computing cluster and parallelized using MPI. Each core is responsible for a batch of the data. This training process is done for each of the three datasets mentioned above, leading to the three neural networks NN1, NN2, and NN3.

\subsection{Flow Solver Implementation}
The machine learning code is implemented directly into the open-source Interface Reconstruction Library (IRL) \citep{Chiodi2}, and translation and deformation advection test cases are run within IRL. The open-source NGA2 flow solver \citep{Desjardins} is then used in combination with IRL to test PLIC-Net in multiphase flow simulations. In order to increase computational efficiency, the code to execute the forward pass of PLIC-Net is also directly written into NGA2. To prepare the data for PLIC-Net, the phase is flipped if the central cell volume fraction is larger than 0.5, the volume fractions and barycenters at any given time step are rotated to the first octant as necessary, and the barycenters are scaled by the cell sizes in each direction. This first octant data is then passed to PLIC-Net. The resulting normal vector is returned from the neural network, rotated back to the original octant, flipped as needed, and the PLIC plane with that normal vector is then located in its cell so that the original volume fraction is conserved exactly. Test cases are run for a falling drop into a pool, the head-on collision between two drops, and the aerobreakup of a droplet exposed to a cross-flow. NGA2 solves the two-phase Navier–Stokes equations with a one-fluid formulation. Away from the interface, the flow solver is second order in time and space and discretely mass, momentum, and kinetic energy conserving. At the interface, local discontinuities slightly degrade the methods and although mass is still discretely conserved and momentum is approximately conserved, kinetic energy conservation is lost. The surface tension force is calculated using the continuous surface force approach \citep{Brackbill} while the interface curvature is obtained from parabolic surface fits \citep{Jibben}. As studied in \cite{Han2}, this curvature estimation provides a good compromise between low errors and computational cost and is shown to have lower errors than the Height Function method \citep{Hirt} for dynamic tests. In all cases (pure advection and NGA2 cases), advection of the volume fraction field is performed using the unsplit geometric advection method of \cite{Owkes}. Note that, as mentioned in Section 2.1, phasic barycenters for each cell are also transported along with the volume fractions, thus making the barycenter data directly available for PLIC-Net to use.

\section{Results and Discussion}
\label{Results}
\subsection{Neural Network Training and Testing Performance}

\begin{figure}[t]
   \centering
       \includegraphics[width=\linewidth]{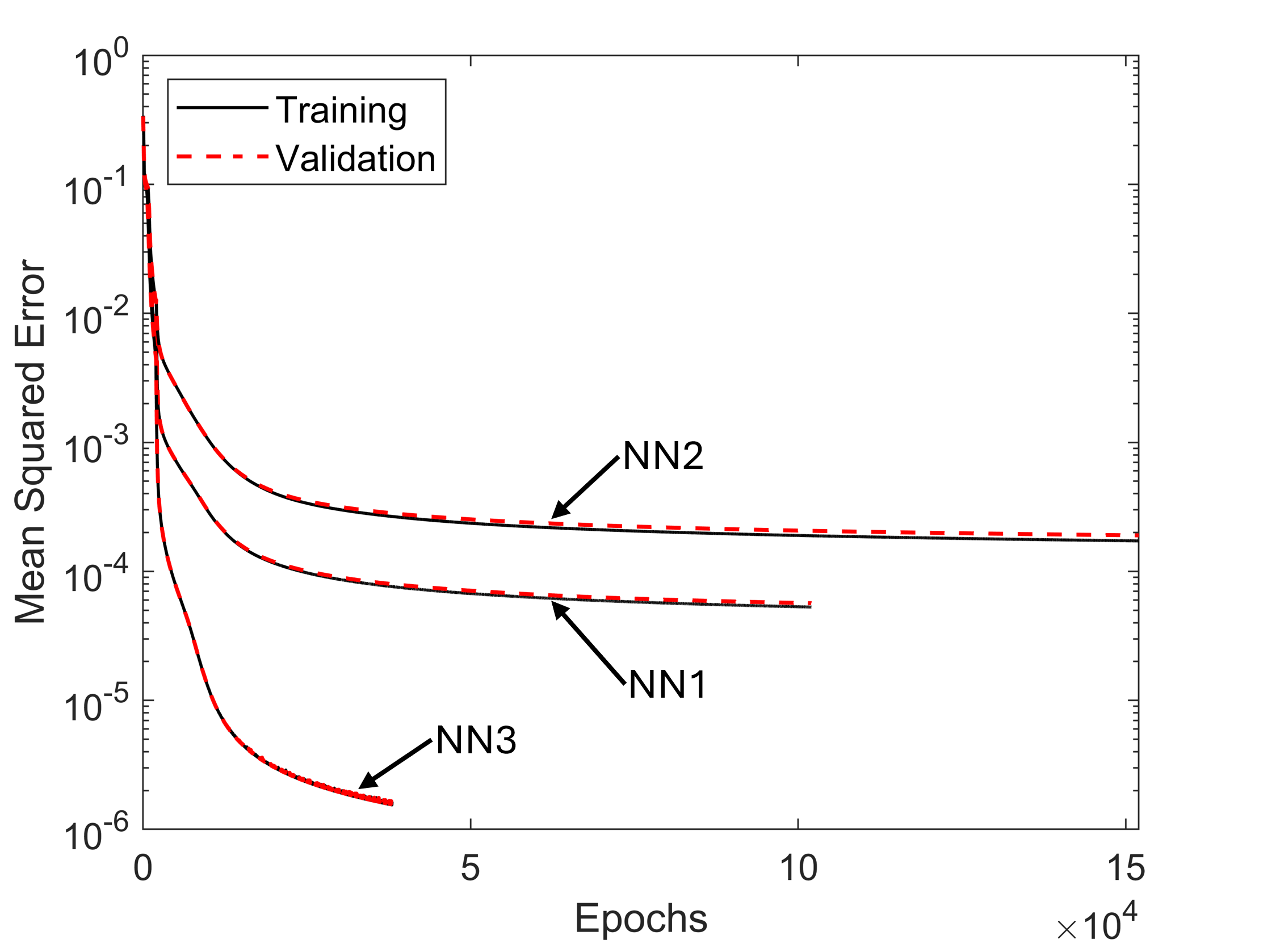}
  \caption{Convergence plot of PLIC-Net's error (MSE) over the training epochs for the training and validation datasets}
  \label{training_plot}
\end{figure}

Figure~\ref{training_plot} shows the MSE loss over the epochs for the training and validation datasets, which are very close to each other during training for all three of the neural networks. In NN1, the training loss converges to \num{5.3e-5} after 102,000 epochs. NN2 converges to \num{1.7e-4} after 152,000 epochs, and NN3 converges to \num{1.6e-6} after 38,000 epochs. It is not surprising that NN3 converges to the lowest error in the fewest epochs since it has purely planar training data, which results in significantly simpler data to learn. Conversely, NN2 has the highest loss due to its wider range of input data.

\begin{figure*}[t]
   \centering
   \begin{subfigure}{\textwidth}
       \centering
       \includegraphics[width=.33\linewidth]{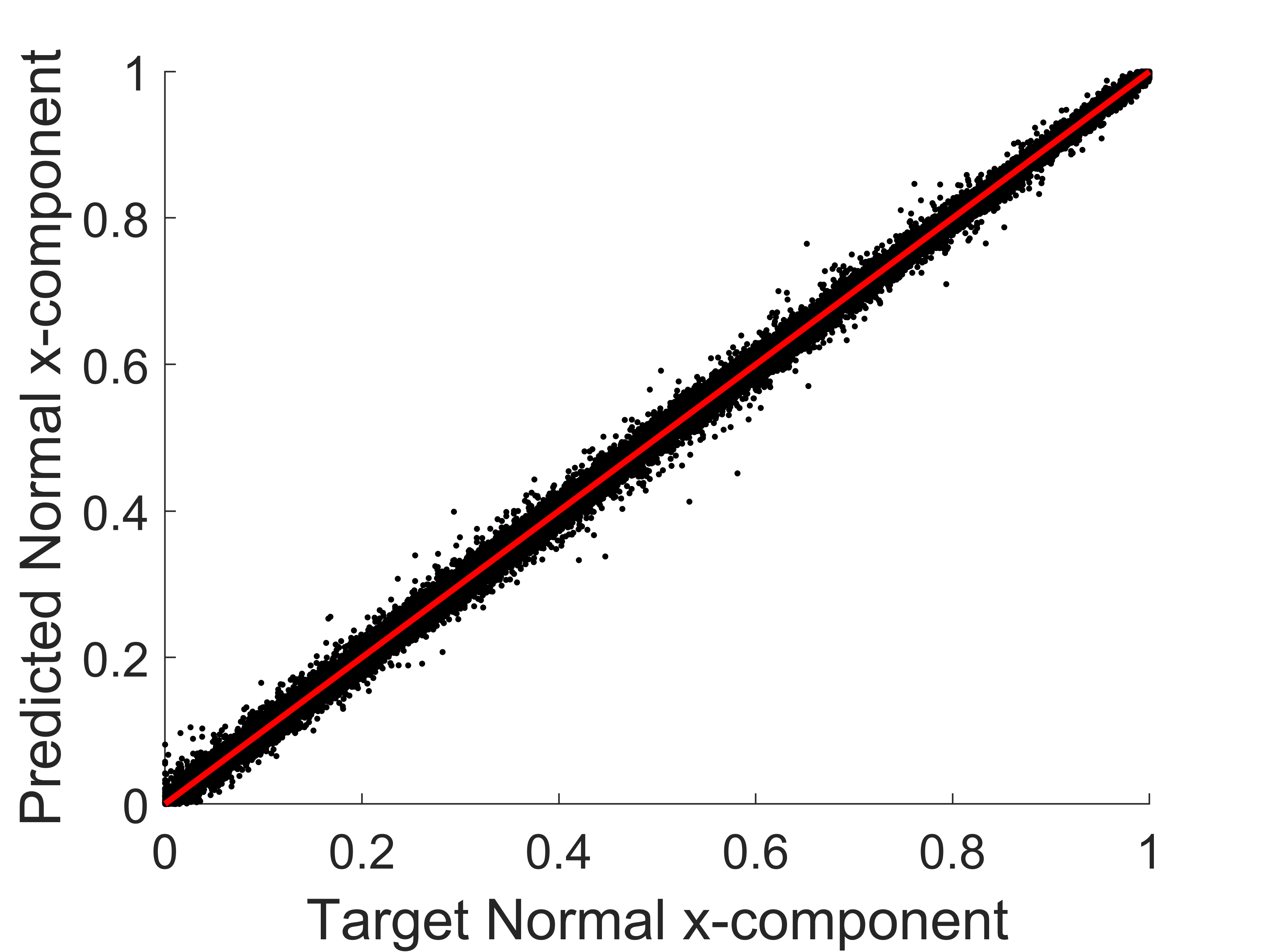}
       \includegraphics[width=.33\linewidth]{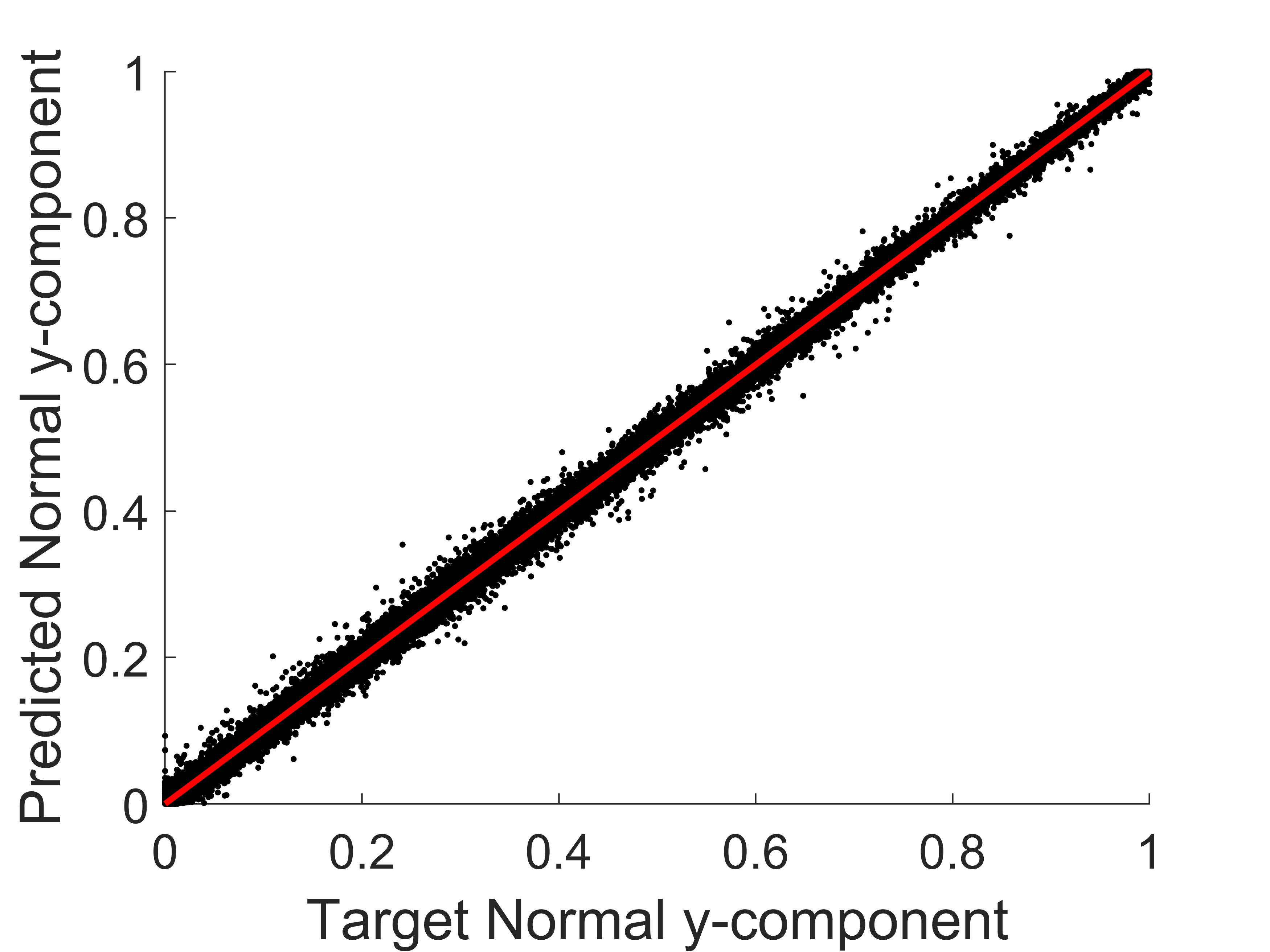}
       \includegraphics[width=.33\linewidth]{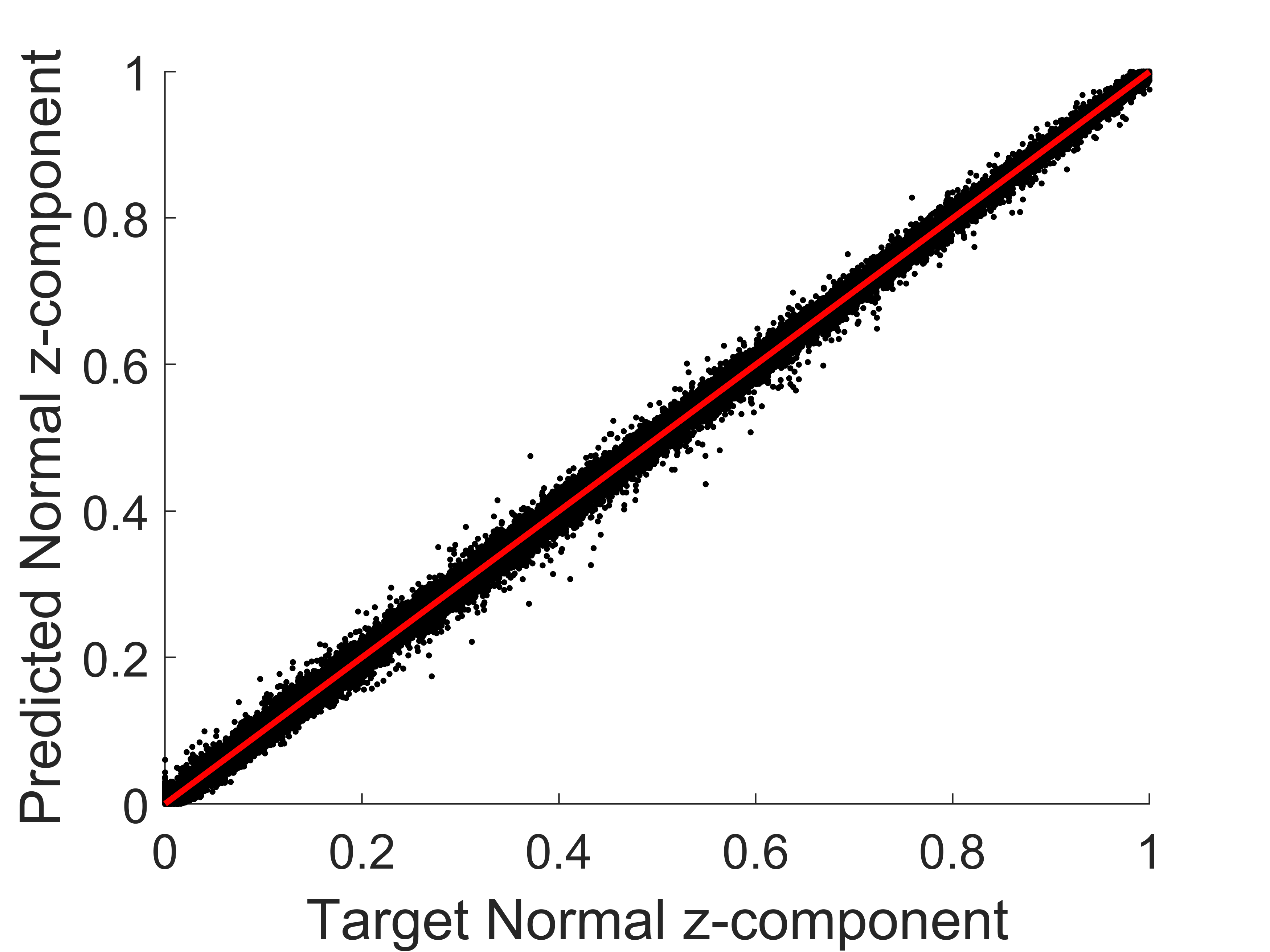}
       \caption{NN1}
       \label{testing_plot:sub1}
   \end{subfigure}%
   \\
   \begin{subfigure}{\textwidth}
       \centering
       \includegraphics[width=.33\linewidth]{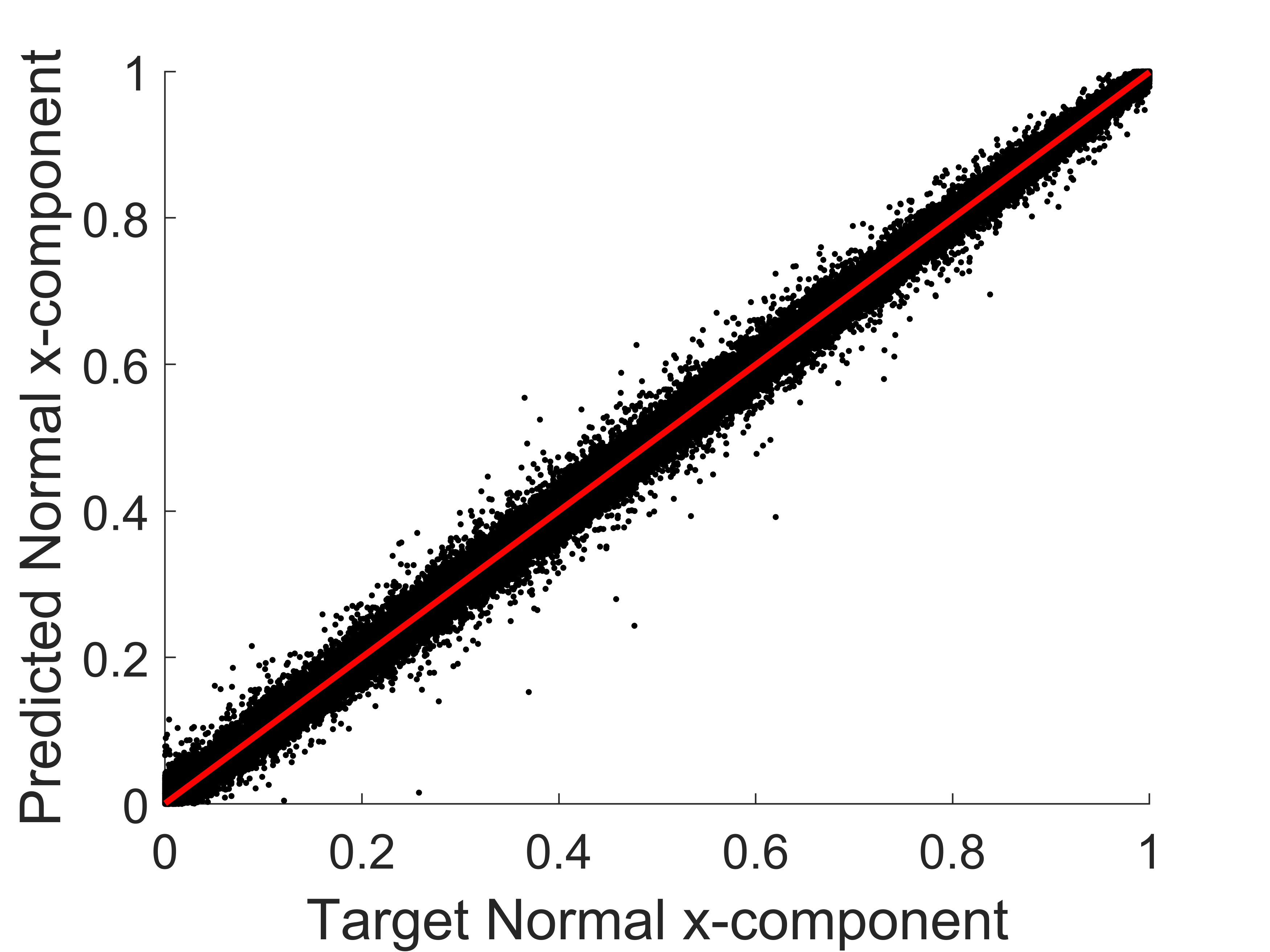}
       \includegraphics[width=.33\linewidth]{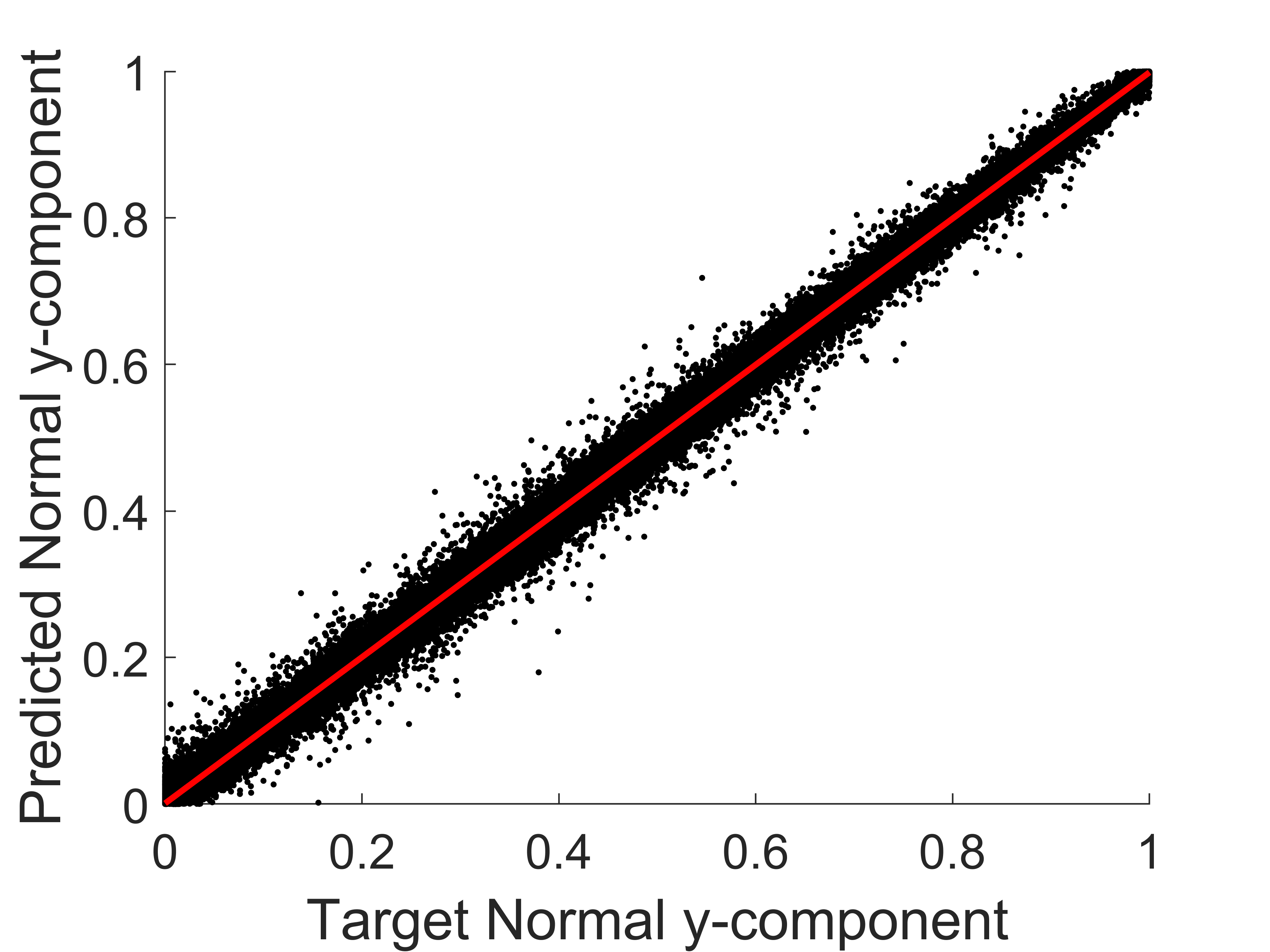}
       \includegraphics[width=.33\linewidth]{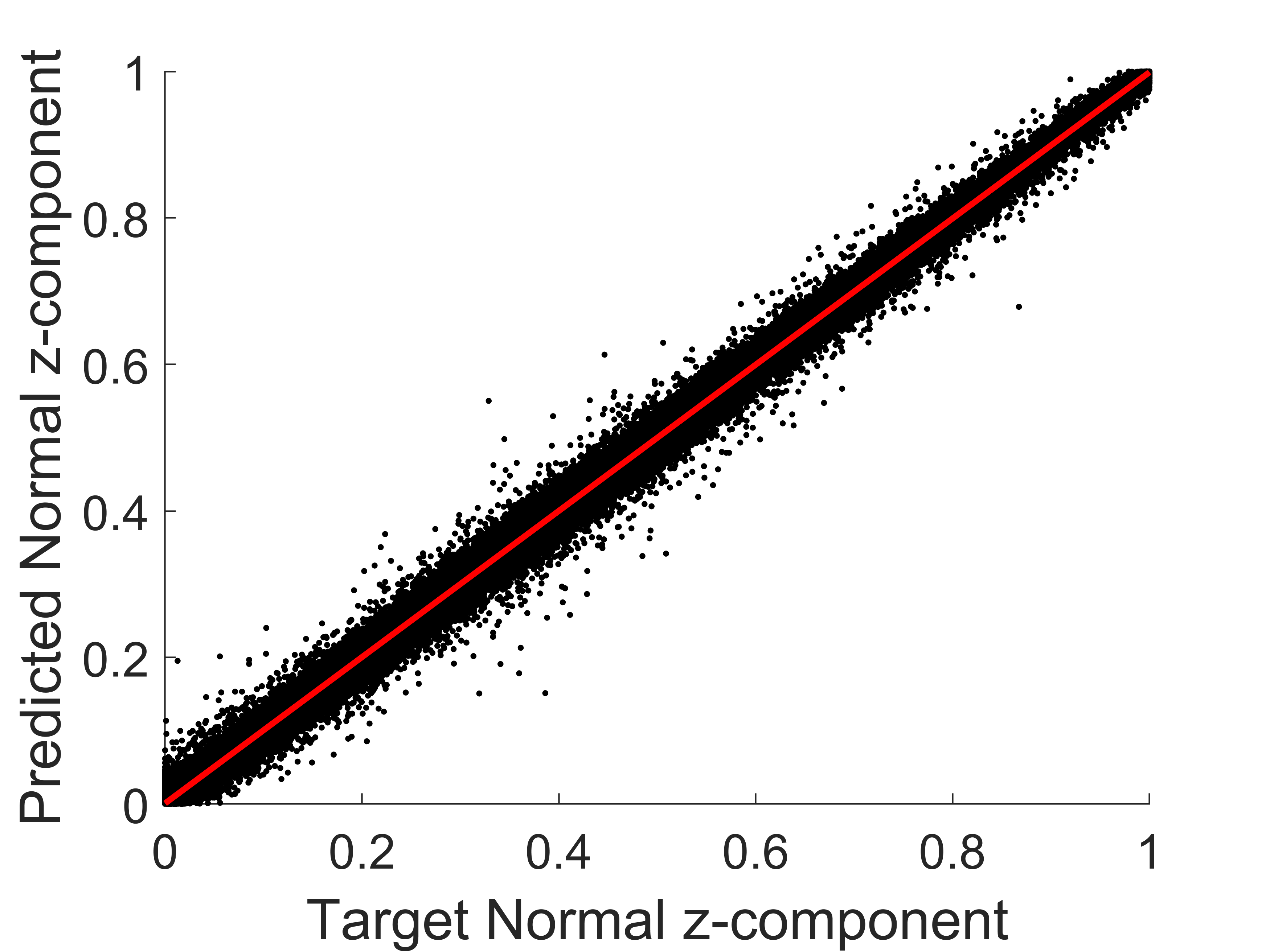}
       \caption{NN2}
       \label{testing_plot:sub2}
   \end{subfigure}%
  \\
   \begin{subfigure}{\textwidth}
       \centering
       \includegraphics[width=.33\linewidth]{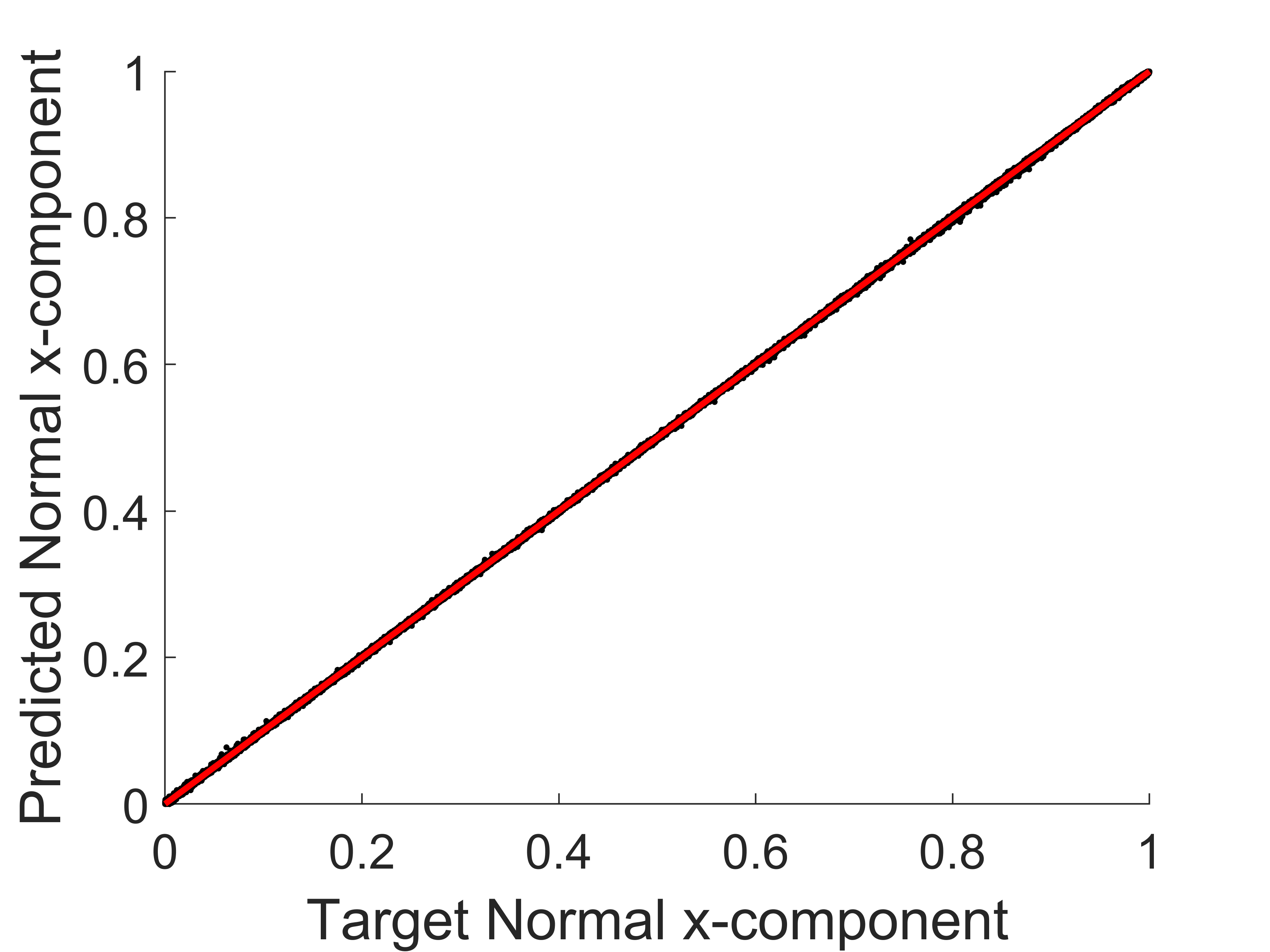}
       \includegraphics[width=.33\linewidth]{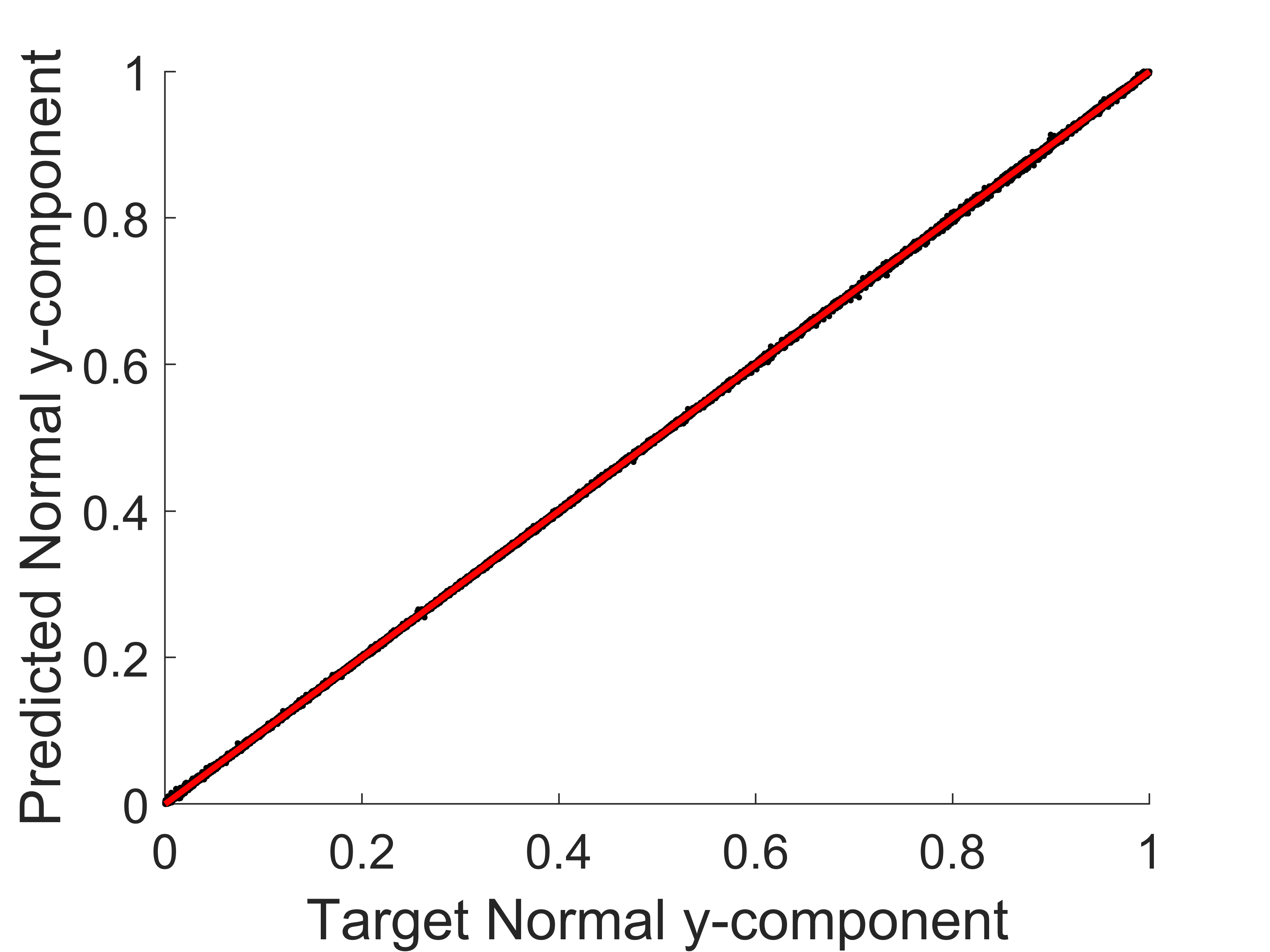}
       \includegraphics[width=.33\linewidth]{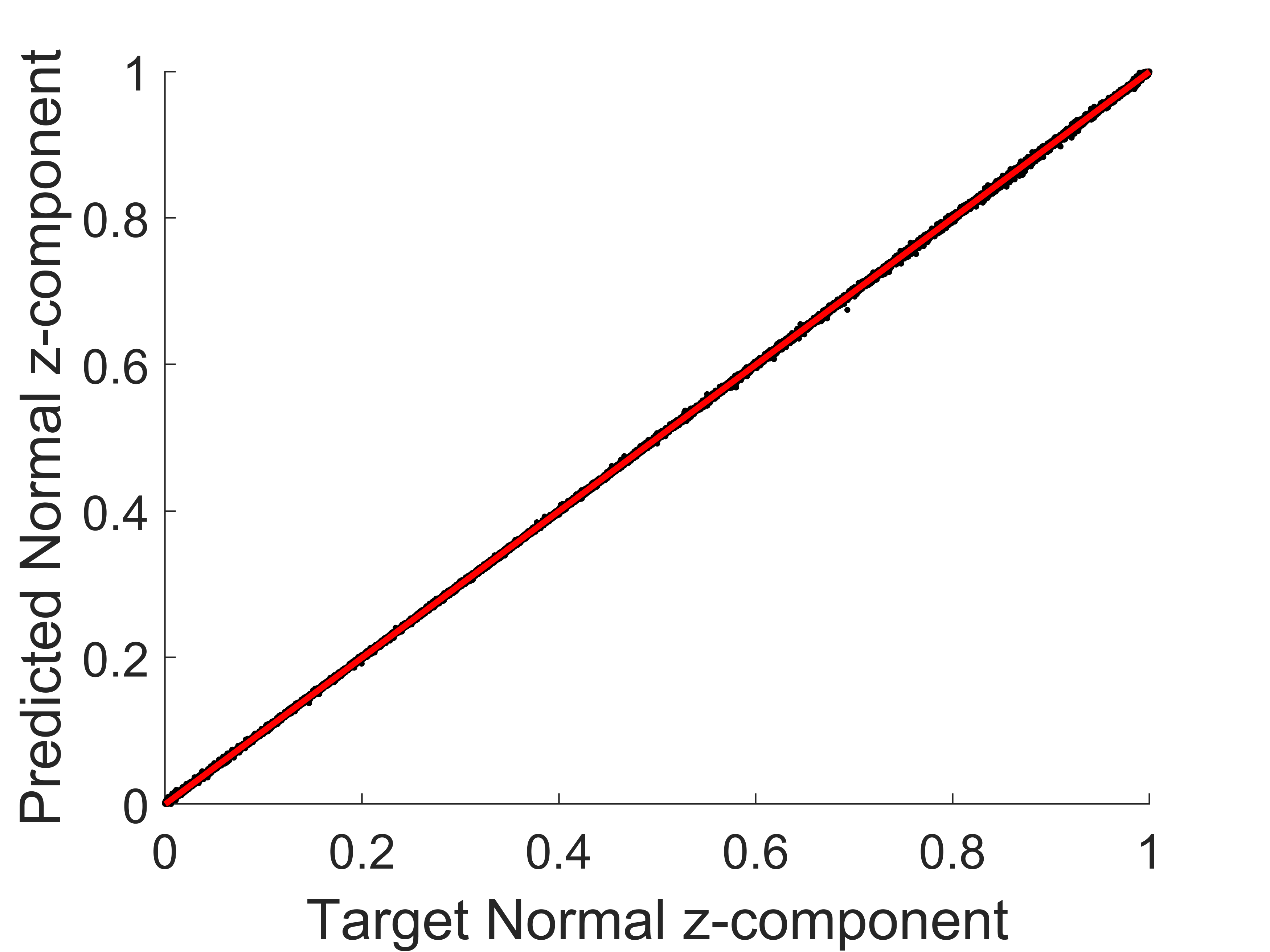}
       \caption{NN3}
       \label{testing_plot:sub3}
   \end{subfigure}%
  \caption{Scatter plots of the predicted normal vector components vs. the target normal vector components from the test datasets (black dots). The line of best linear fit is shown in red in each plot.}
  \label{testing_plot}
\end{figure*}

Figure~\ref{testing_plot} shows plots of the predicted normal vectors from the test datasets against the target normal vectors. Tables~\ref{fits1},~\ref{fits2}, and~\ref{fits3} contain the corresponding equations for the lines of best fit and their $R^2$ values, in which $n^p_x$, $n^p_y$, and $n^p_z$ are the $x$, $y$, and $z$ components of the predicted normal vector, respectively, and $n^t_x$, $n^t_y$, and $n^t_z$ are the $x$, $y$, and $z$ components of the target normal vector, respectively.
\begin{table}[H]
\centering
\caption{Linear regression equations and $R^2$ values for the test dataset on NN1.}
\vspace{-0.8em}

\begin{tabular}{|l|l|l|l|}
\hline
           & \makecell{$x$-component} & \makecell{$y$-component} & \makecell{$z$-component} \\ \hline
\rule{0pt}{4ex}   \makecell{Linear\\Fit} & \makecell[l]{$n^p_x=0.9995n^t_x$\\$+\num{2.78e-4}$} & \makecell[l]{$n^p_y=0.9992n^t_y$\\$+\num{3.24e-4}$} &    \makecell[l]{$n^p_z=0.9992n^t_z$\\$+\num{3.18e-4}$}\\ \hline
\rule{0pt}{2.5ex}   \makecell{$R^2$} & \makecell{0.9995}      & \makecell{0.9993}      & \makecell{0.9993}      \\ \hline
\end{tabular}
\label{fits1}
\end{table}

\begin{table}[H]
\centering
\caption{Linear regression equations and $R^2$ values for the test dataset on NN2.}
\vspace{-0.8em}
\begin{tabular}{|l|l|l|l|}
\hline
           & \makecell{$x$-component} & \makecell{$y$-component} & \makecell{$z$-component} \\ \hline
\rule{0pt}{4ex}   \makecell{Linear\\Fit} & \makecell[l]{$n^p_x=0.9983n^t_x$\\$+\num{9.20e-4}$} & \makecell[l]{$n^p_y=0.9979n^t_y$\\$+\num{9.07e-4}$} &    \makecell[l]{$n^p_z=0.9981n^t_z$\\$+\num{9.31e-4}$}\\ \hline
\rule{0pt}{2.5ex}   \makecell{$R^2$} & \makecell{0.9982}      & \makecell{0.9979}      & \makecell{0.9979}      \\ \hline
\end{tabular}
\label{fits2}
\end{table}

\begin{table}[H]
\centering
\caption{Linear regression equations and $R^2$ values for the test dataset on NN3.}
\vspace{-0.8em}
\begin{tabular}{|l|l|l|l|}
\hline
           & \makecell{$x$-component} & \makecell{$y$-component} & \makecell{$z$-component} \\ \hline
\rule{0pt}{4ex}   \makecell{Linear\\Fit} & \makecell[l]{$n^p_x=1.0000n^t_x$\\$-\num{2.10e-4}$} & \makecell[l]{$n^p_y=1.0000n^t_y$\\$-\num{5.31e-5}$} &    \makecell[l]{$n^p_z=1.0000n^t_z$\\$-\num{3.07e-4}$}\\ \hline
\rule{0pt}{2.5ex}   \makecell{$R^2$} & \makecell{1.0000}      & \makecell{1.0000}      & \makecell{1.0000}      \\ \hline
\end{tabular}
\label{fits3}
\end{table}

\begin{figure*}[!t]
   \centering
   \begin{subfigure}{.25\textwidth}
       \centering
       \includegraphics[width=\linewidth, clip]{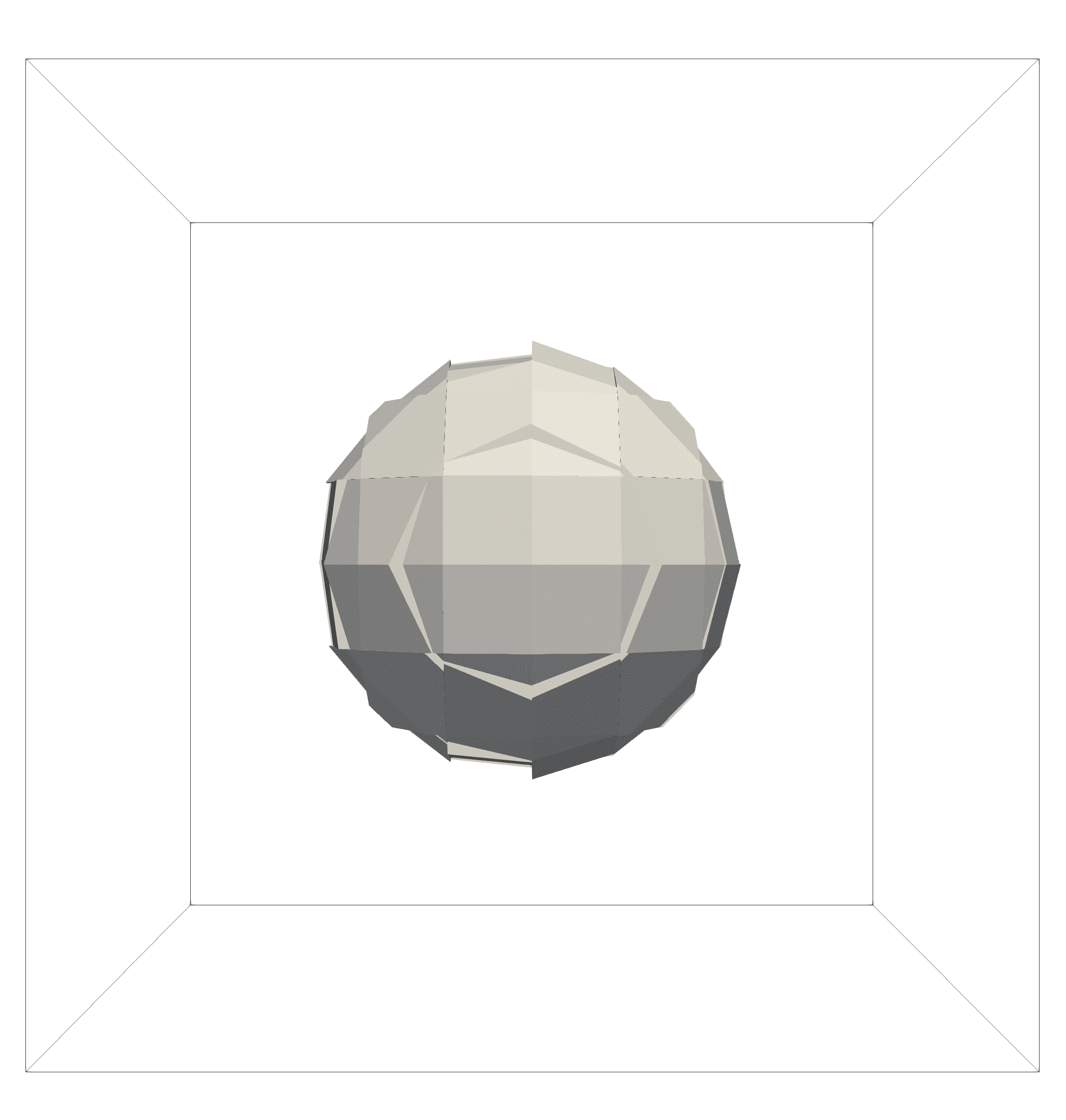}
       \caption{ $t=0$}
       \label{translation_plot:sub1}
   \end{subfigure}%
   \begin{subfigure}{.25\textwidth}
       \centering
       \includegraphics[width=\linewidth, clip]{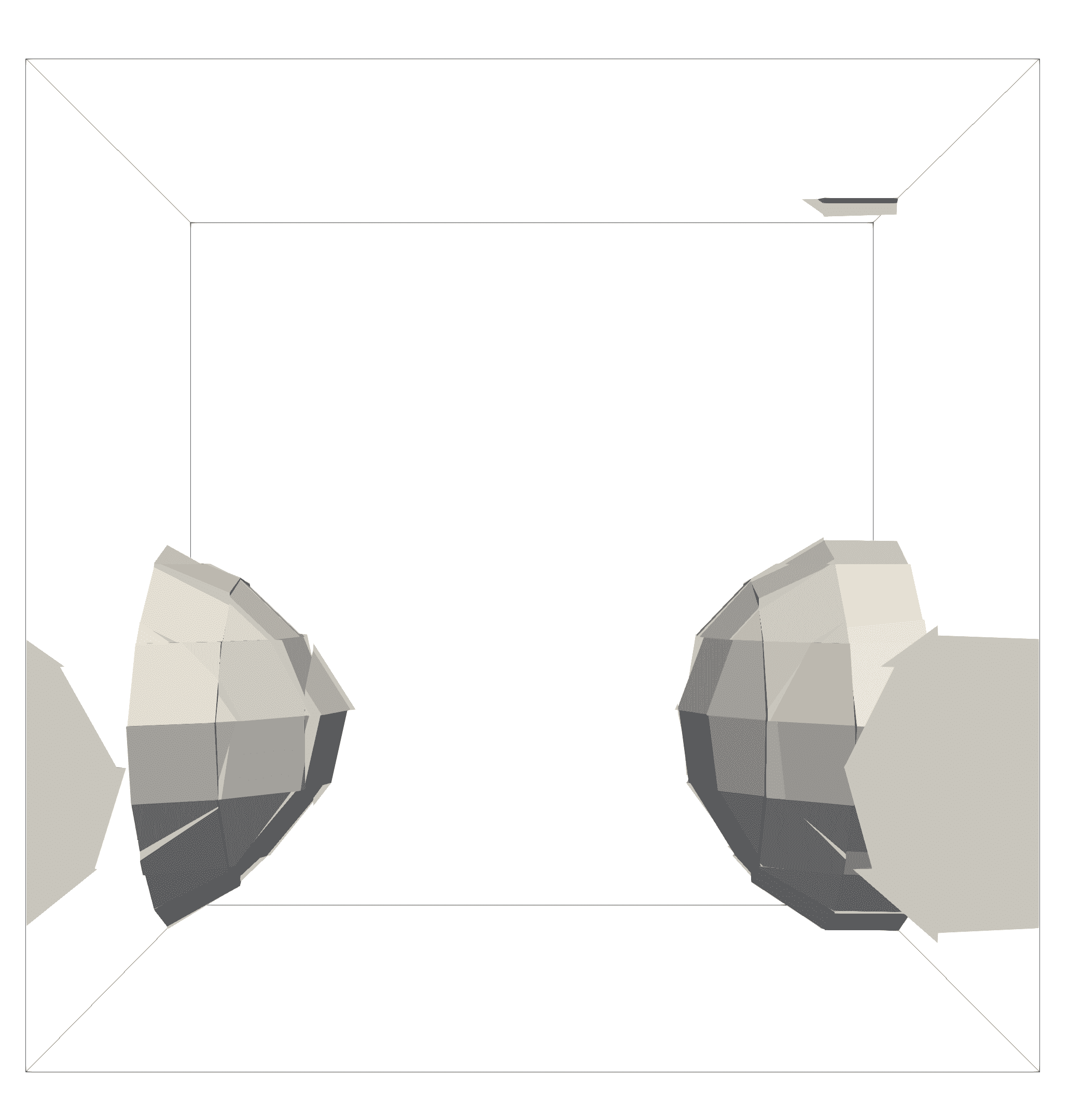}
       \caption{ $t = 2.25$}
       \label{translation_plot:sub2}
   \end{subfigure}%
   \begin{subfigure}{.248\textwidth}
       \centering
       \includegraphics[width=\linewidth, clip]{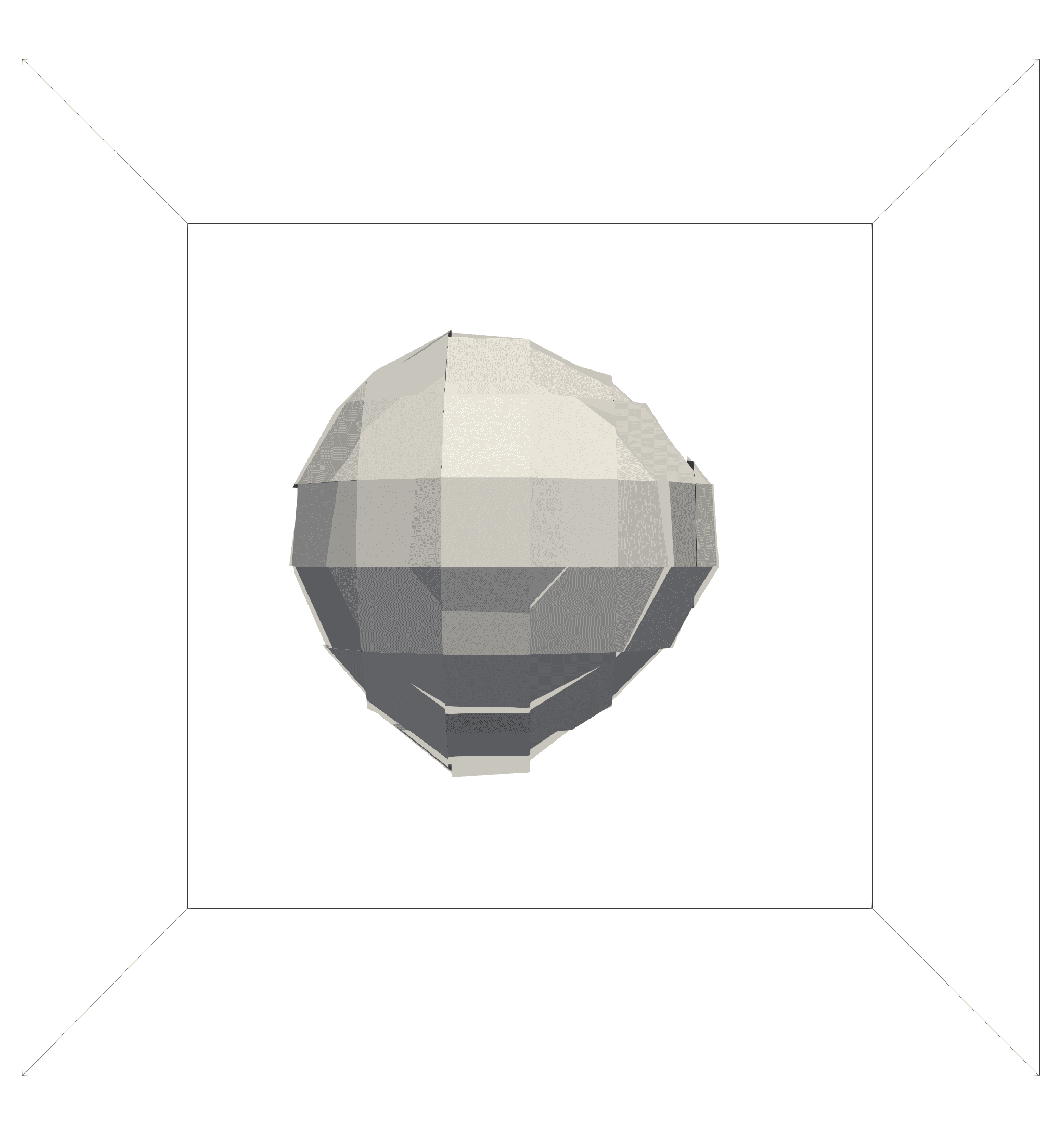}
       \caption{ $t = 3$}
       \label{translation_plot:sub3}
   \end{subfigure}%
   \\
   \centering
   \begin{subfigure}{.25\textwidth}
       \centering
       \includegraphics[width=\linewidth, clip]{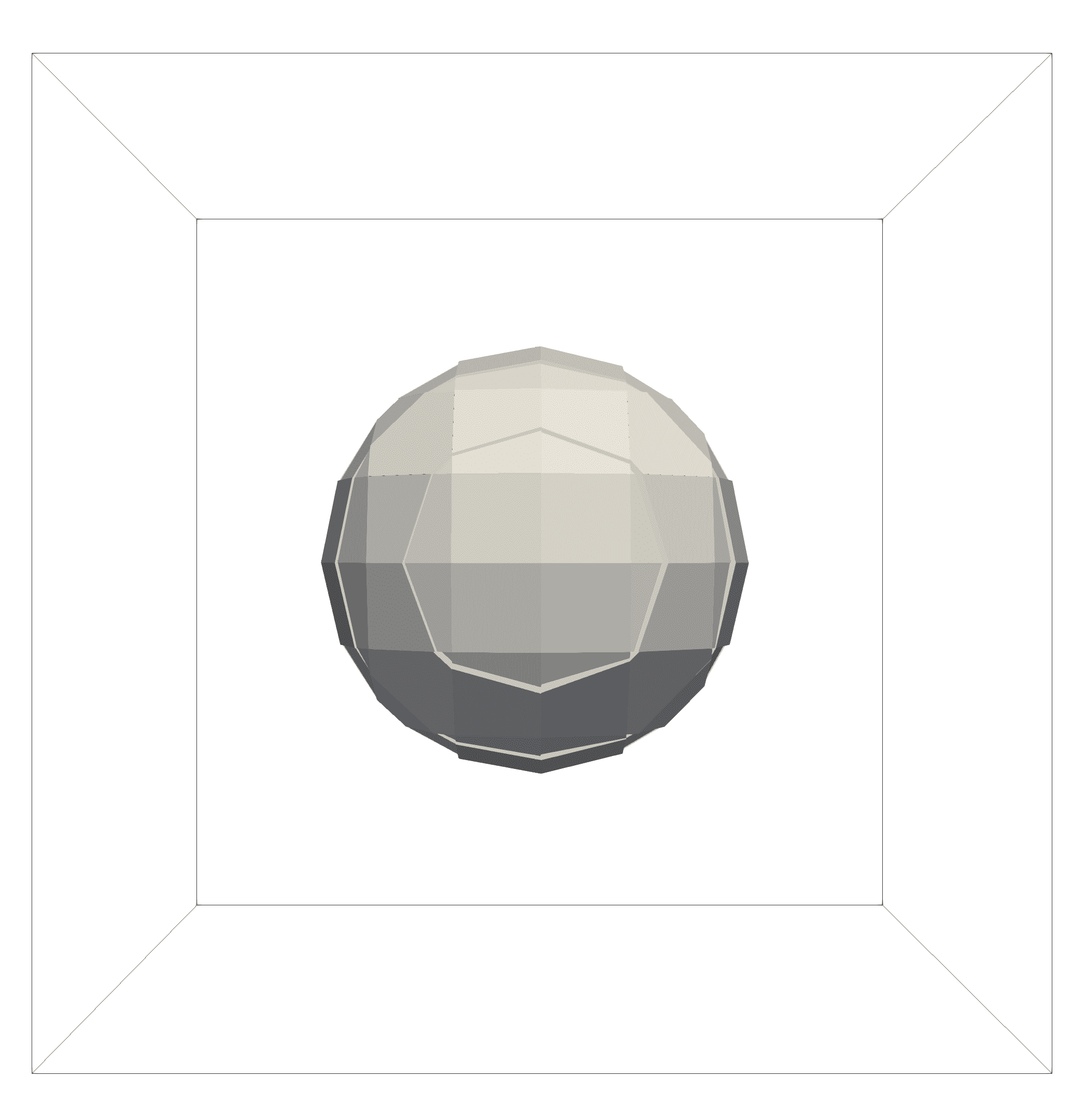}
       \caption{ $t = 0$}
       \label{translation_plot:sub4}
   \end{subfigure}%
   \begin{subfigure}{.25\textwidth}
       \centering
       \includegraphics[width=\linewidth, clip]{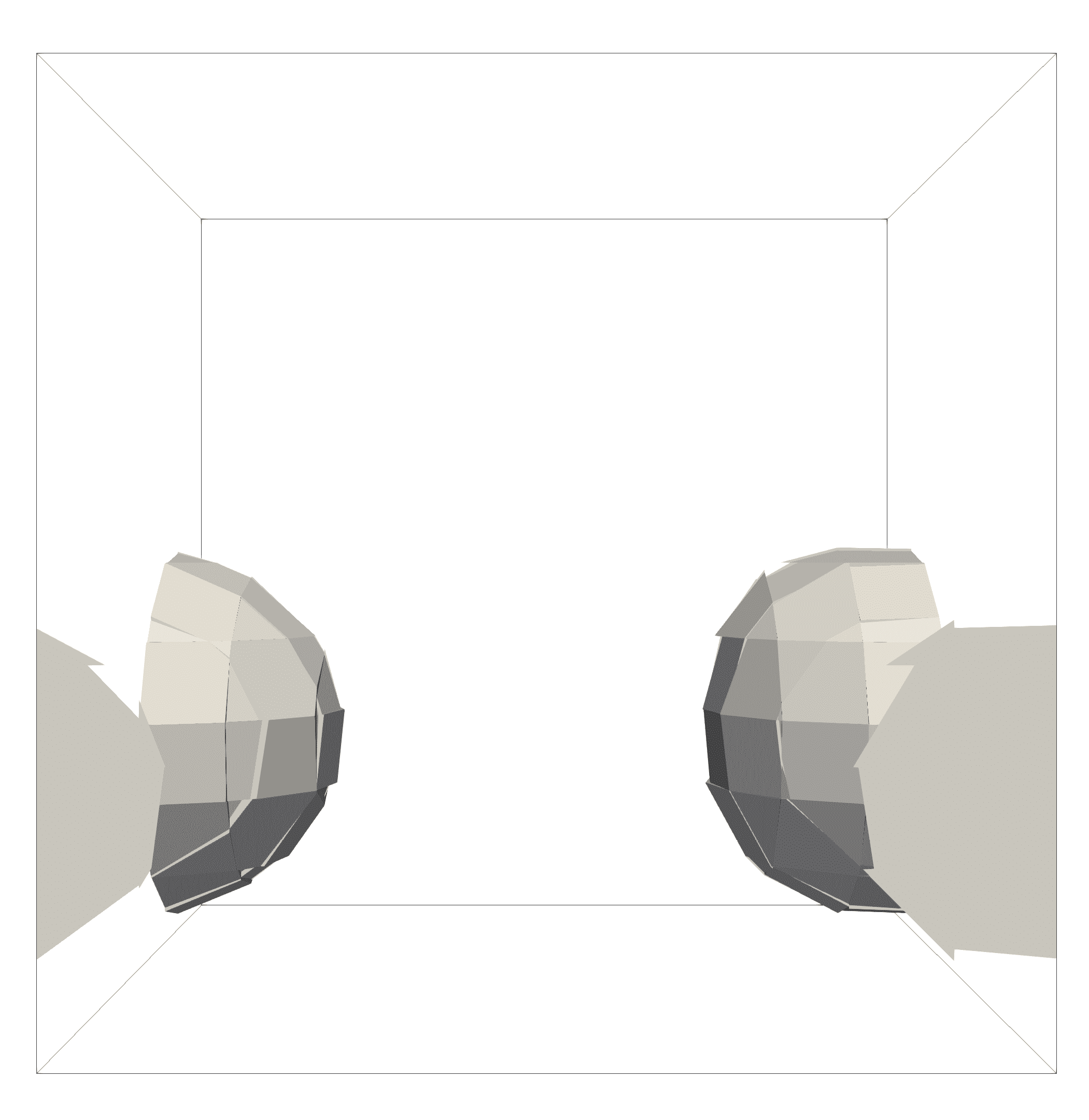}
       \caption{ $t = 2.25$}
       \label{translation_plot:sub5}
   \end{subfigure}%
   \begin{subfigure}{.25\textwidth}
       \centering
       \includegraphics[width=\linewidth, clip]{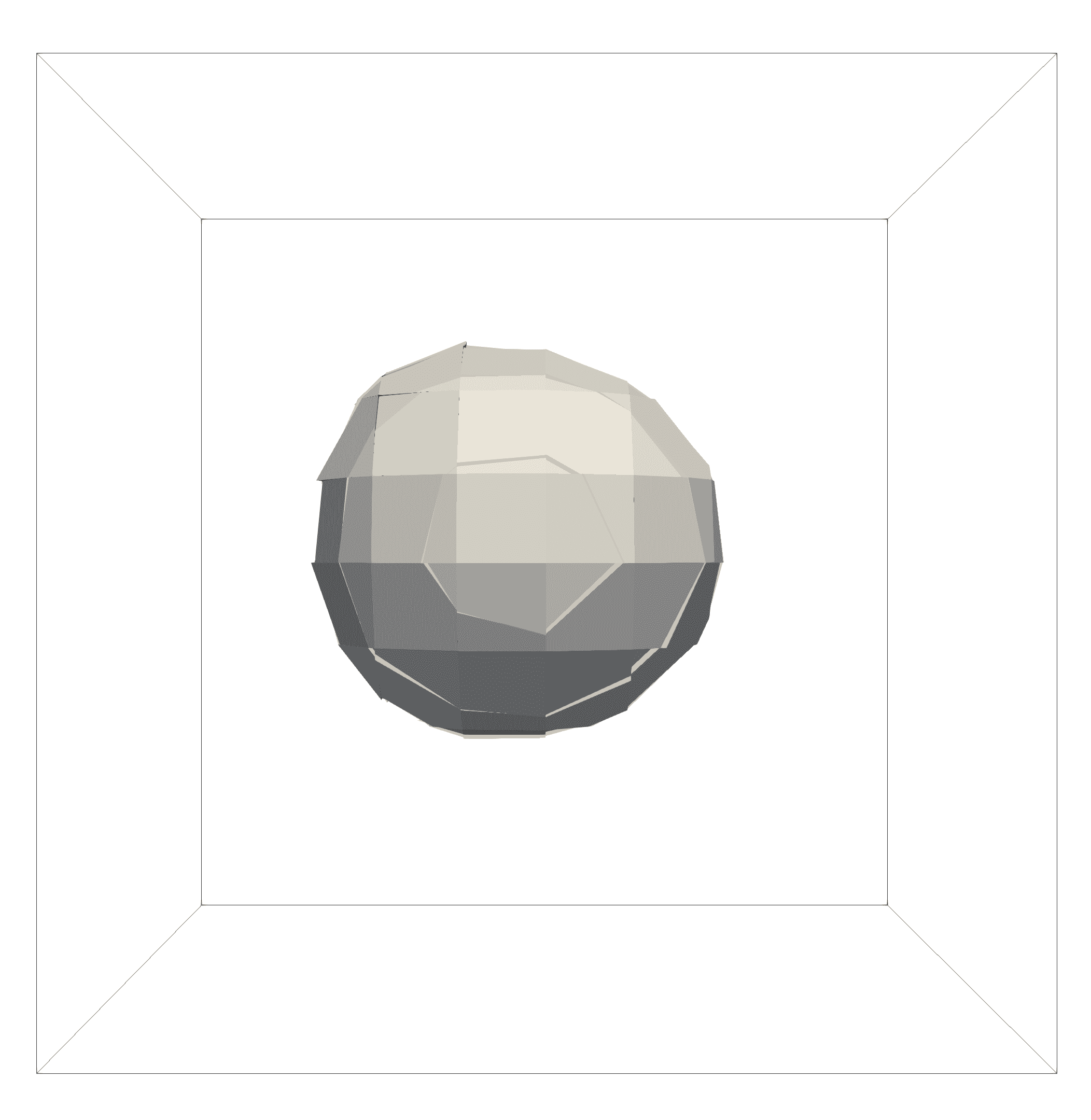}
       \caption{ $t = 3$}
       \label{translation_plot:sub6}
   \end{subfigure}%
  \caption{Time series of the translation advection case. (a), (b), and (c) are with LVIRA and (d), (e), and (f) are with NN1. All times are non-dimensional.}
  \label{translation_plot}
\end{figure*}

\begin{figure}
  \begin{center}
   \includegraphics[width=8.8cm]{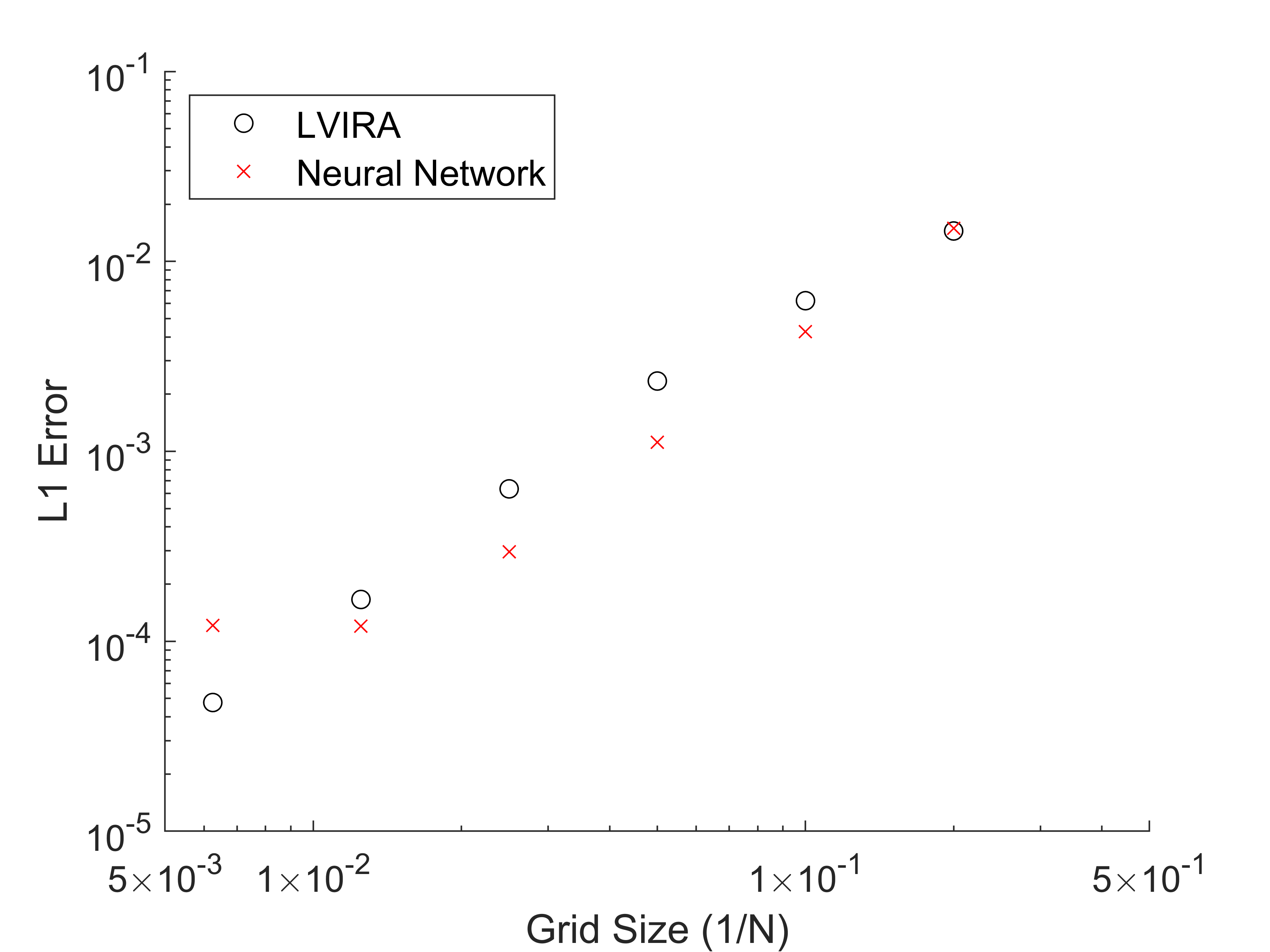}
  \end{center}
  \caption{Convergence plot of the shape error with grid size for the translation advection case with LVIRA and with NN1. $N$ is the number of grid cells in each coordinate direction.}
  \label{convergence_plot}
\end{figure}

\begin{figure}
   \centering
   \begin{subfigure}{.24\textwidth}
       \centering
       \includegraphics[width=\linewidth, clip]{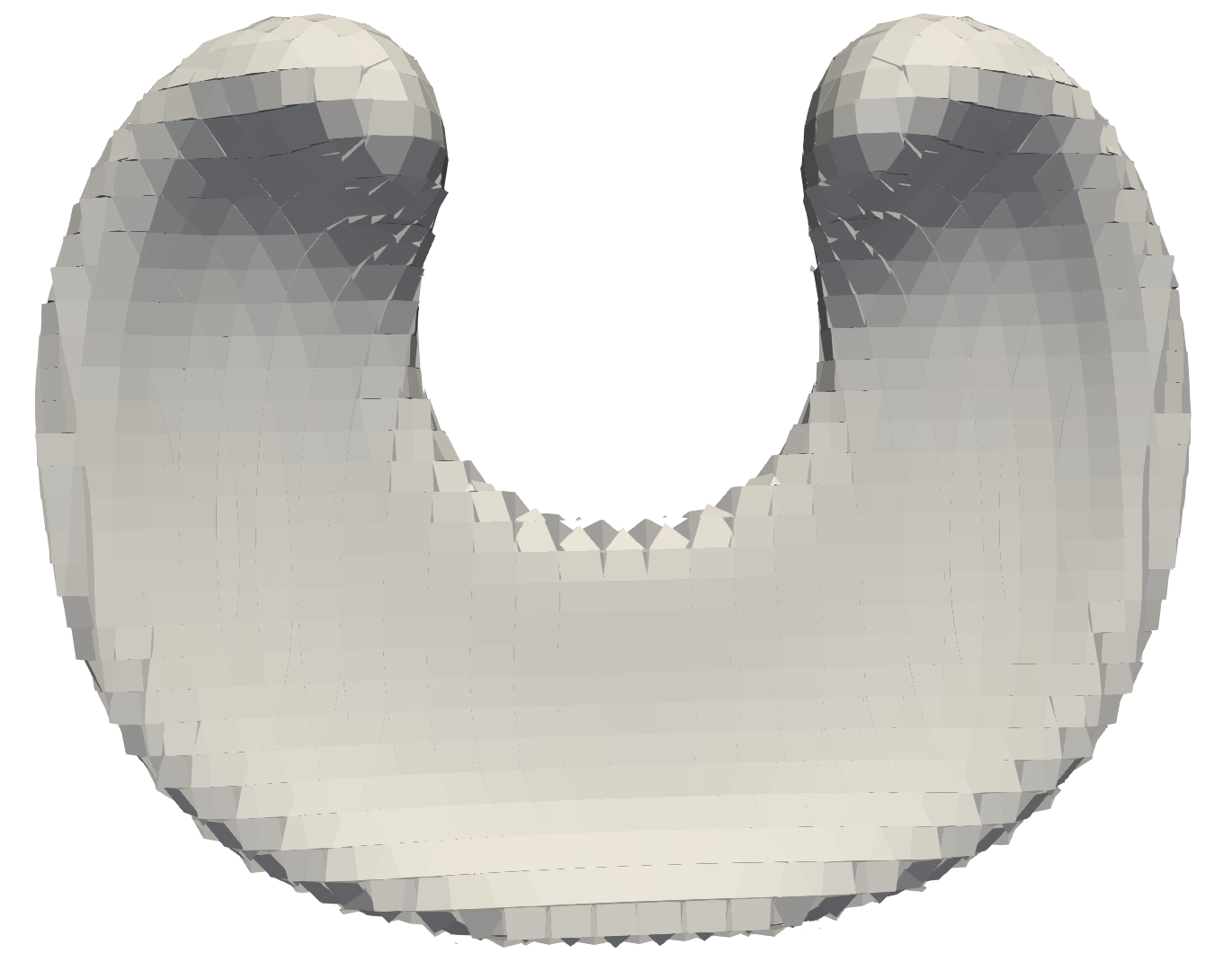}
       \caption{LVIRA, $t = 0.75$}
       \label{deform_plot:sub1}
   \end{subfigure}%
   \begin{subfigure}{.24\textwidth}
       \centering
       \includegraphics[width=\linewidth, clip]{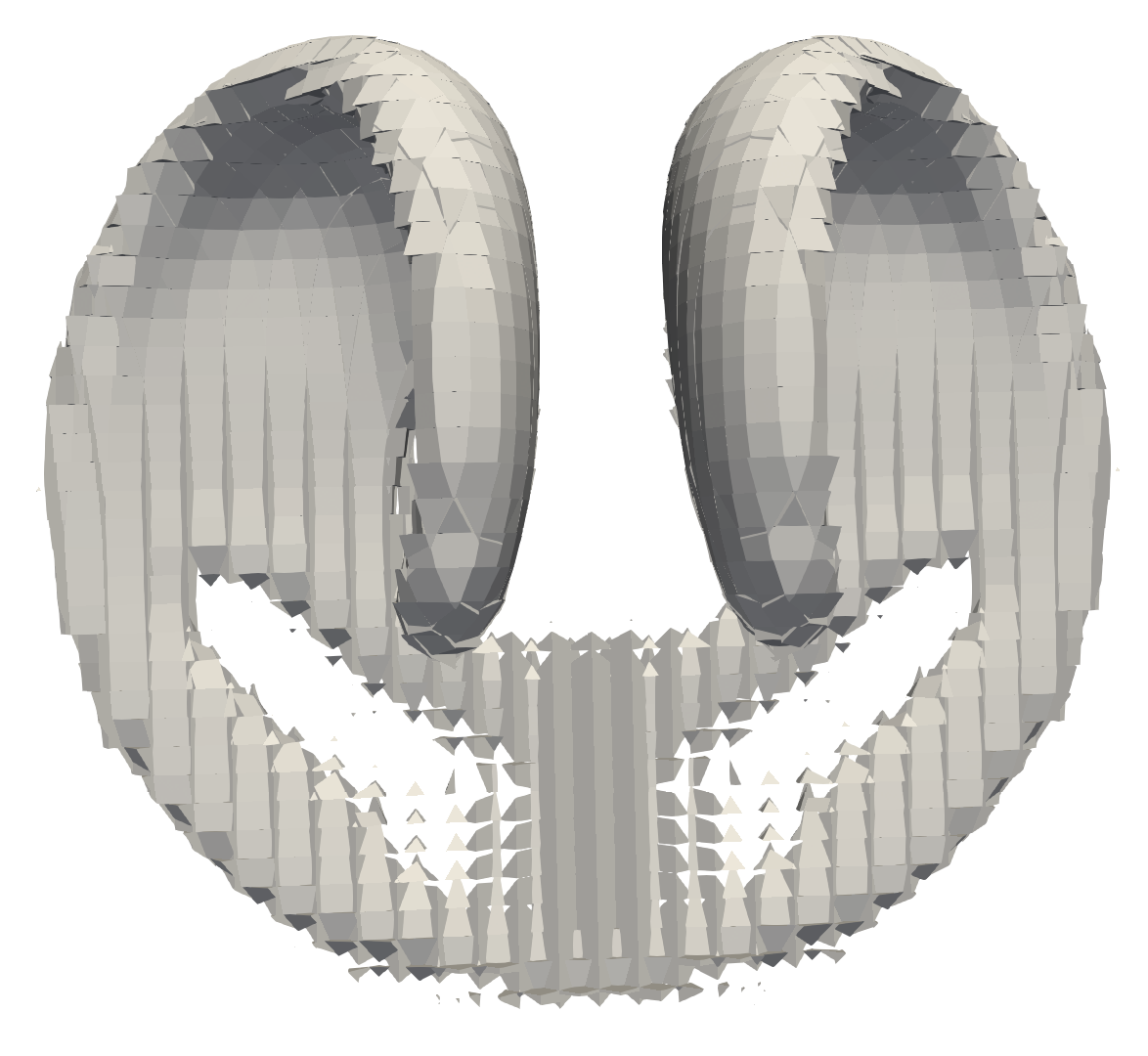}
       \caption{LVIRA, $t = 1.5$}
       \label{deform_plot:sub2}
   \end{subfigure}%
   \\
   \centering
   \begin{subfigure}{.24\textwidth}
       \centering
       \includegraphics[width=\linewidth, clip]{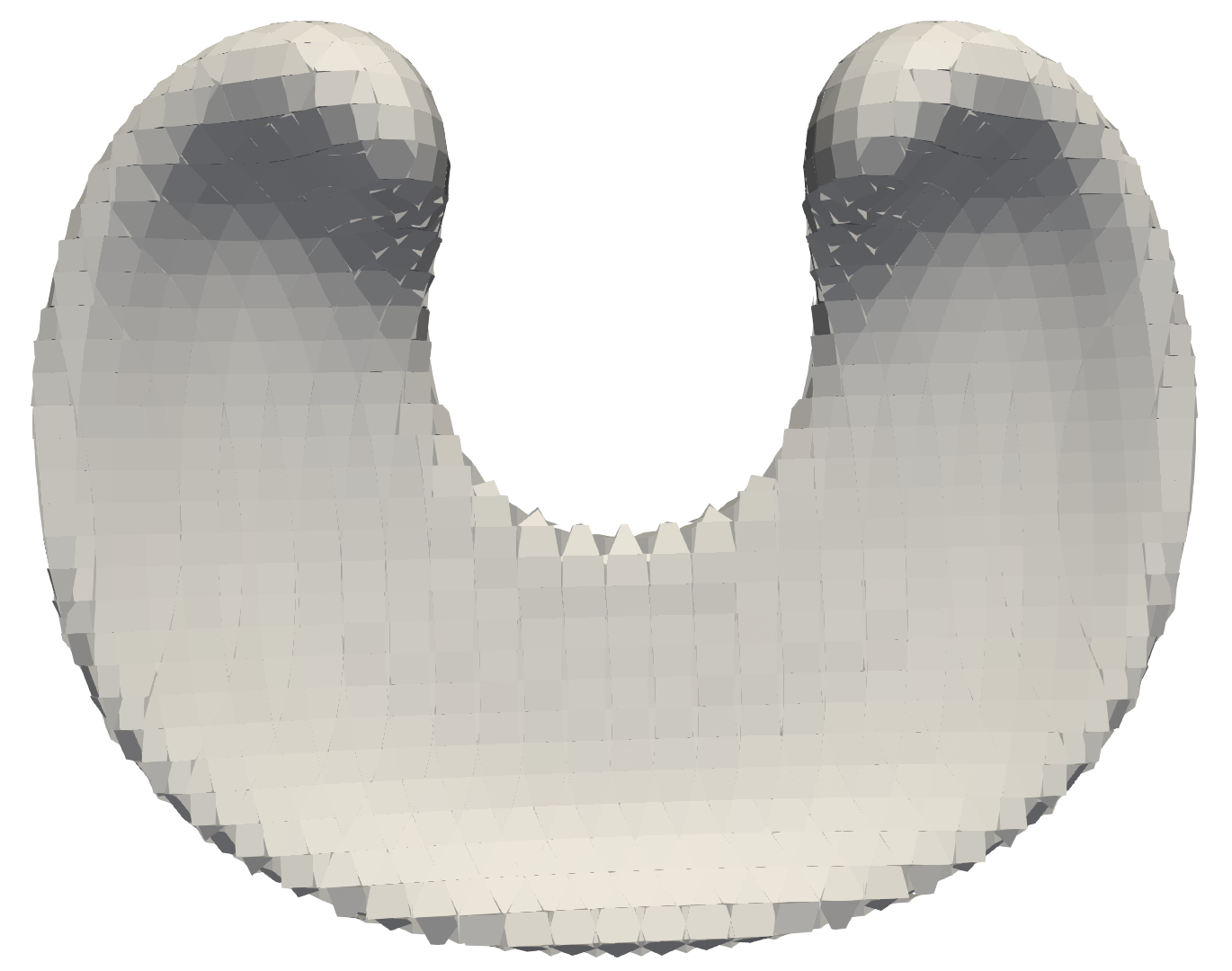}
       \caption{NN1, $t = 0.75$}
       \label{deform_plot:sub3}
   \end{subfigure}%
   \begin{subfigure}{.24\textwidth}
       \centering
       \includegraphics[width=\linewidth, clip]{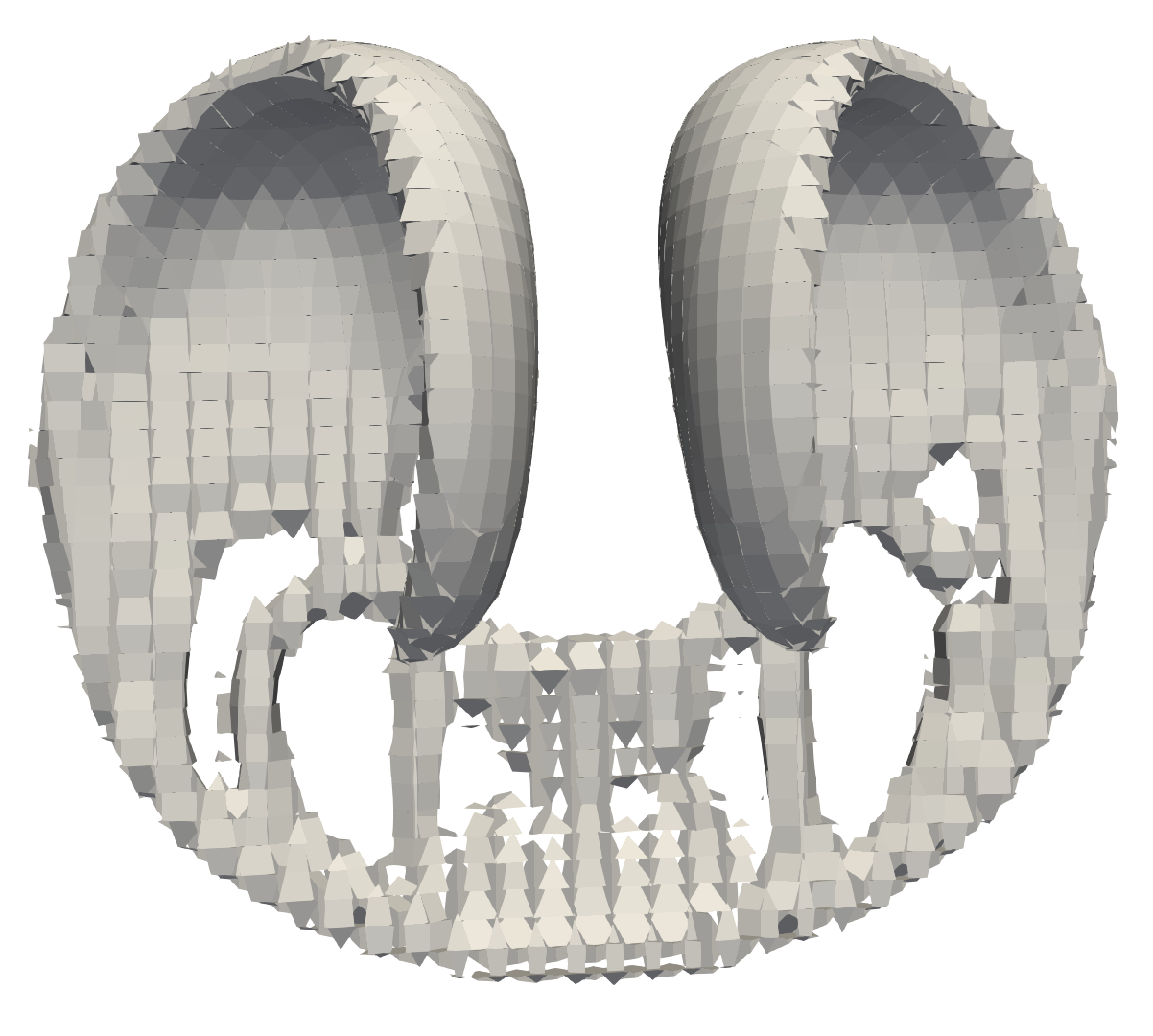}
       \caption{NN1, $t = 1.5$}
       \label{deform_plot:sub4}
   \end{subfigure}%
  \caption{Time series of the deformation advection case.}
  \label{deform_plot}
\end{figure}

The linear fits are all very close to the ideal of $\bm{n^{p}}=\bm{n^{t}}$, and the $R^2$ values are all very close to 1. NN3 has minimal scatter and the best linear fits, which is again not surprising given its simpler data. Meanwhile, NN2 has the largest scatter and worst fits with a non-trivial amount of outliers, although the overall accuracy is still high. NN1 lies between these extremes in terms of scatter and linear fit.

\subsection{Advection Test Cases}
\subsubsection{Translation and Grid Convergence}
Simple advection test cases performed within IRL use a prescribed velocity to advect the volume fraction field. A pure translation case was first run to establish a preliminary benchmark of PLIC-Net against the LVIRA optimization algorithm. Note that while LVIRA or ELVIRA are used as references when assessing the performance of PLIC-Net hereinafter, these methods typically cannot be considered fully ``correct'' solutions. For example, as multiple interfaces are advected into the same cell, it is impossible for any PLIC method to produce an exact solution. Therefore, by comparing PLIC-Net with LVIRA or ELVIRA, the objective is not to replicate their solutions. Rather, it is to show that PLIC-Net can provide alternative solutions that can be considered at least on par with, but potentially substantially different than, an LVIRA or ELVIRA solution.

The domain for the translation case was set to be a unit cube with a sphere of radius 0.25 initialized in the center. It was given a non-dimensional velocity in the $x$ direction of 1, in the $y$ direction of $\frac{2}{3}$, and in the $z$ direction $\frac{1}{3}$. Periodic boundary conditions were imposed such that the sphere would return to its initial position after a non-dimensional time of 3. A constant CFL number of 0.8 was used. Cases were run with 5, 10, 20, 40, 80, and 160 grid cells in each direction. The $L^1$ error of the volume fraction field between the final state and the initial state was calculated to study error convergence with grid size. LVIRA is known to converge with 2$^\text{nd}$ order accuracy. However, there is no expectation that a neural network reconstruction should converge under grid refinement. A time series of the LVIRA and NN1 drop translation cases are shown in Figure~\ref{translation_plot} for the case with 10 grid cells in each direction. PLIC-Net displays a qualitatively smoother reconstruction than LVIRA and appears to have slightly less shape distortion between the initial and final states. This shape distortion was quantitatively analyzed with the convergence study. The resulting convergence plot is shown in Figure~\ref{convergence_plot}. Although PLIC-Net has no analytical proof of convergence, it still demonstrates approximately 2$^\text{nd}$ order convergence for many of the grid sizes before eventually plateauing at the highest resolution. It must be noted that this study does not decouple transport error from reconstruction error. Since the transport error in this simulation was 2$^\text{nd}$ order, a likely interpretation of these results is that transport error dominates over reconstruction error for much of the range of grid sizes. Any convergence of the actual reconstruction error from PLIC-Net plateaus as the grid size decreases, which eventually results in it dominating over the converging transport error. But, for most of the range of grid sizes studied, the errors from PLIC-Net are close to those from LVIRA, and are in fact lower for many of the grid sizes. This shows that machine learning can be used to create a PLIC reconstruction method that performs on par with, if not better than, LVIRA and maintains effective convergence for a useful range of grid sizes.

\begin{figure}
   \centering
    \includegraphics[width=\linewidth, clip]{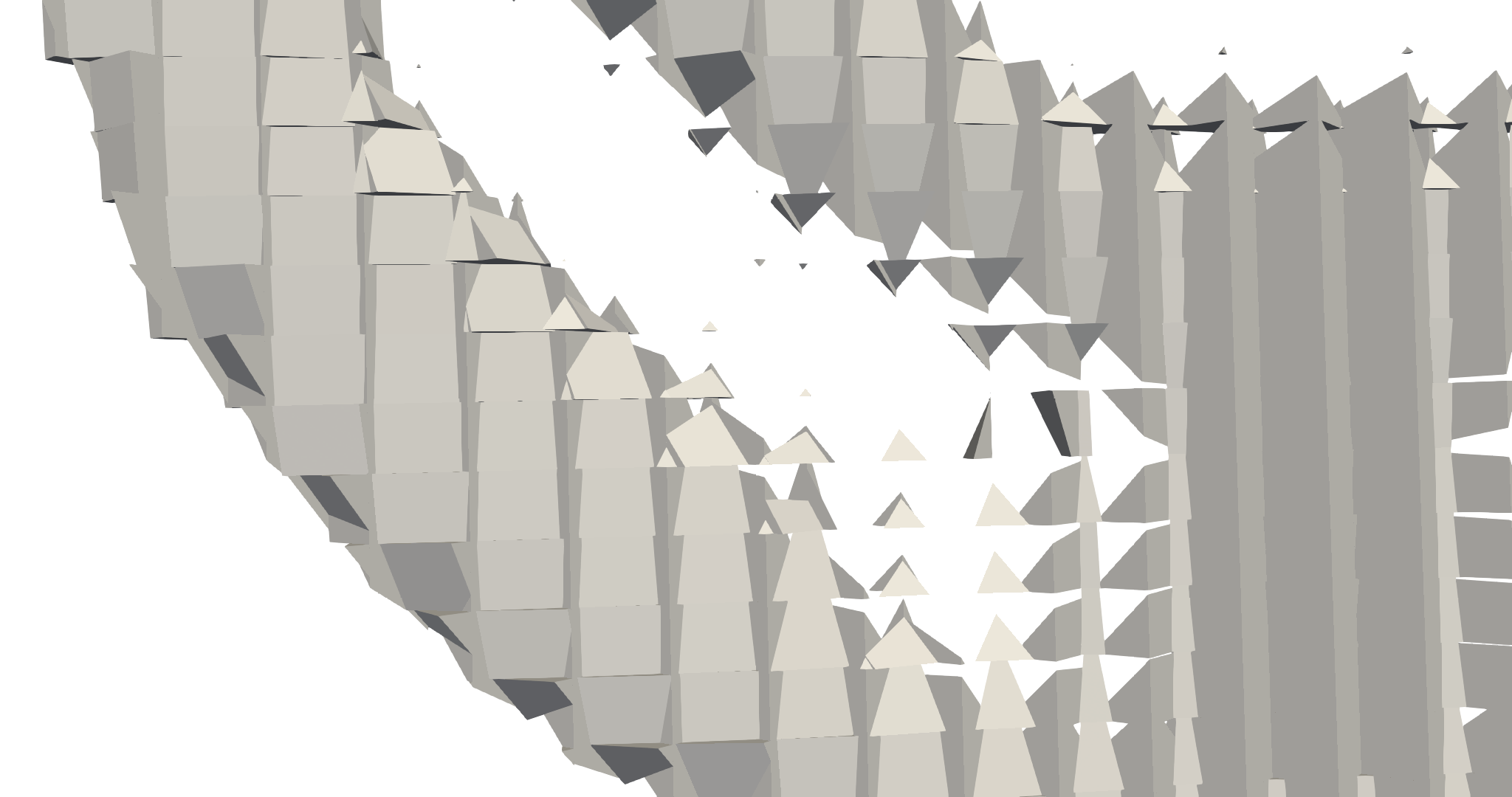}
  \caption{Closeup of the final state of the LVIRA deformation advection case to show the spurious planes developing near the film breakup.}
  \label{deform_LVIRA_close}
\end{figure}
\begin{figure}
   \centering
   \begin{subfigure}{.24\textwidth}
       \centering
       \includegraphics[width=\linewidth, clip]{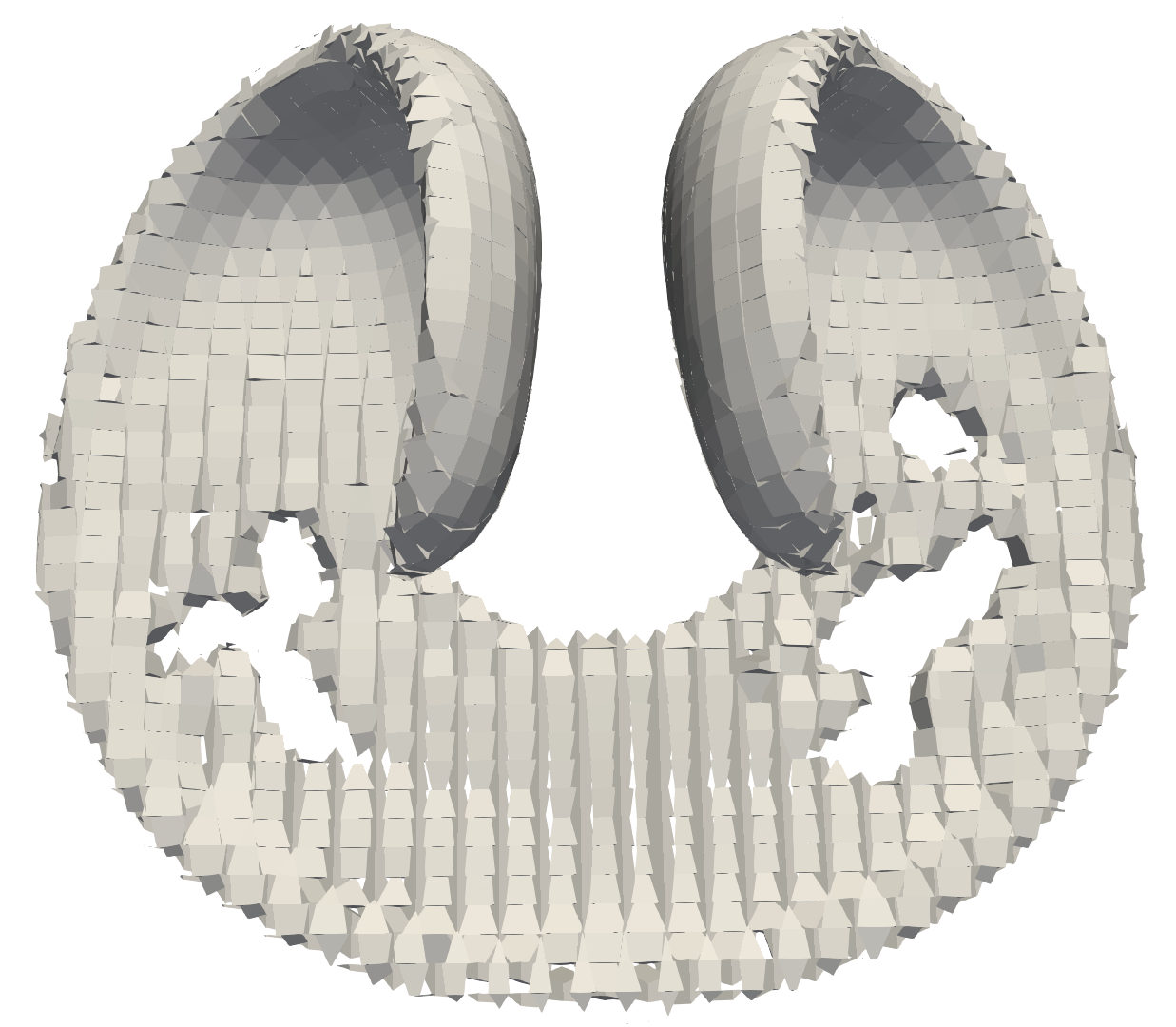}
       \caption{}
       \label{deform_plot2:sub1}
   \end{subfigure}%
   \begin{subfigure}{.24\textwidth}
       \centering
       \includegraphics[width=\linewidth, clip]{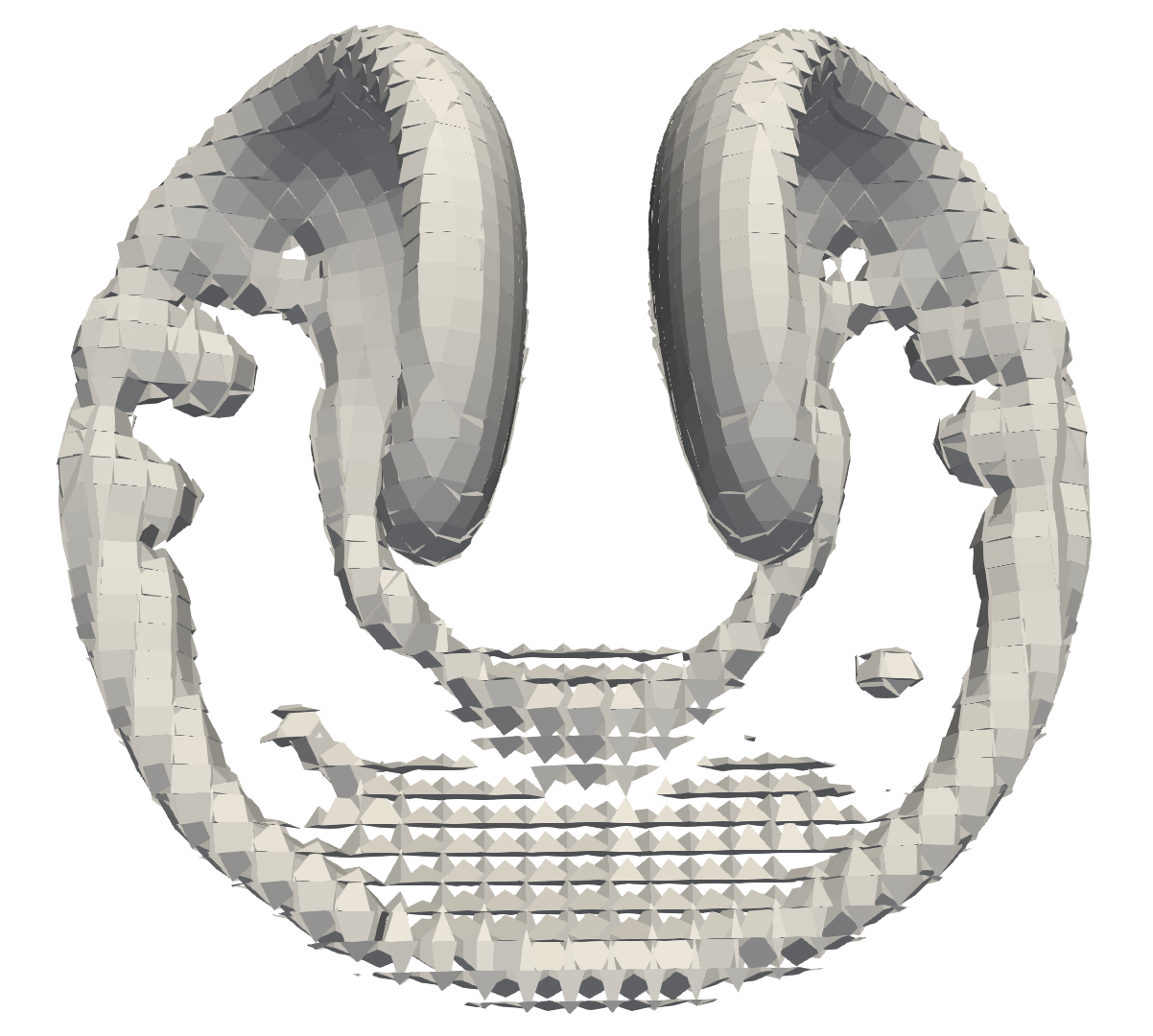}
       \caption{}
       \label{deform_plot2:sub2}
   \end{subfigure}%
  \caption{Final states from the deformation advection case using (a) NN2 and (b) NN3.}
  \label{deform_plot2}
\end{figure}

\subsubsection{Deformation and Numerical Surface Tension}
A deformation advection case was run to test PLIC-Net's performance in a more complex advection problem with higher curvatures. The domain was set to be a unit cube with a sphere of radius 0.15 initialized at $x=0.35$, $y=0.35$, and $z=0.35$, in reference to the center of the cubic domain. Cases were run with the 3D deformation velocity field defined by \cite{LeVeque}, given non-dimensionally by
\begin{equation}
\begin{aligned}
u=2\sin^2\left(\pi x\right)\sin\left(2\pi y\right)\sin\left(2\pi z\right)\cos\left(\frac{\pi t}{3}\right),
\end{aligned}
\end{equation}
\begin{equation}
\begin{aligned}
v=-\sin\left(2\pi x\right)\sin^2\left(\pi y\right)\sin\left(2\pi z\right)\cos\left(\frac{\pi t}{3}\right),
\end{aligned}
\end{equation}
\begin{equation}
\begin{aligned}
w=-\sin\left(2\pi x\right)\sin\left(2\pi y\right)\sin^2\left(\pi z\right)\cos\left(\frac{\pi t}{3}\right),
\end{aligned}
\end{equation}
where $u$, $v$, and $w$ are the velocities in the $x$, $y$, and $z$ directions, respectively, and $t$ is the non-dimensional time. The simulations were run until $t=1.5$ with a time step of 0.008. There were 50 grid cells in each direction, and the maximum CFL number was 0.8.

Time series from the LVIRA and NN1 models are shown in Figure~\ref{deform_plot}. As the drop deforms, a thin film begins to develop. When the film thickness decreases towards the size of the grid cells, numerical break-up occurs. With LVIRA, some spurious planes develop near the high curvature edges that form after break-up. The number of these planes grows as the film becomes thinner and numerical break-up continues. A closeup of these spurious planes is shown in Figure~\ref{deform_LVIRA_close}. Meanwhile, NN1 has a cleaner breakup with fewer spurious planes that do not propagate. However, it is observed that its numerical break-up happens with slightly thicker films than with LVIRA, which implies that NN1 has a higher numerical surface tension. The numerical surface tension can be controlled in PLIC-Net with the range and distribution of curvatures used in the training data. Figure~\ref{deform_plot2} shows the final states of cases run with NN2 and NN3. NN3 was trained with only planar data and thus no interface curvature was considered. The numerical surface tension is very high for this case which causes early break-up. Meanwhile, NN2 had a uniform distribution of paraboloid curvature coefficients between $-2$ and $2$ in the training data, and it seems to delay and reduce numerical break-up. Both NN2 and NN3 also have few spurious planes. This illustrates that the choices when sampling training data directly impact the nature of numerical break-up in PLIC-Net, opening avenues for controlling and fine-tuning that behavior.

\subsection{Flow Solver Test Cases}
\subsubsection{Falling Drop}
\begin{figure*}
   \centering
   \begin{subfigure}{.19\textwidth}
       \centering
       \includegraphics[width=\linewidth, clip]{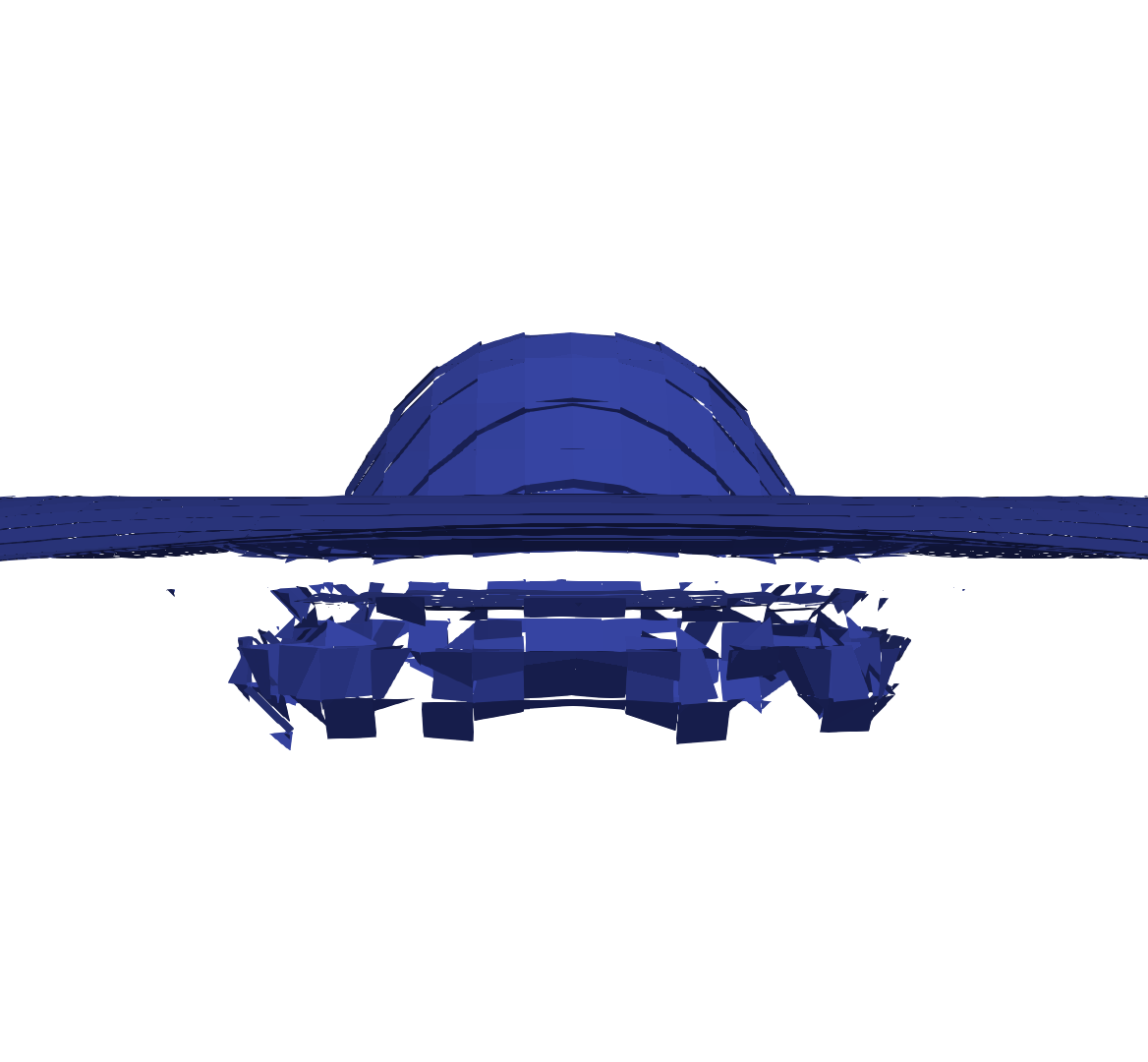}
       \caption{LVIRA}
       \label{falling_plot:sub1}
   \end{subfigure}%
   \hspace{0.3em}
   \begin{subfigure}{.19\textwidth}
       \centering
       \includegraphics[width=\linewidth, clip]{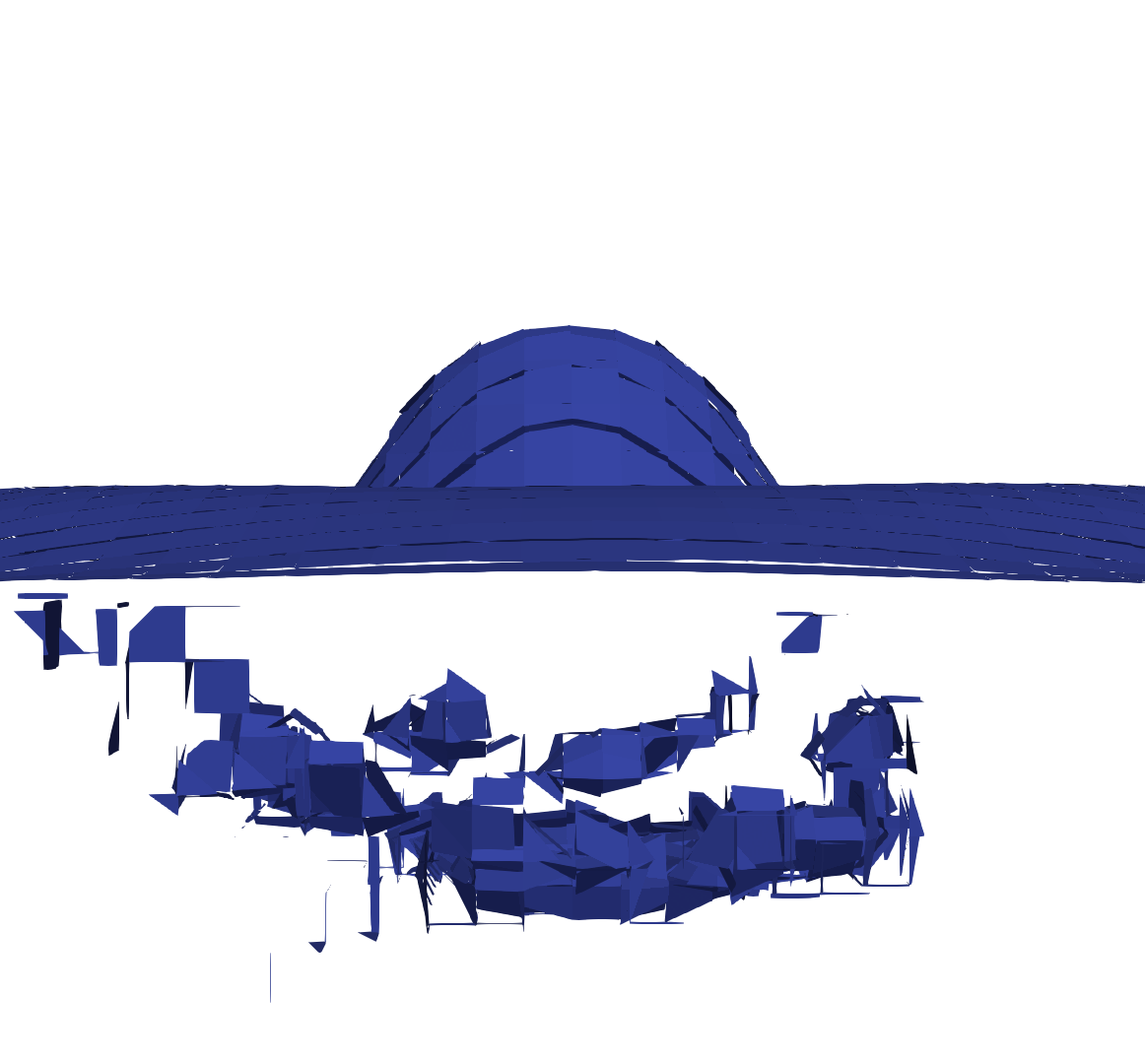}
       \caption{ELVIRA}
       \label{falling_plot:sube}
   \end{subfigure}%
   \hspace{0.3em}
   \begin{subfigure}{.19\textwidth}
       \centering
       \includegraphics[width=\linewidth, clip]{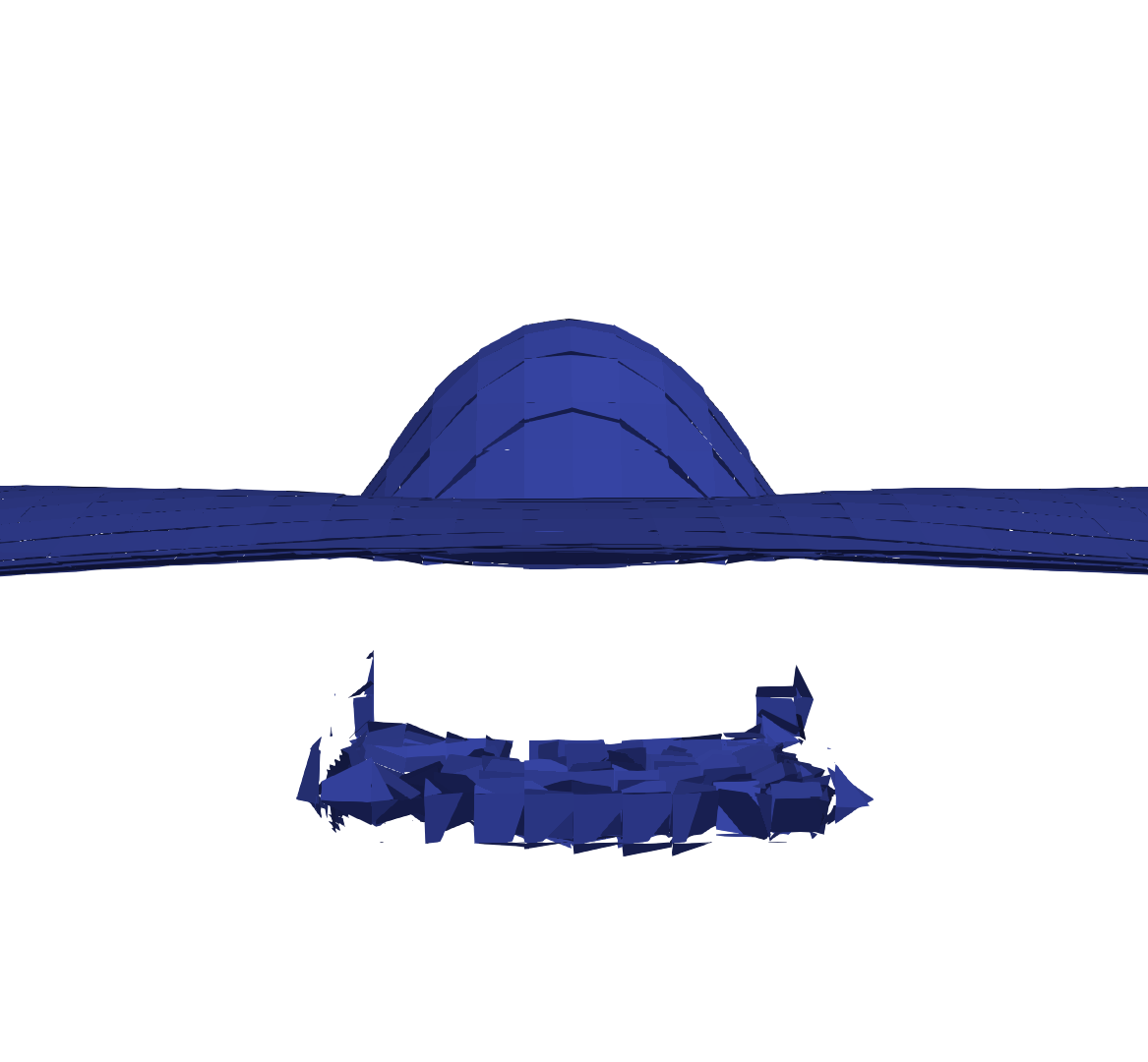}
       \caption{NN1}
       \label{falling_plot:sub2}
   \end{subfigure}%
   \hspace{0.3em}
   \centering
   \begin{subfigure}{.19\textwidth}
       \centering
       \includegraphics[width=\linewidth, clip]{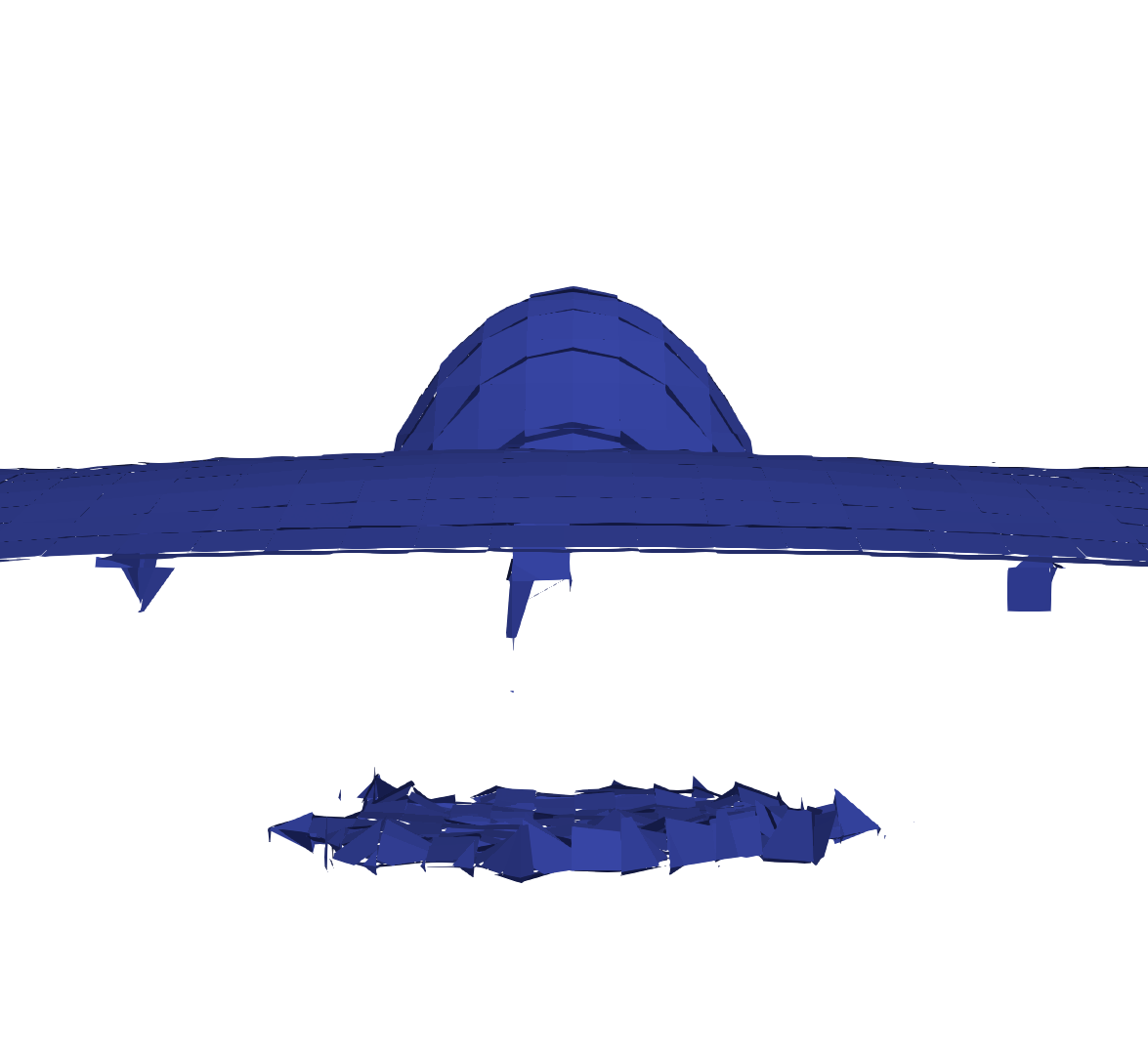}
       \caption{NN2}
       \label{falling_plot:sub3}
   \end{subfigure}%
   \hspace{0.3em}
   \begin{subfigure}{.19\textwidth}
       \centering
       \includegraphics[width=\linewidth, clip]{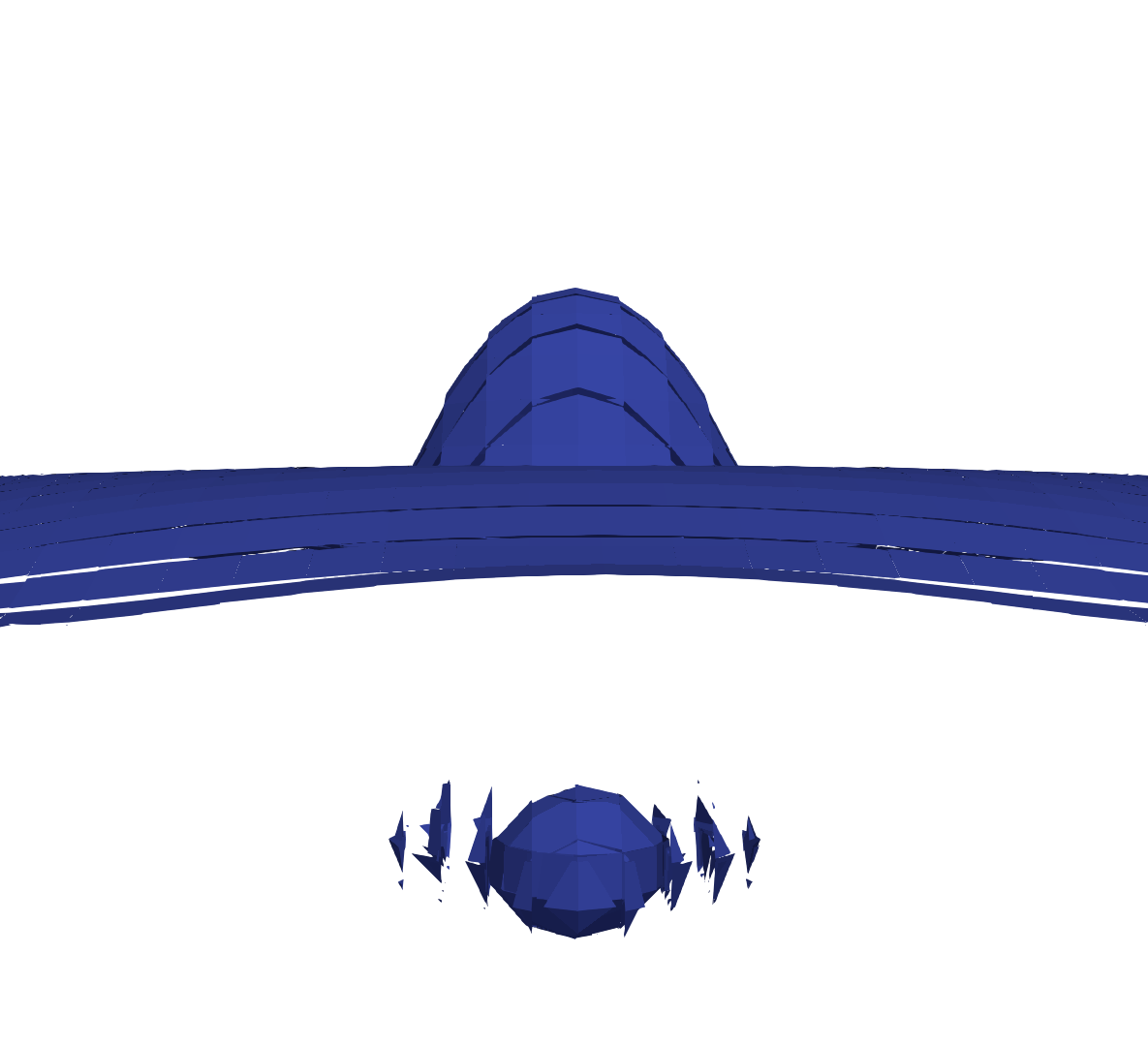}
       \caption{NN3}
       \label{falling_plot:sub4}
   \end{subfigure}%
  \caption{Interface reconstructions from falling drop case at time $t = 0.0365$ s.}
  \label{falling_plot}
\end{figure*}
\begin{figure}[t]
   \centering
   \includegraphics[width=\linewidth, clip]{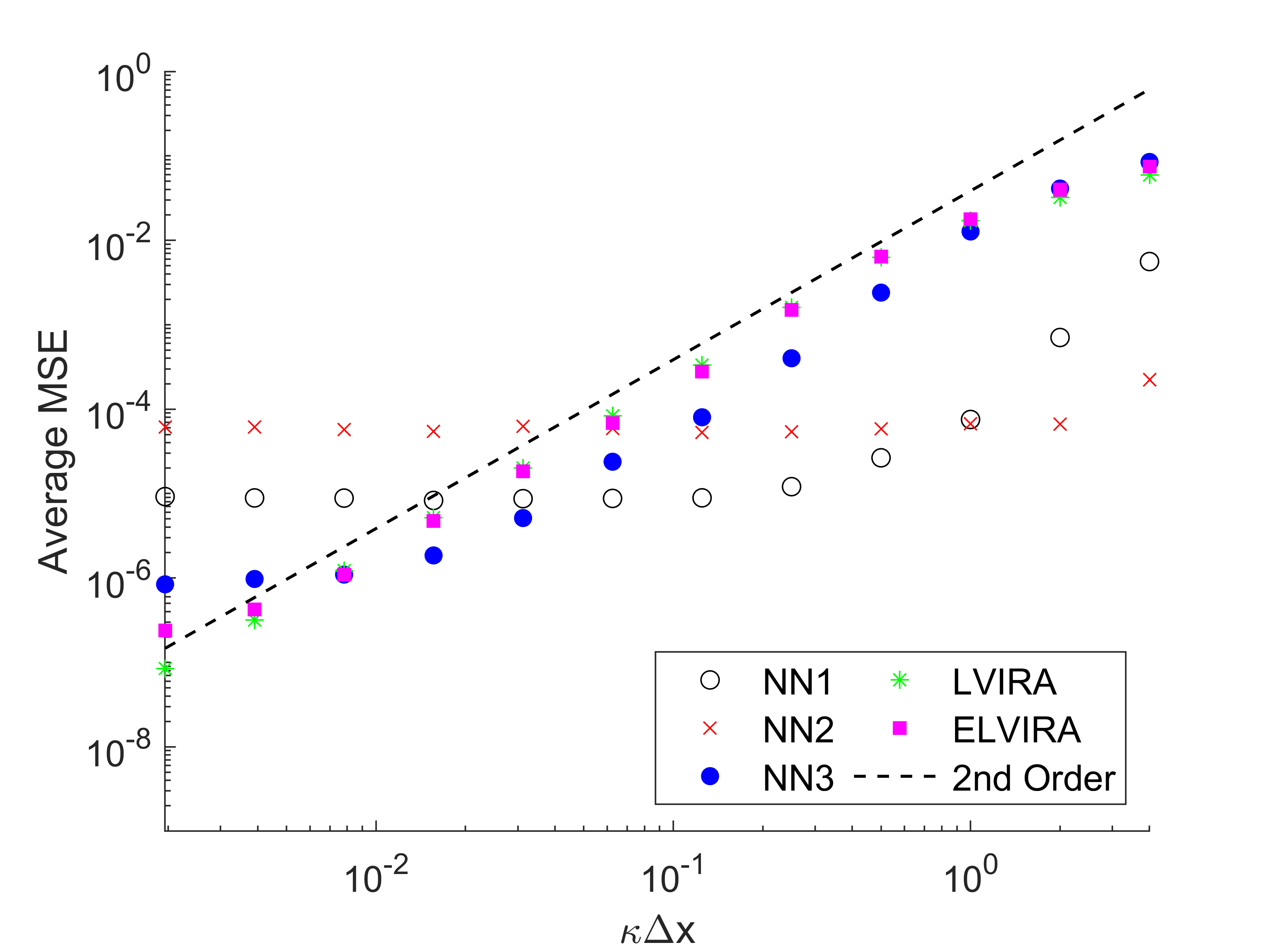}
  \caption{Average of the mean squared errors between predicted and expected normal vector components vs.~paraboloid principal curvature ($\kappa$) (using $a=b$) normalized by the cell size ($\Delta x$) for NN1, NN2, and NN3. LVIRA and ELIVRA are shown for comparison along with the line of second order accuracy.}
  \label{accuracy}
\end{figure}
A simulation of a liquid drop falling into a pool was run in the NGA2 flow solver in conjunction with the IRL. The liquid and gas viscosities were set to \SI{1.137e-3}{\kilogram\per\meter\per\second} and \SI{1.780e-5}{\kilogram\per\meter\per\second}, respectively, while the liquid and gas densities were set to \SI{1000}{\kilogram\per\meter\cubed} and \SI{1.226}{\kilogram\per\meter\cubed}, respectively. The surface tension coefficient was set to \SI{0.0728}{\newton\per\meter}. The domain was set to be 0.01 m with 40 cells in the $x$ direction, 0.02 m with 80 cells in the $y$ direction, and 0.01 m with 40 cells in the $z$ direction. A liquid sphere with radius 0.002 m was initialized at 0.01 m above the $y$ minus domain boundary. The liquid pool had its free surface initialized at $y = 0.00401$ m. Gravity was set in the negative $y$ direction at \SI{9.81}{\meter\per\second\squared}. Periodic boundary conditions were used in the $x$ and $z$ directions, and walls at the top and bottom in the $y$ direction. A maximum time step of \num{1e-4} s was used with a maximum allowed CFL number of 0.9, which included the capillary time step constraint. Cases were run with LVIRA, ELVIRA, and PLIC-Net. The simulations were stopped after 0.1 s of simulation time. During this time, the drop accelerates downward and violently impacts the pool, leading to an intense splash, gas film entrapment, and Worthington jet formation.

The main difference between the reconstruction methods in this case is the treatment of the change of topology when the drop collides with the pool. Figure~\ref{falling_plot} shows the aftermath of the collision for each reconstruction. LVIRA creates a thin and slightly disorganized cluster of planes below the surface that separates from the drop and then stays relatively close to the surface. ELVIRA produces a large number of spurious planes and has a very disorganized appearance. Meanwhile, the behavior of PLIC-Net once again depends on the training dataset used. As discussed in section 3.3.2, the effective numerical surface tension in PLIC-Net is directly controlled by the distribution of interface curvatures in the training data. NN3, trained only on planar data, exhibits a quick separation of a single gas bubble under the interface, which is better organized than with LVIRA. Unlike with LVIRA, this bubble continues to move downwards in the pool before rising again due to buoyancy and colliding with the interface. NN1 and NN2 both capture a subsurface gas film that resembles the thin structure captured by LVIRA. But, both are qualitatively more organized except for a few spurious planes. 

This example case highlights how the distribution of training data can cause PLIC-Net's performance to vary significantly. Since NN2 was shown a relatively wide and uniform distribution of curvature data, it struggles to accurately reconstruct the simple flat interface of the free surface before the drop collides with it. This is qualitatively apparent in the simulation results since LVIRA, ELVIRA, and NN3 both reproduce the flat interface very well (LVIRA and ELVIRA exactly reproduces it by design), while NN2 has a disjointed interface with tilted planes. Conversely, NN3 struggles with higher curvature data when compared to NN2. This point can be illustrated quantitatively by measuring the loss of PLIC-Net when given paraboloid data for a range of small to large curvatures with $a$ and $b$ set to be equal. For this study, 1000 paraboloids are created for each dataset, with random orientations and locations in the center cell of the stencil, taken to be of size $\Delta x=1$. The paraboloid curvature coefficients $a$ and $b$ are initially set to 2, and then halved in each successive dataset. The MSE of each normal vector component is calculated (compared to the paraboloid surface normal), and then those values are themselves averaged. Note that this study is equivalent to a static grid refinement error convergence test with constant curvature. Figure~\ref{accuracy} contains the results of this comparison, and LVIRA and ELIVRA are included for reference. The neural networks generally perform better on lower curvature data than higher curvature data since it is easier to learn the simpler data. However, out of the three training sets, NN3 shows the lowest error for very low curvatures and the highest error for most higher curvature data, while NN2 is the worst at low curvature data and the best at the highest curvatures. Note that NN2 exhibits little variation in its error over the different curvatures when compared to NN1 and NN3. NN1 tends to fall between NN2 and NN3, except at moderate curvatures where it is the best. It is also interesting to note the very similar errors of LVIRA, ELVIRA, and NN3, which shows that LVIRA and ELVIRA have some similarities to a neural network that has never been shown curved interface data.

The fact that NN2 has a consistently low error for most of these curvatures suggests that a uniform distribution of data is largely effective. However, its performance for very low curvatures is a notable disadvantage. NN1, as shown in the previous results, tends to perform in-between NN2 and NN3 by most metrics, which suggests its normal distribution of curvature data prepared it well for a variety of situations at low and high curvatures. This makes NN1 a suitable choice for many simulations. Still, there is no expectation that the actual distribution of curvatures in any given simulation is normal, especially with the chosen mean and standard deviation. On the contrary, there are plenty of example cases that have highly custom distributions that are not close to being normal, such as this particular falling drop case which has a disproportionate amount of flat interfaces. Therefore, it is desirable to train PLIC-Net with a well-sampled dataset that reflects reality. Determining what might constitute this optimal dataset requires more attention and will be the subject of future work, and the use of the normal distribution in this example was simply to demonstrate how different distributions can lead to reconstructions with improved results in certain simulations and worse results in others. Still, the falling drop simulation has shown that PLIC-Net is a viable method in practical multiphase flow simulations with violent dynamics and non-trivial topology changes.

\subsubsection{Drop Collision}
This example case involved the head-on collision of two identical drops. The fluid properties are the same as in the falling drop case. The triply-periodic domain was set to be 0.01 m with 40 cells in the $x$ direction, 0.02 m with 80 cells in the $y$ direction, and 0.01 m with 40 cells in the $z$ direction. Two liquid spheres with radii of 0.002 m were initialized at 0.015 m and 0.005 m above the $y$ minus domain boundary, respectively. The drop velocities were initialized to be \SI{0.52}{\meter\per\second} (one in the $+y$ direction, the other in the $-y$ direction). There was no gravity in this case. A maximum time step of \num{1e-4} s was used with a maximum allowed CFL number of 0.9, which included the capillary time step constraint. Cases were run with LVIRA, ELVIRA, and PLIC-Net. The simulations were stopped after 0.1 s of simulation time.

\begin{figure*}
   \centering
   \begin{subfigure}{\textwidth}
       \centering
       \includegraphics[width=.11\linewidth, clip]
       {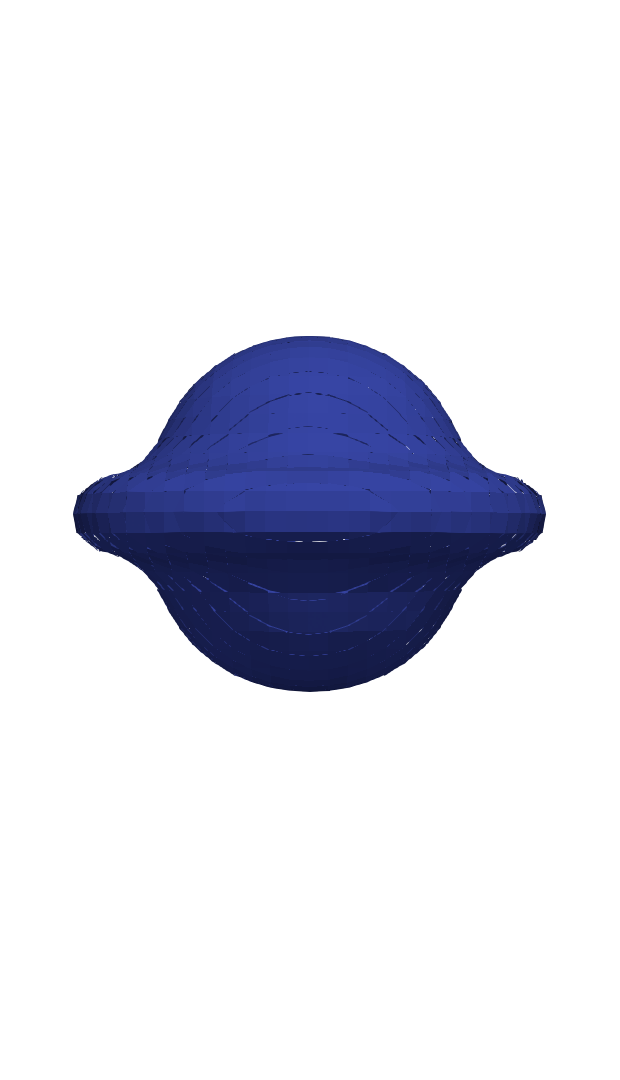}
       \includegraphics[width=.11\linewidth, clip]{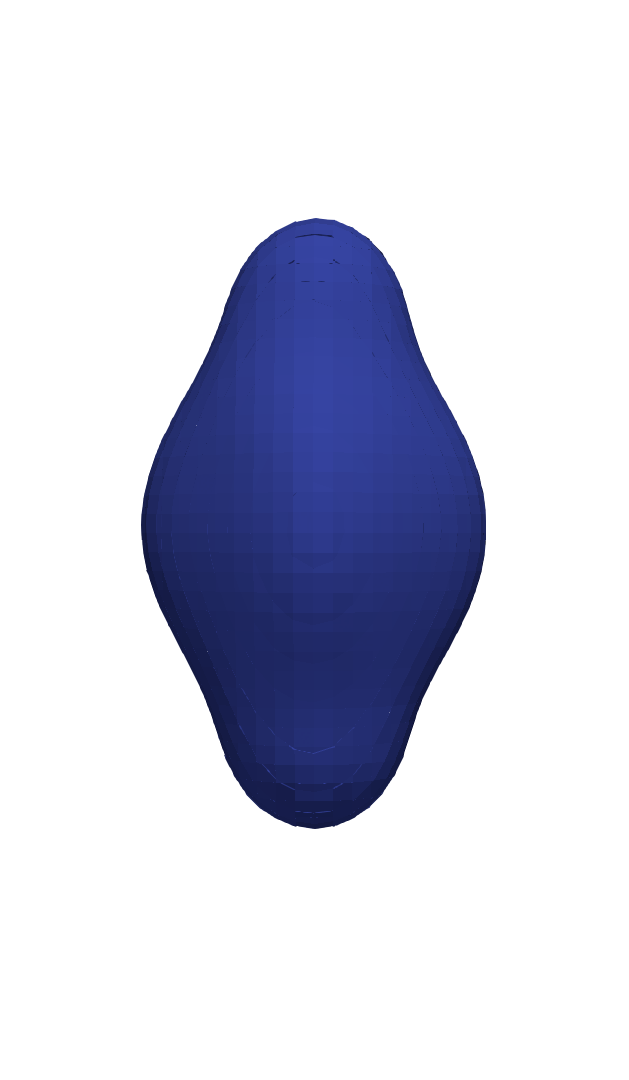}
       \includegraphics[width=.11\linewidth, clip]{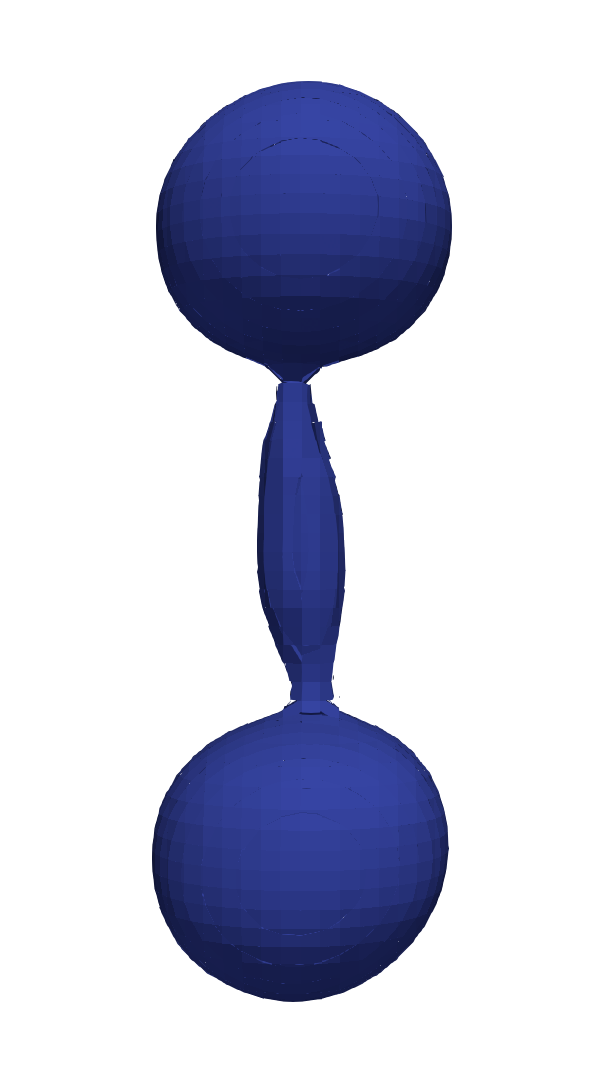}
       \includegraphics[width=.11\linewidth, clip]{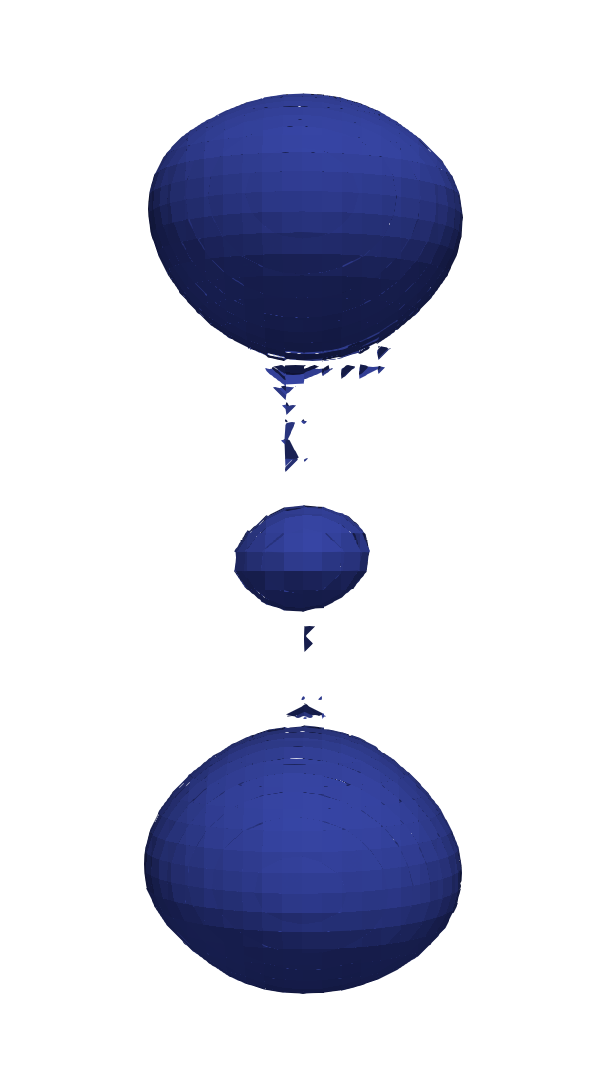}
       \caption{LVIRA}
       \label{collide1_plot:sub1}
   \end{subfigure}%

   \begin{subfigure}{\textwidth}
       \centering
       \includegraphics[width=.11\linewidth, clip]{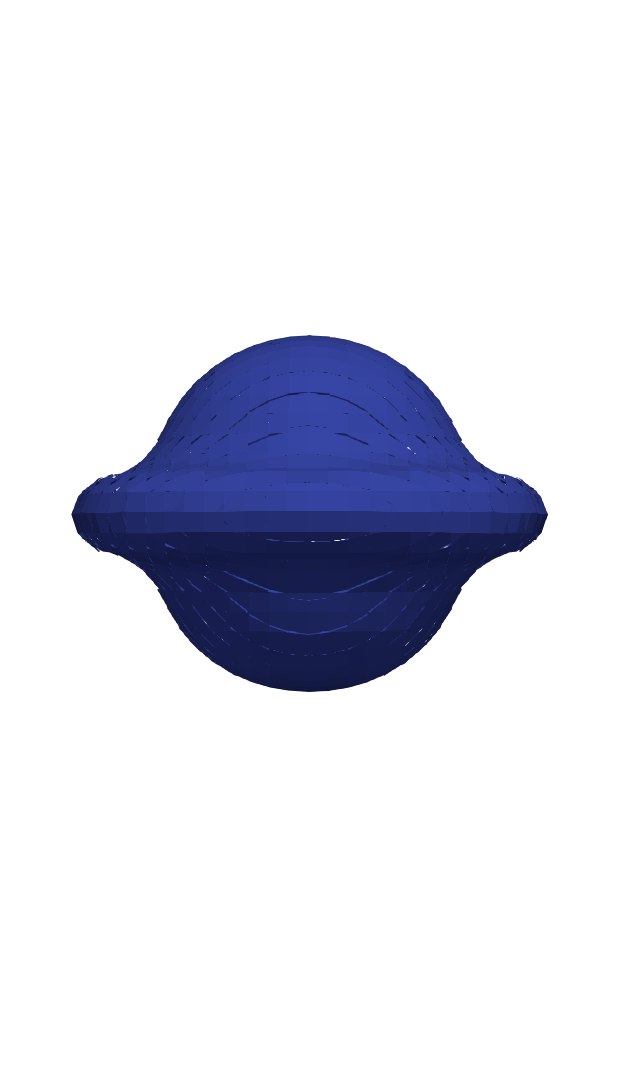}
       \includegraphics[width=.11\linewidth, clip]{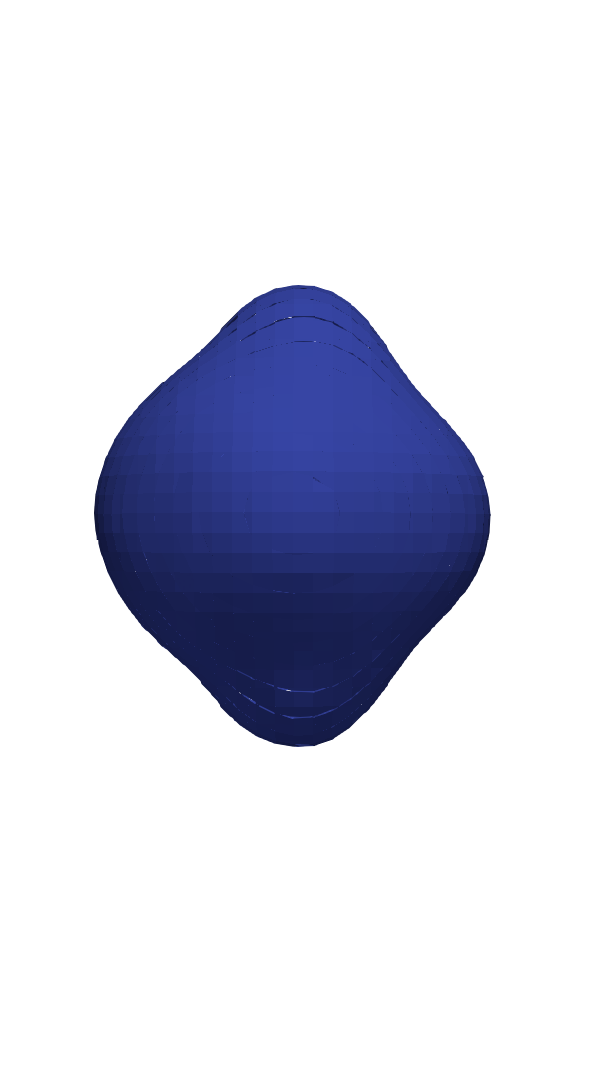}
       \includegraphics[width=.11\linewidth, clip]{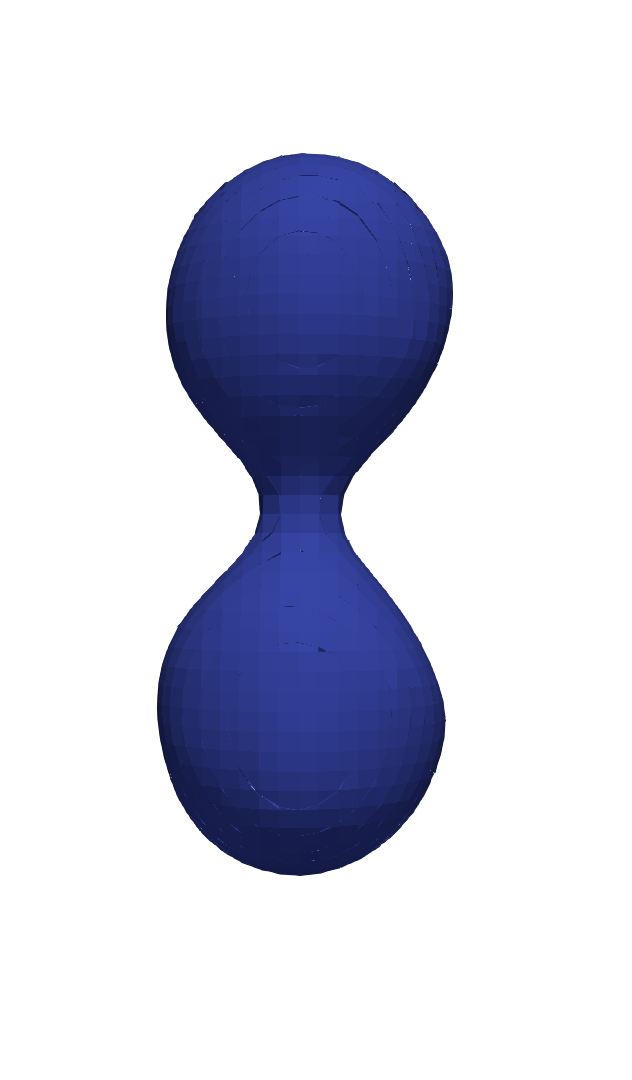}
       \includegraphics[width=.11\linewidth, clip]{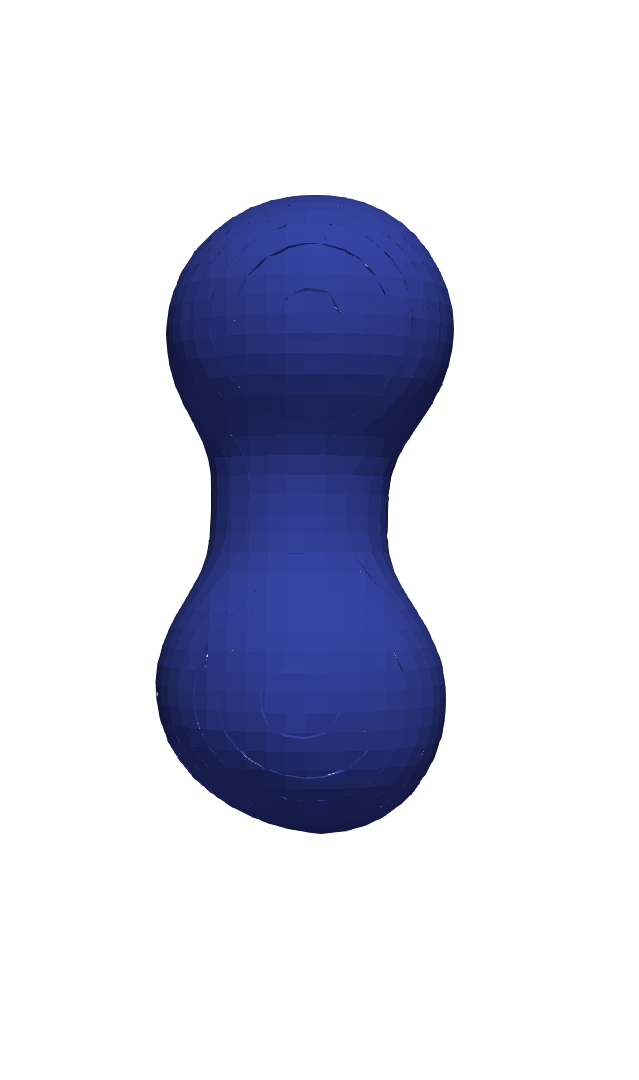}
       \caption{ELVIRA}
       \label{collide1_plot:sub2}
   \end{subfigure}%
   
   \begin{subfigure}{\textwidth}
       \centering
       \includegraphics[width=.11\linewidth, clip]{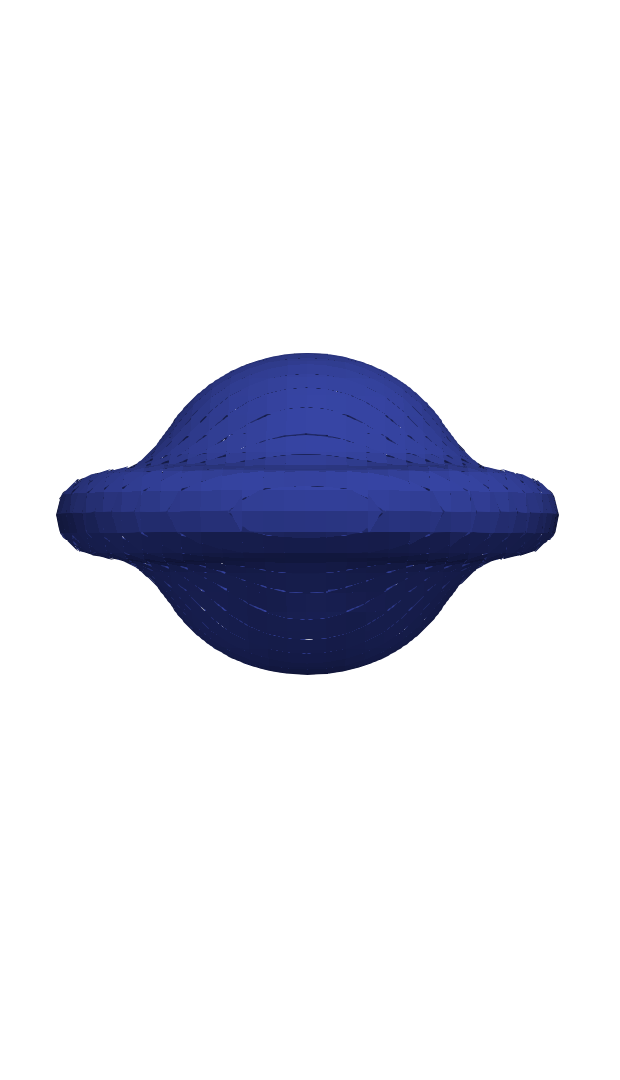}
       \includegraphics[width=.11\linewidth, clip]{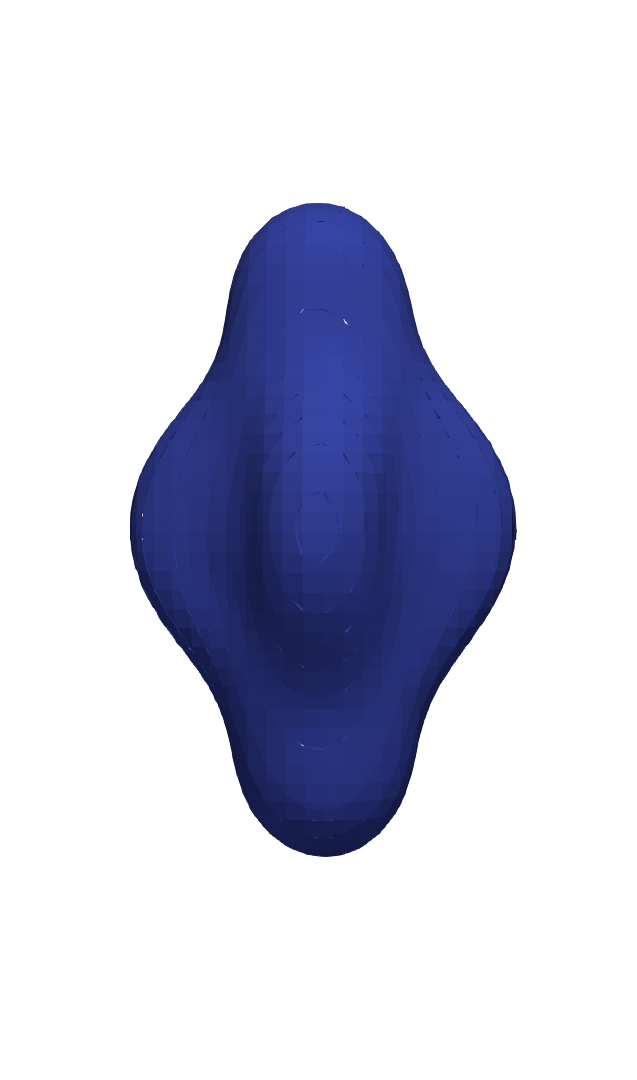}
       \includegraphics[width=.11\linewidth, clip]{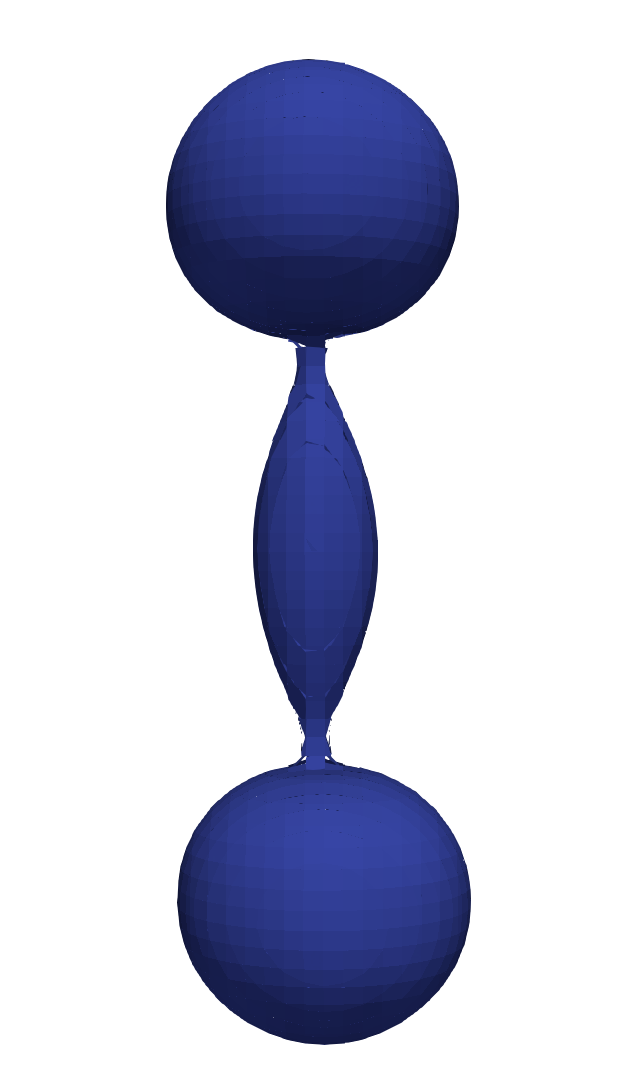}
       \includegraphics[width=.11\linewidth, clip]{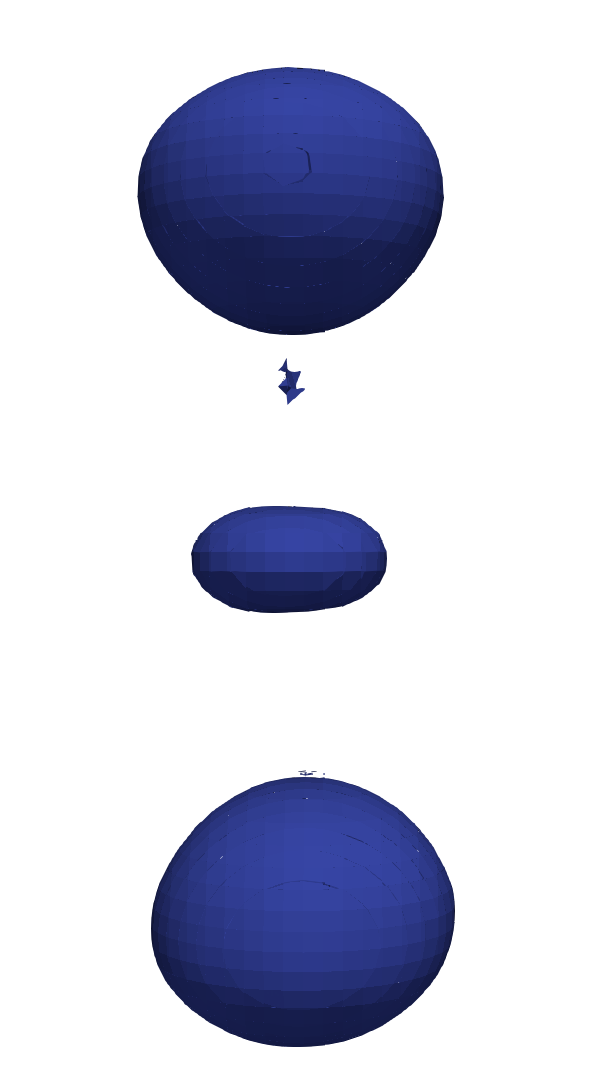}
       \caption{NN1}
       \label{collide1_plot:sub3}
   \end{subfigure}%
\vspace{1em}
   \begin{subfigure}{\textwidth}
       \centering
       \includegraphics[width=.11\linewidth, clip]{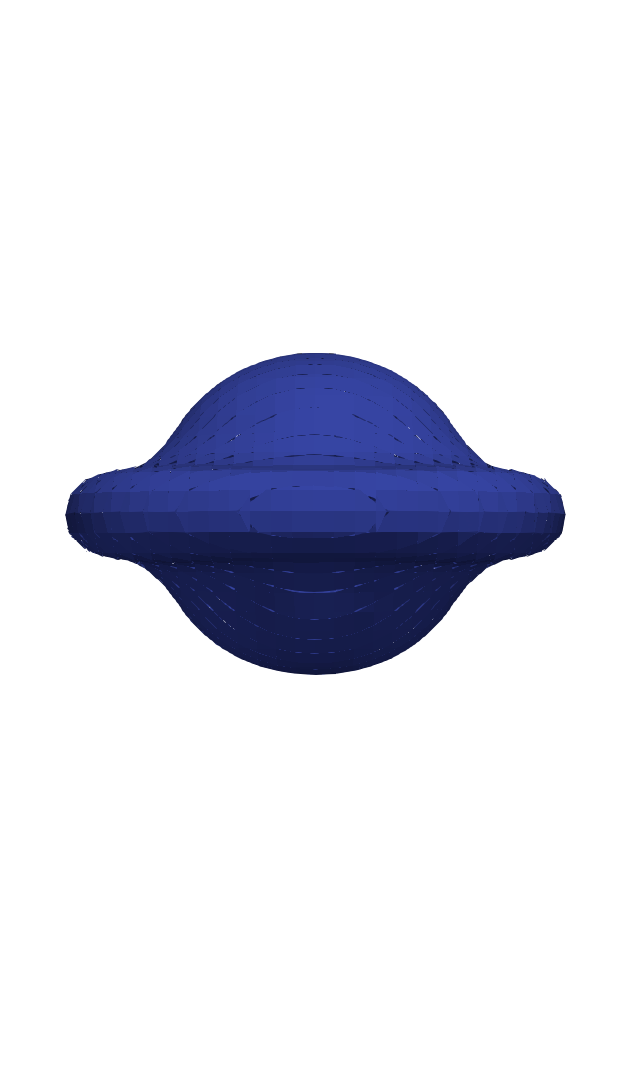}
       \includegraphics[width=.11\linewidth, clip]{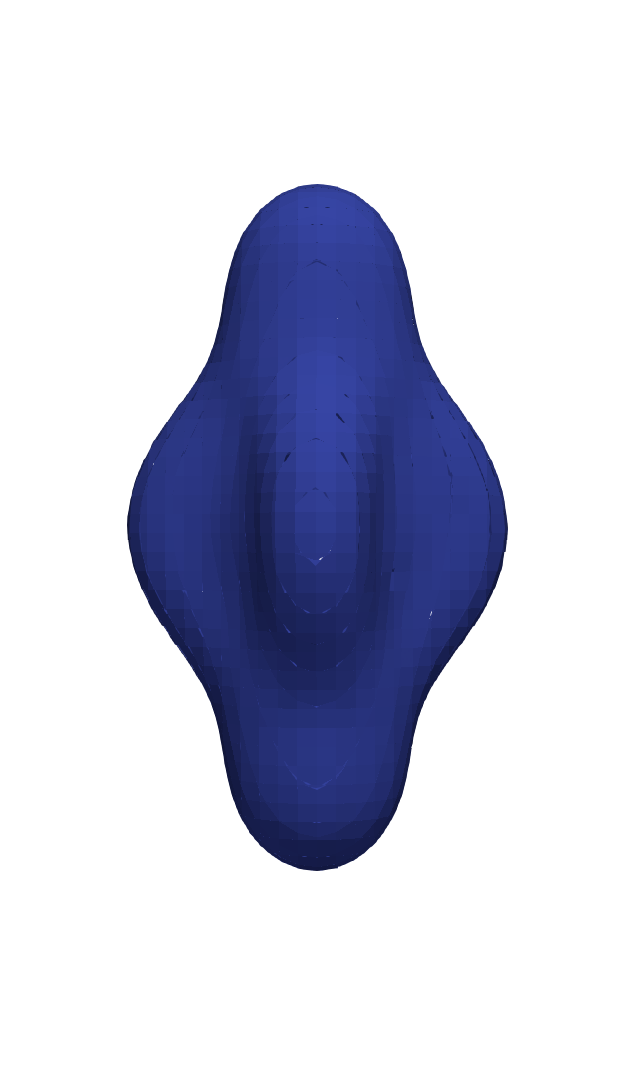}
       \includegraphics[width=.11\linewidth, clip]{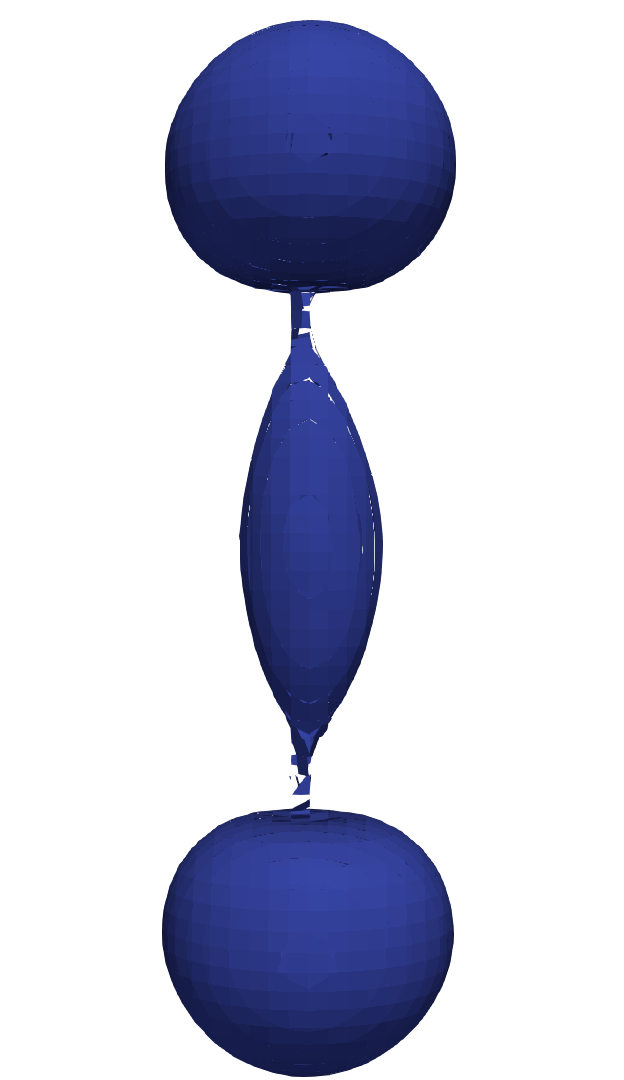}
       \includegraphics[width=.11\linewidth, clip]{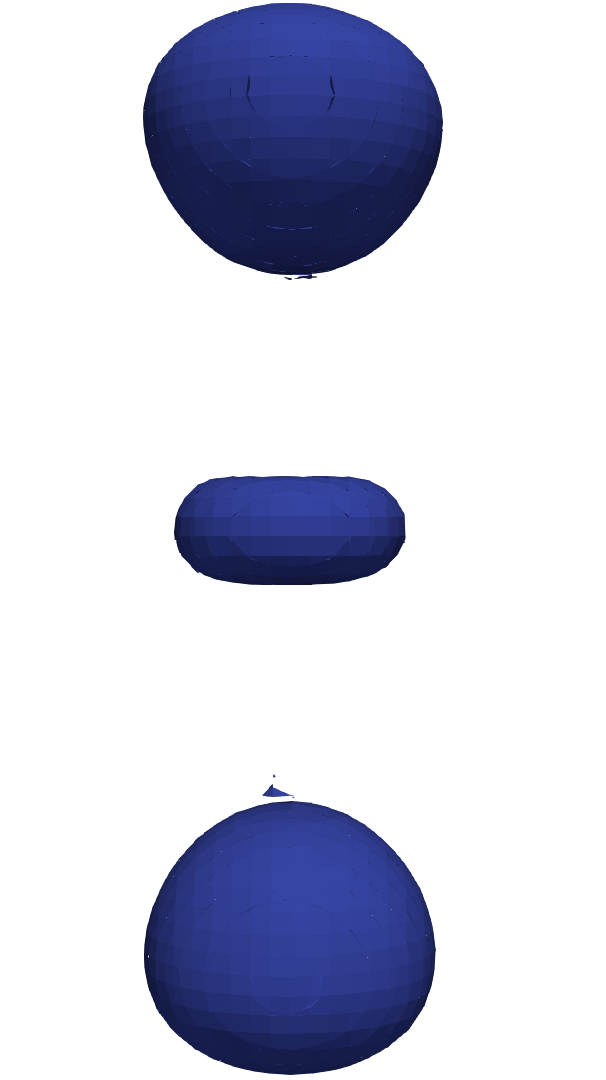}
       \caption{NN2}
       \label{collide1_plot:sub4}
   \end{subfigure}%

   \begin{subfigure}{\textwidth}
       \centering
       \includegraphics[width=.11\linewidth, clip]{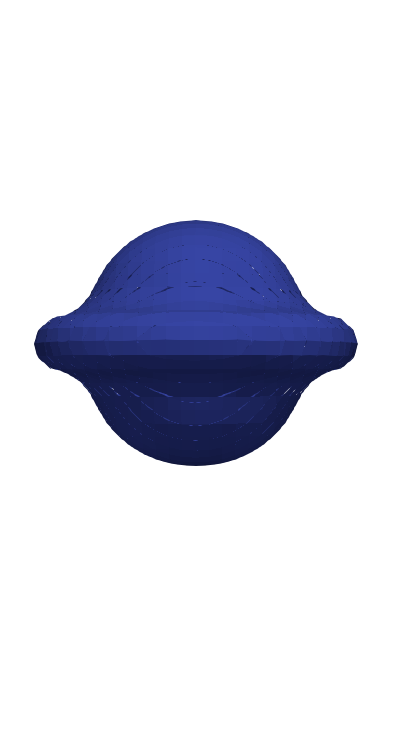}
       \includegraphics[width=.11\linewidth, clip]{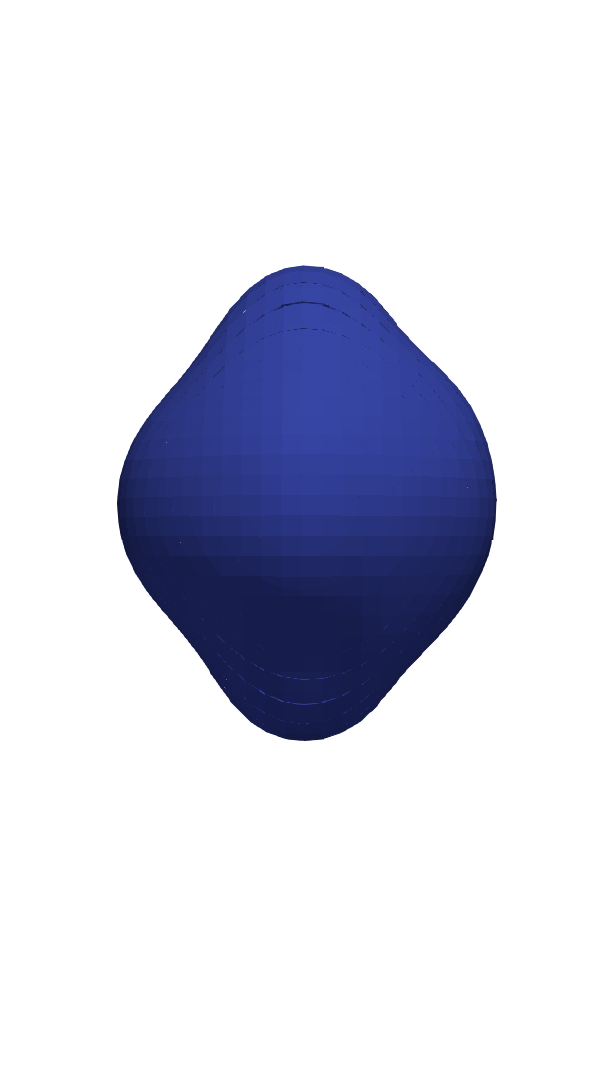}
       \includegraphics[width=.11\linewidth, clip]{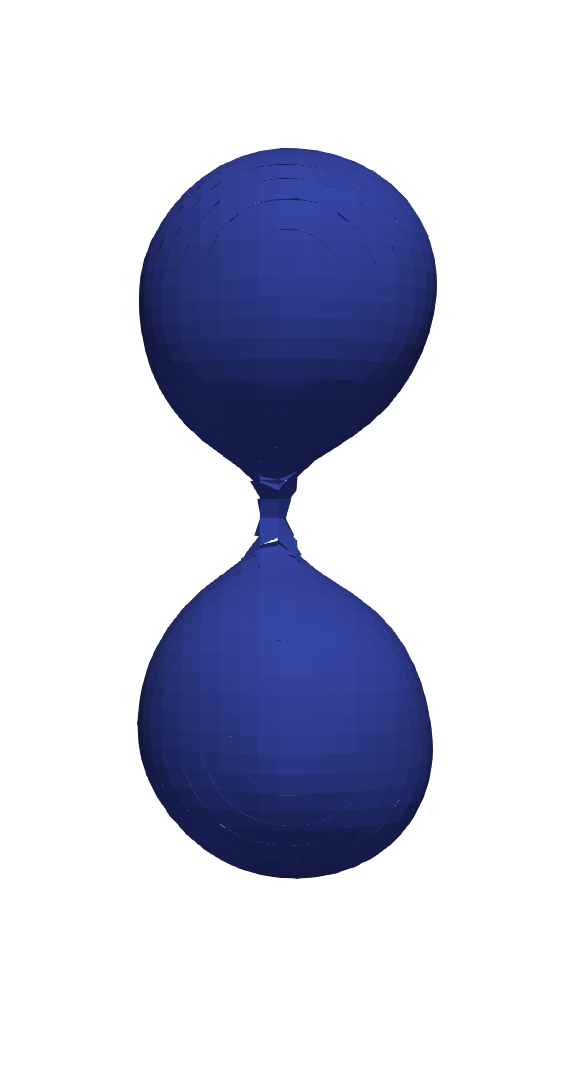}
       \includegraphics[width=.11\linewidth, clip]{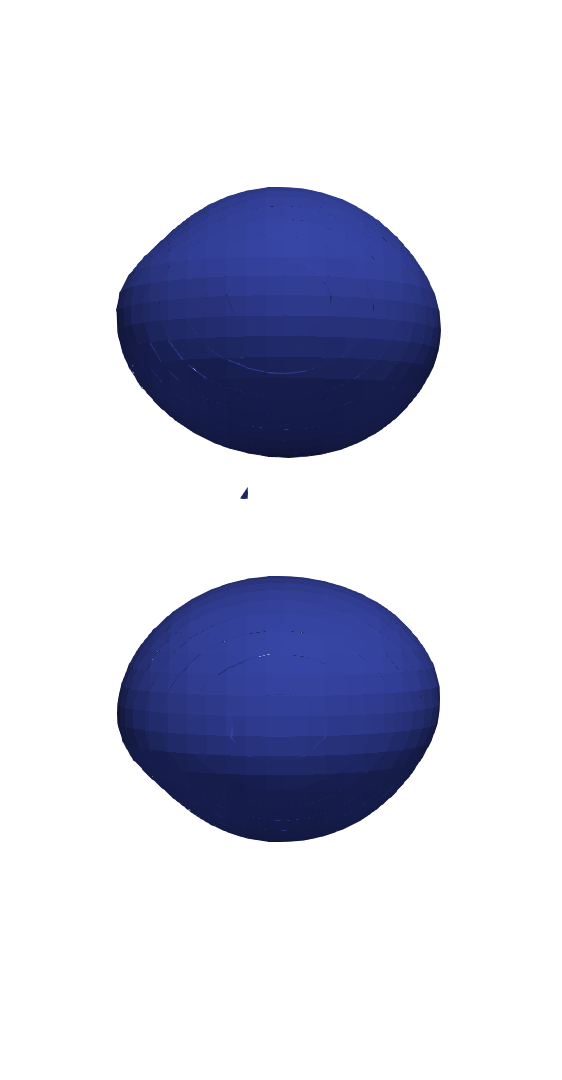}
       \caption{NN3}
       \label{collide1_plot:sub5}
   \end{subfigure}%
  \caption{Time series of the colliding drops case at times $t=0.009$ s, $t=0.03$ s, $t=0.0575$ s, and $t=0.064$ s (from left to right).}
  \label{collide1}
\end{figure*}

\begin{figure*}
   \centering
       \includegraphics[width=\linewidth, clip]{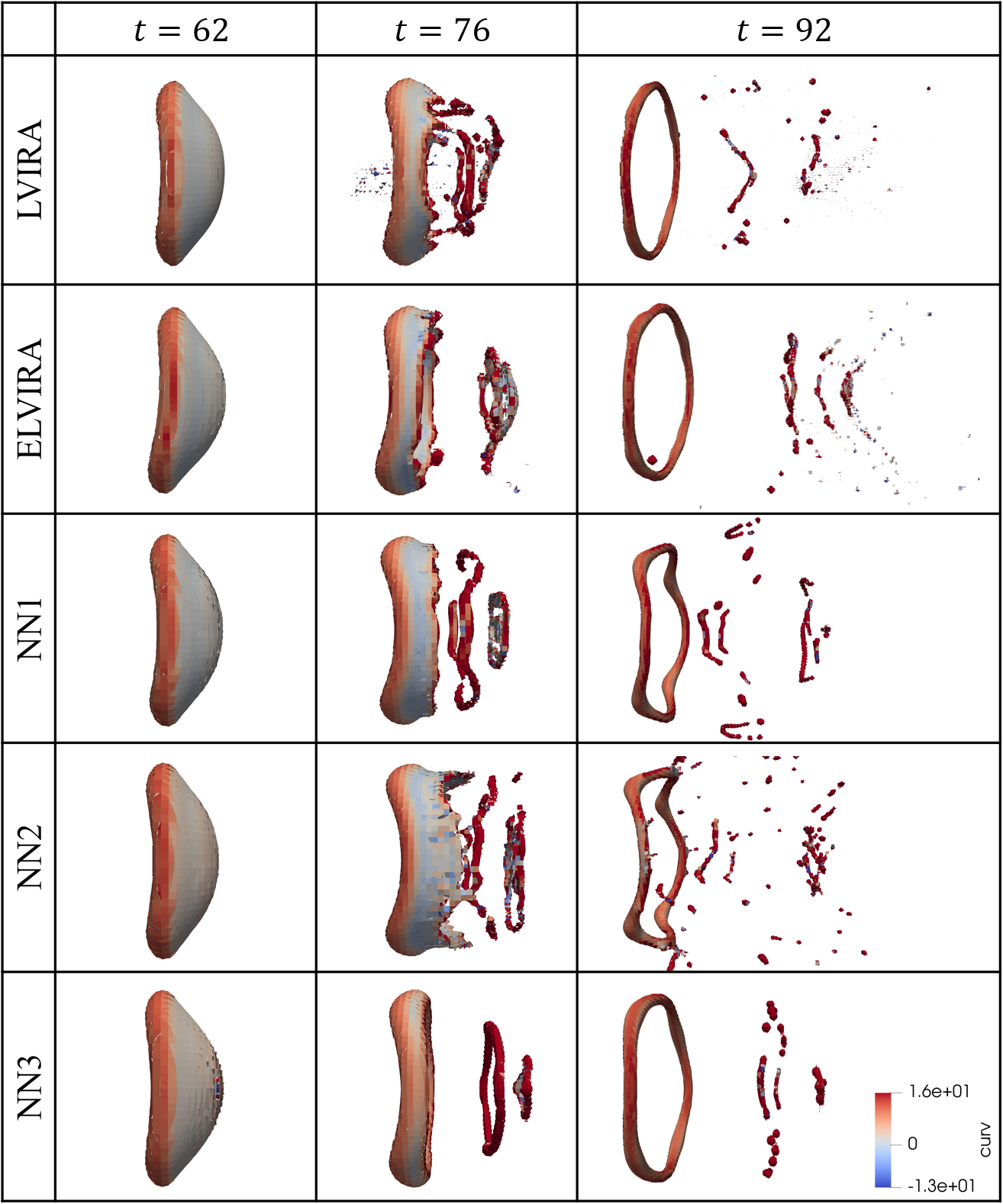}
  \caption{Time series from the drop in a cross-flow simulations. Interface is colored by its curvature. All times are non-dimensional.}
  \label{lig1}
\end{figure*}

All reconstruction methods predict that the drops merge and flatten before rebounding and stretching into a narrow structure. As the stretching continues, the middle of the structure begins to neck until the drops separate again, except with ELVIRA where no separation occurs. Figure~\ref{collide1} shows the time series for all of the reconstructions. LVIRA produces clusters of spurious planes near the points of separation along with a third drop where the necking occurred. It is also noted that LVIRA and ELVIRA lose symmetry during the simulation. Since the drops are initialized as symmetrical and the collision is head-on, it would be expected that the necking and separation would also be symmetrical. However, the numerical nature of the drop merging process led to LVIRA and ELVIRA having the rebound and necking occur in an asymmetrical manner. Meanwhile, NN1 and NN2 produce very similar overall results to each other and are similar to LVIRA, except with very few spurious planes and a slightly larger third drop. The primary drops also rebound slightly further than with LVIRA. Almost all of the few spurious planes that do form are quickly reabsorbed into the resolved drops, and symmetry is mostly preserved. NN3, by contrast, behaves more similarly to ELVIRA by having a much smaller necking region and shorter rebound of the primary drops. However, NN3 still predicts that the drops separate, although it does not produce a third drop. As commented on previously, this is likely due to its high numerical surface tension, which resulted from its planar training data. These results are in general agreement with the falling drop simulation and advection cases, in which NN3 exhibited very high numerical surface tension with quick break-up while NN1 and NN2 were more similar to LVIRA. In this case, NN1 and NN2 behave much more similarly to each other than in the previous results, which suggests that the level of numerical surface tension in NN1 and NN2 are similar in certain simulations. As previously observed, it is found that PLIC-Net limits the formation of spurious planes and generally produces cleaner interfaces and numerical break-up than LVIRA and ELVIRA.

\subsubsection{Drop in a Cross-Flow}
\begin{figure*}
   \centering
       \includegraphics[width=\linewidth, clip]{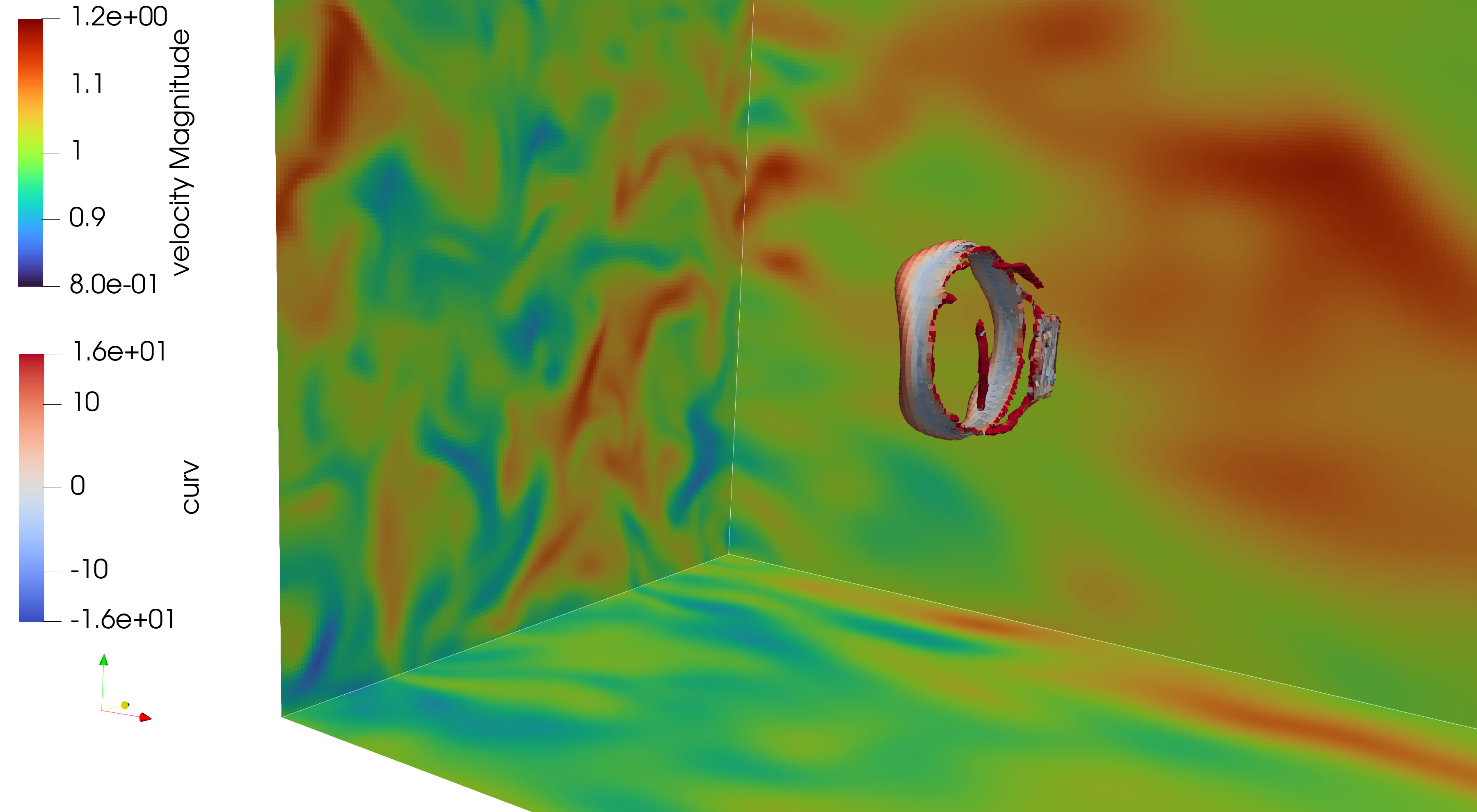}
  \caption{NN1 used in a case with a drop in a turbulent cross-flow undergoing numerical breakup. The velocity magnitude is shown at the domain boundaries.}
  \label{lig6}
\end{figure*}
To test PLIC-Net in a complex flow with catastrophic interface break-up, a drop in a cross-flow was finally considered. Using the cross-flow velocity, drop diameter, and gas properties as reference, the Reynolds number was set to 100 and the gas Weber was set number to 20. The liquid-to-gas viscosity and density ratios were 50 and 1000, respectively. The domain was set to be 20 diameters with 320 cells in the $x$ direction, 10 diameters with 160 cells in the $y$ direction, and 10 diameters with 160 cells in the $z$ direction. A liquid drop was initialized two diameters away from the left side of the domain (i.e., the $x$ minus domain boundary). A uniform cross-flow velocity is applied using Dirichlet on the left, a clipped Neumann outflow is applied on the right. The $y$ and $z$ directions use periodic conditions. A maximum time step of 0.04 was used with a maximum CFL number of 1, which included the capillary time step constraint. Cases were run with LVIRA, ELVIRA, and PLIC-Net until a non-dimensional time of 100.

Current research \citep{Han} demonstrates how two-plane reconstructions and physics-based break-up modeling can be successfully used for flow problems similar to this one to create results in agreement with experiments. However, this work considers only single-plane reconstructions without physics-based break-up modeling. Thus, numerical break-up dominates as soon as interfacial scales fall below the cell size. In this case, all of the reconstruction methods predict that as the drop is carried downstream, it flattens and then a thin bag begins to expand in the downstream direction. As the bag expands, it becomes thinner until numerical break-up is triggered. This results in a ligament forming from the rim of the bag and many other smaller drops and ligaments forming from the thinner part of the bag. Figure~\ref{lig1} shows the reconstructions with LVIRA, ELVIRA, NN1, NN2, and NN3. With LVIRA and ELVIRA, there are many spurious planes created during numerical bag break-up. LVIRA creates a field of smaller spurious planes near the center of the bag, while ELVIRA creates larger spurious planes that develop later than with LVIRA. In contrast, PLIC-Net once again generally exhibits much cleaner numerical break-up with few, if any, spurious planes. In agreement with the previous cases, the numerical surface tension controls how quickly break-up occurs, with NN3 breaking up with thicker films than NN2 or NN1. This higher numerical surface tension also leads to fewer droplets and ligaments downstream with NN3. NN1 and NN2 once again lead to similar reconstructions, although the lower numerical surface tension with NN2 is apparent in this case.

The work by \cite{Han} modified this case to have a turbulent cross-flow instead of a uniform flow. In order to prevent run-to-run variance caused by the turbulence when comparing results, no turbulence was considered in the above cases. However, PLIC-Net is capable of handling the addition of turbulence without any reduction in the quality of the reconstruction. While LVIRA and ELVIRA still produce many spurious planes in the turbulent case, PLIC-Net maintains its clean break-up without spurious planes. An example is shown with NN1 in Figure~\ref{lig6}. Overall, in both the uniform and turbulent inflow velocity cases, PLIC-Net performs very well and produces solutions that are comparable to the expected behavior from experiments \citep{Opfer}, except for the inability to capture the thin liquid film. These results demonstrate that PLIC-Net can be used in complex flow simulations as an alternative to LVIRA and ELVIRA. Thus, this work successfully establishes the groundwork to continue the development and deployment of neural networks in multiphase flow reconstructions.

\subsection{Computational Cost}
One of the potential benefits of using neural networks for interface reconstruction is a reduction in computational cost when compared to optimization-based techniques. However, the single-plane PLIC reconstruction performed by LVIRA or ELVIRA can be implemented to be very efficient, therefore it was not necessarily expected for PLIC-Net to achieve lower cost than these methods. In contrast, newer PLIC alternatives such as the Reconstruction with 2 Planes (R2P) \citep{Chiodi} and the Piecewise Parabolic Interface Calculation (PPIC) \citep{Evrard} are more complex and expensive, and could benefit greatly from a machine learning formulation. While the primary objective was to demonstrate the viability of machine learning approaches for interface reconstruction, this section provides a brief discussion of timing results.

The time spent performing the interface reconstruction for each mixed-phase cell was measured for the falling drop simulation discussed in Section 3.3.1. For these time comparisons, the simulations were run on a single core of an Intel i7-10750H CPU. The average CPU time for NN1 to reconstruct a plane was \SI{9.13e-6}{\second} with a standard deviation of \SI{4.64e-6}{\second}, while the average CPU time with LVIRA was \SI{1.04e-5}{\second} with a standard deviation of \SI{7.51e-6}{\second}. It must be noted that this implementation of LVIRA was highly optimized for computational efficiency. In fact, in this case LVIRA was found to be faster than ELVIRA. ELVIRA had an average CPU time of \SI{1.91e-5}{\second} with a standard deviation of \SI{6.42e-6}{\second}. Therefore, not only is the quality of the PLIC-Net reconstruction often superior to LVIRA and ELVIRA, but it also benefits from a slightly lower computational cost.

\section{Conclusion}
 
\label{Conclusion}

In this work, a machine learning approach called PLIC-Net was successfully used to reconstruct interfaces in multiphase flow simulations. A 3 layer neural network used the volume fractions and phasic barycenters in a $3\times3\times3$ cells stencil to predict the normal vector of a PLIC plane. PLIC-Net was made equivariant to reflections about the Cartesian planes by invoking symmetries with respect to the global phase barycenter in the stencil. The training data was constructed using the analytically-calculated volume moments of paraboloids clipping the cells, which allowed for two principal curvatures to be varied in the training data. When testing PLIC-Net in simulations, it was observed that the distribution of the principal curvatures in the training data is directly correlated with the apparent numerical surface tension of the reconstruction. Meanwhile, the error convergence with grid size in a sphere advection case was studied and it was found that PLIC-Net maintained the convergence order of the transport scheme for a wide range of resolution. Thus, despite not having a proof of convergence on its own, there was no significant impact on the overall convergence when using PLIC-Net. In fact, PLIC-Net's error was lower than LVIRA's error for most of the grid sizes considered. When deployed in more complex multiphase flow simulations, PLIC-Net limited the generation of spurious planes in the reconstruction, which are numerous with LVIRA and ELVIRA in many cases. The overall reconstructions were always at least comparable with LVIRA, and often were qualitatively smoother and displayed cleaner numerical break-up. In addition, the computational cost of PLIC-Net was found to be lower than LVIRA and ELVIRA. Therefore, not only are machine learning approaches viable and at least on par with traditional methods, they can offer notable benefits.

Future work will continue to determine the best distributions of curvatures in the training data, which has the potential to improve PLIC-Net's general performance. However, it must be noted that the curvature data's influence is mostly on a purely numerical process: the break-up of under-resolved PLIC interfaces. Although it is important to have a neural network that can accurately reconstruct to the limit of grid resolution, any single plane reconstructions at smaller scales are inherently inaccurate on their own. Thus, future work will also expand the machine learning approach to two-plane and paraboloid reconstructions and physics-based break-up modeling. It must also be noted that the architecture of the neural network itself was not varied during this work, and finding an ideal number of hidden layers, neurons per layer, and proper activation functions will be investigated in future work.

\section*{Acknowledgements}
F. Evrard was funded by the European Union's Horizon 2020 research and innovation program under the Marie Skłodowska-Curie Grant Agreement No. 101026017.

\end{document}